\documentclass[11pt]{article}
\usepackage[left=1.5cm,right=1.5cm]{geometry}
\usepackage{cite}
\usepackage[T1]{fontenc}
\usepackage{tikz-cd}
\usepackage[british]{babel}
\usepackage{layout}
\usepackage{amsmath}
\usepackage{amssymb,amstext, amsthm, simplewick, amsfonts}
\usepackage{framed}
\usepackage{amsfonts}
\usepackage{multirow}
\usepackage{color} 
\usepackage{braket}
\usepackage{multicol} 
\usepackage{epsfig}
\usepackage{cancel}
\usepackage{caption}
\usepackage{epsf}
\usepackage{feynmf} 
\usepackage{frontespizio}
\usepackage{rotating}
\usepackage{subfigure}
\usepackage{pstricks}
\usepackage{type1ec}
\usepackage{lettrine}
\usepackage{bbold}
\usepackage{calligra}
\usepackage{tikz}
\usepackage{subfigure}
\usepackage{slashed}
\usepackage{mathrsfs}
\usepackage{curve2e}
\usepackage{setspace}
\usepackage{indentfirst}
\usepackage{emptypage}
\usepackage{relsize}
\usepackage{mathrsfs}
\usepackage{stackengine}
\usepackage{calc}
\usepackage{hyperref}
\hypersetup{
	colorlinks,
	citecolor=green,
	filecolor=black,
	linkcolor=blue,
	urlcolor=black
}
\numberwithin{equation}{section} 
\makeatletter
\@addtoreset{equation}{section}
\makeatother
\newcommand{\beqa}{\begin{eqnarray}}
\newcommand{\eeqa}{\end{eqnarray}}

\newcommand{\sm}{\mathcal{S}}

\newcommand{\be}{\begin{equation}}
\newcommand{\ee}{\end{equation}}

\newcommand{\dpar}[2]{ \frac{\partial #1}{\partial #2}}
\newcommand{\dfun}[2]{ \frac{\delta #1}{\delta #2}}
\def\tor{\leftrightarrow}

 \newcommand{\beq}{\begin{equation}}
\newcommand{\eeq}{\end{equation}}
\newcommand{\figref}[1]{Fig.~\ref{#1}}			
\newcommand{\tabref}[1]{Tab.~\ref{#1}}			
\newcommand{\secref}[1]{Section~\ref{#1}}		
\newcommand{\appref}[1]{Appendix~\ref{#1}}		

\newcommand{\Tr}{\text{Tr}}

\usepackage{accents}
\makeatletter
\newcommand{\xLine}[2][]{\ext@arrow 0359\Rightarrowfill@{#1}{#2}}
\makeatother
\xdefinecolor{darkgreen}{RGB}{102, 204, 70}
\xdefinecolor{darkblue}{RGB}{0, 0, 153}
\usepackage{amssymb}
\usepackage{pifont}
\newcommand{\cmark}{\ding{51}}%
\newcommand{\xmark}{\ding{55}}%
\begin{document}
	\begin{center}
		\vspace{1.5cm}
		{\Large \bf Four-Point Functions of Gravitons and Conserved Currents of CFT in Momentum Space: Testing the Nonlocal Action with the TTJJ \\}
		\vspace{0.3cm}
		\vspace{1cm}
		{\bf $^{(1)}$Claudio Corian\`o, $^{(2)}$Matteo Maria Maglio and $^{(1)}$Riccardo Tommasi\\}
		\vspace{1cm}
		{\it  $^{(1)}$Dipartimento di Matematica e Fisica, Universit\`{a} del Salento \\
			and INFN Sezione di Lecce, Via Arnesano 73100 Lecce, Italy\\
		National Center for HPC, Big Data and Quantum Computing\\}
		\vspace{0.5cm}
		{\it  $^{(2)}$ Institute for Theoretical Physics (ITP), University of Heidelberg\\
			Philosophenweg 16, 69120 Heidelberg, Germany}

	\end{center}
	\begin{abstract}
We present an analysis of the perturbative realization of the $TTJJ$ correlator, with two stress energy tensors and two conserved currents,
using free field theory, integrating out conformal sectors in the quantum 
corrections. This allows to define, around flat space, an exact perturbative expansion of the complete anomaly effective action - up to 4-point functions -  whose predictions can be compared against those of the anomaly induced action. The latter is a variational solution of the conformal anomaly constraint at $d=4$ in the form of a nonlocal Wess-Zumino action. The renormalization procedure and the degeneracies of the tensor structures of this correlator are discussed, valid for a generic conformal field theory, deriving its anomalous trace Ward identities (WIs).
In this application, we also illustrate a general procedure that identifies the minimal number of tensorial structures and corresponding form factors for the $TTJJ$ and any $4$-point function. The approach is implemented for three, four and five dimensions, addressing the tensor degeneracies of the expansion in momentum space. 
We show that the renormalized $TTJJ$ can be split into two contributions, a non anomalous and an anomalous part, each separately conserved. The first satisfies ordinary trace WIs, while the second satisfies anomalous trace WIs. The result of the direct computation is compared against the expression of the same 4-point function derived from the nonlocal anomaly induced action. We show that the prediction for the anomalous part of the $TTJJ$ derived from such action, evaluated in two different conformal decompositions, the Riegert and Fradkin-Vilkovisky (FV) choices, differ from the anomaly part identified in the perturbative $TTJJ$, in the flat spacetime limit. The anomaly part of the correlator computed with the Riegert choice is affected by double poles, while the one computed with the FV choice does not satisfy the conservation WIs. We present the correct form of the expansion of the anomaly induced action at the second order in the metric perturbations around flat space that reproduces the perturbative result.

	\end{abstract}

	\newpage
	\section{Introduction}
The study of multi-point tensor correlators in conformal field theory (CFT) plays a significant role in several contexts, with applications to nonlocal cosmology, condensed matter theory and particle phenomenology. \\
Conformal symmetry describes the properties of systems at energy scales at which the dynamics of the correlation functions are constrained by the generators of the conformal group, whose number varies according to the spacetime dimensions $(d)$. In $d=4$, a conformally invariant theory is characterized by a partition function that satisfies all the 15 constraints coming from the conformal group $SO(2,4)$, modulo the presence of conformal anomalies. \\
Conformal anomalies \cite{Duff:1993wm} are associated, for a given CFT, with the need to renormalize the theory, 
according to a regularization procedure which, at least around a flat spacetime, can be implemented directly using ordinary Dimensional Regularization (DR).
Indeed, all the conformal constraints can be derived, for Lagrangian CFTs, by resorting to the formalism of the effective action via its functional expansion with respect to the external background metric \cite{Coriano:2017mux, Coriano:2021nvn}. In the pure gravity sector, its renormalization can be performed using only two counterterms: the Euler-Poincar\`e density ($E$) 
(Gauss-Bonnet) and the Weyl tensor squared $(C^2)$. Both terms will play an important role in our analysis. \\
All the quantum corrections and the conformal constraints can be generated by investigating such renormalized partition functions in the form of trace and special conformal Ward identities either in coordinate \cite{Osborn:1993cr, Erdmenger:1996yc}, or in momentum space \cite{Coriano:2013jba, Bzowski:2013sza} 
\cite{Bzowski:2020kfw,Bzowski:2017poo}. The breaking of scale invariance 
by the renormalized correlator is present in its traceless part. \\
The derivation of such constraints is general and formulated in curved spacetime, and subsequently can be specialized around flat Minkowski space. This approach bypasses the traditional operator formalism used in deriving the same constraints in the standard CFT literature.\\
The analysis around flat space allows to address quite clearly several issues left open by the renormalization procedure, such as the breaking of scale invariance due to the inclusion of a renormalization scale, which is missing in both local and nonlocal effective actions \cite{Coriano:2022ftl} of Wess-Zumino forms, commonly discussed in the anomaly literature. Scale violations are expected to be part of the complete effective action, which is at the centre of our work. However, they are not present in those actions identified as possible variational solutions of the anomaly constraints. These do not necessarily include the quantum corrections discussed in this work, that will be taken into account for the $TTJJ$ correlator.\\ 
While all the correlators of a certain CFT are of interest, some play a crucial role in the possible implications of such theories in realistic physical contexts. Conformal anomaly actions, investigated in their Weyl-invariant and anomaly terms, provide definitive information about the correct effective field theory description that results once a conformal sector is integrated out of a specific partition function. This procedure modifies the background gravitational metric in a very special way. The extra terms  
induce a gravitational backreaction that can be studied around flat space with great accuracy and quite explicitly in specific renormalization schemes \cite{Coriano:2022ftl}.

\subsection{Implications for topological matter and nonlocal cosmology}
Among all the correlators, those containing stress-energy tensors and classically conserved vector and chiral currents, besides the scalar CFT primaries, for the reasons mentioned above, are certainly among the most important ones. They can be studied in ordinary field theories of conformally coupled scalars, fermions and spin-1 gauge fields in free field theory realizations \cite{Giannotti:2008cv,Armillis:2009pq,Armillis:2010qk} \cite{Donoghue:2015xla,Donoghue:2015nba}. Such realizations match exactly, at least for those correlators containing stress energy tensors $(T)$ and conserved currents $(J)$, the general tensor structure predicted for them by the conformal Ward Identities (CWIs) of a CFT, with significant simplifications 
\cite{Coriano:2020ees,Coriano:2018bsy,Coriano:2018bbe}.
In condensed matter theory, they find application in the study of topological materials, such as Dirac and Weyl semimetals, where anomalies are expected to play an important role. Due to the presence of an effective linear dispersion relation in the band structure of such materials and to their topological character (see \cite{Chernodub:2017jcp,Chernodub:2021nff,Chernodub:2019tsx,Tutschku:2020rjq,Fruchart:2013tza,Arouca:2022psl}), chiral \cite{Landsteiner:2013sja, Mottola:2019nui} and conformal anomaly actions can be used to describe their response functions and anomalous transport under external sources. One link, for example, is provided by Luttinger's formula, that relates a thermal solicitation of such materials to an external gravitational field \cite{Luttinger:1964zz,Fruchart:2013tza}. \\
A second important sector in which these analyses play a role is in the physics of the early universe and the production of gravitational waves for defining special corrections to gravity which may either take the form of local scalar tensor theories or a nonlocal form (see for instance \cite{Capozziello:2021krv, Belgacem:2019lwx}).

\subsection{Momentum space analysis}  
The constraints imposed by the conformal group for $d >2$, up to 3-point functions, are sufficient to identify the correlation functions of a certain CFT only modulo few constants \cite{Osborn:1993cr}. The constants appearing in the general solutions of correlation functions, such as the $TTT$ or $TJJ$, are reproduced by combining a certain number of sectors with arbitrary particle multiplicities 
$(n_S,n_f,n_V)$ of scalars, fermions and gauge fields, respectively.\\
We recall that for higher point functions, CWIs do not predict the exact form of such correlators, since arbitrary functions of the conformal invariant ratios - which depend on their coordinate points - are part of their general expressions. This arbitrariness has also been discussed in momentum space \cite{Bzowski:2019kwd}, at least for scalar correlators.    
The analysis of tensor correlators is far more involved and has been formulated in momentum space for 3-point functions of $T$'s $J$'s and scalar operators in \cite{Bzowski:2013sza}. However, it remains valid also for $n$-point functions \cite{Coriano:2021nvn,Coriano:2019nkw} (see \cite{Bzowski:2020kfw,Caloro:2022zuy}), covering also the Minkowski signature \cite{Bautista:2019qxj, Gillioz:2019lgs,Gillioz:2018mto} as well as applications to cosmology \cite{Arkani-Hamed:2018kmz,Arkani-Hamed:2017fdk,Baumann:2020dch,Benincasa:2022gtd}.

We will also be using the same approach in our case, similarly to the case discussed in \cite{Coriano:2021nvn}, but working directly with a free field theory realization. \\
A lot of insight and essential information about the structure of correlation functions can be uncovered by working directly in momentum space \cite{Coriano:2020ees}. Indeed, one of the limitations of CFT in coordinate space is the difficulty of describing the implications of the conformal anomaly in a complete way and, in particular, the anomaly action. 
In this approach, the anomaly is introduced by hand, in each correlation function, by extending the solutions of the (non anomalous) CWI's with the addition of ultralocal terms. For the rest, the anomaly contributions are absent in any application based on the operator product expansion (OPE). The OPE is an operatorial expansion at short distances that necessarily avoids spacetime regions where all the points of a certain correlator coalesce. On the other end, the momentum space analysis allows us to derive the anomaly contributions of a certain correlator in a very natural way since integration over momentum space obviously also covers the contact regions in the external coordinate points of the correlator. 
	
\section{The partition function}
We start our discussion by defining the unrenormalized partition function of the theory $\mathcal{Z}_B(g)$,
identified by the bare functional 
\begin{equation}
	\label{partition}
	\mathcal{Z}_B(g)=\mathcal{N}\int D\chi\,e^{-S_0(g,\chi)},
\end{equation} 
where $\mathcal{N}$ is a normalization constant. We have denoted by $\chi$, a collection of conformal fields that in $d=4$ correspond to scalars, fermions. Our analysis is set in the Euclidean case. The transition to Minkowski space can be performed by a simple analytical continuation of the correlation functions, since we will be dealing with free field theory realizations. The bare effective action will be defined as 
\begin{equation}
	\label{defg}
	e^{-\mathcal{S}_B(g)}=\mathcal{Z}_B(g) \leftrightarrow \mathcal{S}_B(g)=-\log\mathcal{Z}_B(g). 
\end{equation}

Quantum matter fields are assumed to be in a conformal phase at $d=4$. 
The bare effective action $\mathcal{S}_B(g)$ includes all the multiple insertions of the stress energy tensor (pure gravity sector) and mixed graphs with photons and gravitons. 

As usual, $\mathcal{Z}_B(g)$, which is the semiclassical effective action (see for instance \cite{Shapiro:2008sf})
in the Feynman diagrammatic expansion, will contain both connected and disconnected graphs, while $\mathcal{S}_B(g)$ collects only connected graphs. In the gravitational sector, the expansion provides all the pure graviton vertices, the mixed graviton/gauge and pure gauge vertices, defined by the insertions of the stress energy tensor and of the gauge current $J^\mu$. For instance, the quantum averages of $1$-point functions are defined as
\begin{equation}
	\label{anomx0}
	\langle T^{\mu\nu}\rangle = \frac{2}{\sqrt{-g}}\frac{\delta \sm_B}{\delta g_{\mu\nu}}
	\qquad \langle J^{\mu}\rangle = \frac{1}{\sqrt{-g}}\frac{\delta \sm_B}{\delta A_{\mu}},
\end{equation} 
where the metric is taken to be flat after the variation.  
Similarly, correlation functions of higher order are defined in a metric background $\bar{g}$ and 
with a vanishing gauge field $A_\mu$, by varying both external fields 
\begin{align}
\label{exps2}
\sm(g)_B &\equiv\sm(\bar{g})_B+\sum_{n=1}^\infty \frac{1}{2^n n!} \int d^d x_1\ldots d^d x_n \sqrt{-g_1}\ldots \sqrt{-g_n}\,\langle T^{\mu_1\nu_1}\ldots \,T^{\mu_n\nu_n}\rangle_{\bar{g} B}\delta g_{\mu_1\nu_1}(x_1)\ldots \delta g_{\mu_n\nu_n}(x_n),\notag\\
& + \sum_{n,k =1}^\infty \frac{1}{2^{n} n! k!} \int d^d x_1\ldots d^d x_n d^d x_{n+1}\ldots d^d x_{n+k}\sqrt{-g_1}\ldots \sqrt{-g_{n+k}}\,\langle T^{\mu_1\nu_1}\ldots \,T^{\mu_n\nu_n}J^{\mu_{n+1}}\ldots J^{\mu_{n+k}}\rangle_{\bar{g} B}\times \notag \\
&\times \delta g_{\mu_1\nu_1}(x_1)\ldots \delta g_{\mu_n\nu_n}(x_n)\delta A_{\mu_{n+1}}(x_{n+1}) 
\ldots \delta A_{\mu_{n+k}}(x_{n+k}) +\ldots
\end{align}
where the dots refer to contributions from the  pure gauge sector in the expansion. The covariant normalization of the correlation functions is given by
\begin{equation}
	\langle T^{\mu_1\nu_1}(x_1)\ldots T^{\mu_n\nu_n}(x_n)\rangle \equiv\frac{2}{\sqrt{-g_1}}\ldots \frac{2}{\sqrt{-g_n}}\frac{\delta^n \sm_B(g)}{\delta g_{\mu_1\nu_1}(x_1)\delta g_{\mu_2\nu_2}(x_2)\ldots \delta g_{\mu_n\nu_n}(x_n)} 
\end{equation}
for the n-graviton sector, with
with $\sqrt{-g_1}\equiv \sqrt{-|\textrm{det} \, g_{{\mu_1 \nu_1}}(x_1)|} $ and so on, and by 
\begin{align}
&\langle T^{\mu_1\nu_1}(x_1)\ldots T^{\mu_n\nu_n}(x_n)J^{\mu_{n+1}}(x_{n+1})\ldots J^{\mu_{n+k}}(x_{n+k})\rangle  \equiv \notag\\
&\equiv\frac{2}{\sqrt{-g_1}}\ldots \frac{2}{\sqrt{-g_n}}\frac{1}{\sqrt{-g_{n+1}}}\ldots \frac{1}{\sqrt{-g_{n+k}}}\frac{\delta^n \sm_B(g)}{\delta g_{\mu_1\nu_1}(x_1)\ldots \delta g_{\mu_n\nu_n}(x_n)\delta A_{\mu_{n+1}}(x_{n+1})\ldots \delta A_{\mu_{n+k}}(x_{n+k})}
\end{align}
for the graviton/gauge sector. 

Diagrammatically, the pure gravitational sector is identified, in free field theory realizations, by an infinite sum of 1-loop diagrams with an arbitrary number of external graviton lines. 
The mixed sector will include the $TTJ$, the $TJJ$ and the $TTJJ$ correlation functions.  
The diagrammatic expansion of the $\braket{TTJJ}$ for the fermionic and scalar cases are given below in \figref{DiagramF} and  \figref{DiagramS}. In Dimensional Regularization (DR) the renormalized effective action is defined by the inclusion of three counterterms 
\begin{equation}
	{Z}_R(g)=\, \mathcal{N}\int D\Phi\,e^{-S_0(g,\chi) + \frac{b' }{\epsilon}V_E(g,d) +  \frac{b}{\epsilon}V_{C^2}(g,d) - S_{count}(g,A)},
\end{equation} 
where $\mathcal{N}$ is a normalization constant, $\epsilon=d-4$, and 
\begin{equation}
\begin{split}
	V_{C^2}(g, d)\equiv & \mu^{\varepsilon}\int\,d^dx\,\sqrt{-g}\, C^2,  \\
	V_{E}(g,d)\equiv &\mu^{\varepsilon} \int\,d^dx\,\sqrt{-g}\,E , 
	\end{split}	\label{ffr}
\end{equation}
are the counterterms corresponding to the Gauss-Bonnet  
\begin{equation}
\label{GB1}
E = R^2 - 4 R^{\mu \nu} R_{\mu \nu} + R^{\mu \nu \rho \sigma} R_{\mu \nu \rho \sigma} 
\end{equation}
and Weyl tensor squared densities 
\begin{equation}
C^{(d)}_{\alpha\beta\gamma\delta} = R_{\alpha\beta\gamma\delta} -
\frac{1}{d-2}( g_{\alpha\gamma} \, R_{\delta\beta} + g_{\alpha\delta} \, R_{\gamma\beta}
- g_{\beta\gamma} \, R_{\delta\alpha} - g_{\beta\delta} \, R_{\gamma\alpha} ) +
\frac{1}{(d-1)(d-2)} \, ( g_{\alpha\gamma} \, g_{\delta\beta} - g_{\alpha\delta} \, g_{\gamma\beta}) R.\, 
\end{equation}
and the Weyl terms respectively. 
In order to remove these divergences of the mixed graviton/gauge correlators we add to the action the counterterm 
\begin{align}
	S_{count}(g,A)\equiv -\frac{1}{\varepsilon}V_{F^2}(g, d)\equiv -\frac{\mu^{\varepsilon}}{\varepsilon}\sum_{I=f,s}\,n_I\int d^dx\,\sqrt{-g}\left(\beta_c(I)\,F^2\right),
\end{align}
corresponding to the field strength $F^2=F^{\mu\nu}F_{\mu\nu}$ where the coefficients $\beta_c(I)$ refer to the scalar and fermion contributions.

\subsection{Local and nonlocal actions and the $TTJJ$ test}

Most of the analysis of anomaly actions in the literature deals with the problem of the identification of a functional whose variation with respect to the metric generates the conformal anomaly. The correlation functions extracted from the renormalized action satisfy hierarchical CWIs that allow the identification of two contributions, denoted as the Weyl variant and the Weyl invariant parts, respectively.\\
As just mentioned, in the case of pure graviton vertices, the Weyl variant part of the renormalized effective action is related to the two counterterms $V_E$ and $V_{C^2}$, while in the $TTJJ$ a third counterterm, $V_{F^2}$, is needed. Their variations reproduce the anomaly since 
\begin{align}
2 g_{\mu\nu}\frac{\delta}{\delta g_{\mu\nu}}V_{E}(g,d)&=\epsilon \sqrt{g} E  \notag\\
2 g_{\mu\nu}\frac{\delta}{\delta g_{\mu\nu}}V_{C^2}(g,d)&=\epsilon \sqrt{g} C^2 \notag \\
2 g_{\mu\nu}\frac{\delta}{\delta g_{\mu\nu}}V_{F^2}(g,d)&=\epsilon \sqrt{g} F^2,
\end{align}
while the bare effective action, corresponding to the Weyl invariant part, in $d$ dimensions satisfies the condition
\begin{equation}
2 g_{\mu\nu}\frac{\delta}{\delta g_{\mu\nu}}\sm_B(g,d)=0.
\end{equation}
The Weyl variant part can be summarized by the functional
 \begin{equation}
\sm_{v}\equiv\frac{\beta_C}{\epsilon}V_{F^2}(g,d)+ \frac{b'}{\epsilon}V_E(g,d) + \frac{b}{\epsilon}V_{C^2}(g,d),
\end{equation}
\textit{i.e.} the counterterm action (with $\beta_C=\sum_{I=f,s}\,n_I\beta_c(I)$) and this separation between $\sm_B$ and $\sm_v$, with 
 \begin{equation}
\label{decomp2}
\sm_R=\sm_B +\sm_v,
\end{equation}
is perfectly well defined as far as $d\neq 4$. $\sm_B$ becomes singular at $d=4$ and the renormalization procedure consists in expanding around four spacetime dimensions both $\sm_B$ and the counterterm. The expansion is performed using a fiducial metric $\bar{g}$ and its fluctuations, with $\bar{g}$ taken to be the Minkowski metric. Using the fact that the singularities of $\sm_B$ are removed by the singular parts of the counterterms, 
$\sm_R$ gets effectively re-expressed in the form 
\begin{align}
\sm_R(4)&=\lim_{d\to 4}\left(\sm_B(g,d) +\frac{b'}{\epsilon}V_E(g,d) + \frac{b}{\epsilon}V_{C^2}(g,d) + \frac{\beta_C}{\epsilon} V_{F^2}(g,d)\right)
\nonumber \\
&=\sm_f(4) +b' \,V'_E(\bar g,\phi, 4) + b\, V'_{C^2}(\bar g,\phi, 4) + \beta_C\,V'_{F^2}(\bar g,\phi, 4) \label{ren1}
\end{align}
with 
\begin{equation}
\label{ps}
V'(\bar g,\phi,4)= \lim_{d\to 4}\left(\frac{1}{\epsilon}\left(V(g, d)- V(\bar g, 4)\right)\right)= \lim_{d\to 4}\left(\frac{1}{\epsilon}\left(V(g, d)- V(\bar g, d)\right)\right)
\end{equation}
and the finite contribution coming from the loops contained in $\sm_f$
\begin{equation}
\sm_f(4) = \lim_{d\to 4}\left(\sm_B(d) +\frac{b'}{\epsilon}V_E(g,4) + \frac{b}{\epsilon}V_{C^2}(g,4)+ \frac{\beta_C}{\epsilon}V_{F^2}(g,4)\right).
\end{equation}
The anomaly action generated by this regularization can then be defined in the form 
\begin{equation}
\label{SA}
\sm_A=b' \,V'_{E}(\bar g,\phi,4) + b\,V'_{C^2}(\bar g,\phi,4) + \beta_C \,V'_{F^2}(\bar g,\phi,4),
\end{equation}
In standard approaches in which one tries to solve the constraint
 \begin{equation}
g_{\mu\nu}\frac{\delta \mathcal{S}_R}{\delta g_{\mu\nu}}=\frac{\sqrt{g}}{2}\left[b C^2+b' E+ \beta_C F^2\right],
 \end{equation}
ignoring the renormalization process implicit in the extraction of the effective action at $d=4$, which requires to take a singular limit, the classification of $\sm_R$ in terms of the two parts does not introduce any extra scale. \\
Obviously, this simply implies that  
we are focusing only on $\sm_v$ in \eqref{decomp2}, neglecting, at the same time, the presence of extra scales generated by the virtual quantum corrections, due to the renormalization procedure.
The idea of using the realizations of free field theory in flat space, as already mentioned, allows us to have a firm grip on the
structure of the expansion, although this is only possible for a simpler background, compared to a curved one
For these reasons, it is not surprising that a functional solution of the anomaly constraint, 
which corresponds to an anomaly induced action, may not reproduce the perturbative result and the Ward identities that come with it. \\
As we are going to show by an explicit computation, there is a perfect agreement between the complete effective action derived by a free field theory realization and the anomaly induced actions that we are going to discuss next, up to 3-point functions. The two actions are derived by selecting two different conformal decompositions, usually termed "gauge choices" in the literature, in which the dilaton field is expressed in terms of the metric $g$ by two different functional constraints. In our example, they correspond to the Riegert 
$(\Sigma_R)$ and the Fradkin-Vilkovisky $(\Sigma_{FV})$ choices. 
We are going to investigate this point performing a direct computation on the two actions, showing that as we move to 4-point functions, the correlators do not satisfy the 
hierarchical Ward identities of the case.

\section{Ward Identities}
The symmetry constraints on $\sm_R$, induced on the coefficients of the expansion \eqref{exps2} take the form of WIs which are hierarchical. The conformal constraints, for instants, are linked to the Weyl 
invariance of the renormalized effective action and to its breaking and a derivation of the corresponding WIs can be performed directly from either $\sm_B$ or $\sm_R$, as shown in \cite{Coriano:2017mux,Coriano:2021nvn}.
We recall that in a curved background, for a certain action $\sm(g)$, Weyl invariance is expressed as a symmetry of the form 
\begin{equation}
\sm(g)=\sm(\bar{g}) \qquad  \textrm{when}\qquad  g_{\mu\nu}=\bar{g}_{\mu\nu} e^{2 \phi}.
\end{equation}
The relation between $g$ and $\bar{g}$ defines a conformal decomposition, which remains valid under the gauge transformation  
\begin{equation}
\bar g \to \bar g\,e^{2 \sigma},\qquad   \phi\to \phi - \sigma,
\end{equation}
where $\sigma(x)$ is a local shift. The renormalization of the quantum corrections, via the counterterms above, breaks this symmetry. In the case of a flat background, one is essentially performing the $\phi\to 0$ limit of 
$\sm_R$ after performing the metric variations, with the dilaton variation $\frac{\delta}{\delta \phi}$ replaced 
by $2 g_{\mu\nu}\frac{\delta}{\delta g_{\mu\nu}}$. A study of the semiclassical effective action in the presence of a dilaton background is in \cite{Asorey:2022ebz}.
In general, on the bare functional $\sm_B(g)$, one derives the relation 
\begin{equation}
\label{anomx}
\frac{\delta \sm_B}{\delta \phi(x)}=\sqrt{-g} \,g_{\mu\nu}\,\langle T^{\mu\nu}\rangle, 
\qquad 
\end{equation} 
and its invariance under Weyl  \begin{equation}
\label{www}
\delta_\phi g_{\mu\nu}= 2  g_{\mu\nu} \delta\phi,
\end{equation}
and diffeomorphisms   
\begin{equation} 
\delta_\epsilon g_{\mu\nu}=-\nabla_{\mu}\epsilon_{\nu}- \nabla_{\nu}\epsilon_{\mu}, 
\end{equation}
are summarised by the constraints
\begin{equation} 
\label{eww}
\delta_\phi \sm_B=0 \qquad \delta_\epsilon \sm_B=0,
\end{equation}
leading to trace and conservation conditions of the quantum averages of $T^{\mu\nu}$
 \begin{equation}
\label{comby}
\langle T^\mu_\mu\rangle=0 \qquad \qquad \nabla_\mu\langle T^{\mu\nu}\rangle=0.
\end{equation}
Ordinary trace and conservation WI's can be derived from the equations above by functional differentiations of $\sm_B(g)$ with respect to the background metric. As far as we stay away from $d=4$ and include in the classical action $\sm_0$ conformal fields, we have exact CWI which are derived from the condition of invariance of the generating functional 
$\sm_B$ with respect to diffeomorphisms and Weyl transformations. Anomalous CWIs are 
derived by replacing the effective action $\sm_B$ with the renormalized one $\sm_R$. Non conformal sectors, such as spin-1 contributions, modify the CWIs by inhomogeneous terms unrelated to the anomaly, which is a pure $4d$ phenomenon.

	We move to discuss the derivation of the conformal and conservation WIs for the correlator. This allows us to 
	illustrate its decomposition, following the approach of \cite{Bzowski:2013sza}, in terms of a transverse traceless sector, a longitudinal sector and a trace sector. Only the trace and conservation WIs will play a role in our analysis.   
	
	Assuming that the generating functional of the theory is invariant under the action of some symmetry groups, then the correlation function $\braket{TTJJ}$ satisfies
	\begin{equation}
	\sum_{j=1}^4\,G_g(x_j)\braket{T^{\mu_1\nu_1}(x_1)T^{\mu_2\nu_2}(x_2)J^{\mu_3}(x_3)J^{\mu_4}(x_4)}=0,
	\end{equation}
	with $G_g$ the generators of the infinitesimal symmetry transformations. These constraints come from the invariance of the generating functional under symmetry transformations 
	\begin{equation}
	\sm_B[g',A']=\sm_B[g,A],
	\end{equation}
	that can be expressed equivalently as
	\begin{equation}
		\int\,d^dx\left(\frac{\delta\sm_B}{\delta g_{\mu\nu}}\delta g_{\mu\nu}+\frac{\delta\sm_B}{\delta A_\mu^a}\delta A^a_\mu\right)=0.\label{invW}
	\end{equation}
	
	Among these constraints, the conservation Ward Identity (WI) in flat space of the energy momentum tensor can be obtained by requiring the invariance of $\sm_B[g,A]$ under diffeomorphisms $x^\mu\to x^\mu+\epsilon^\mu(x)$ for which the variation of the metric and the gauge fields are the corresponding Lie derivatives. In the case of a nonabelian $SU(N)$ gauge field $A_{\mu}^a$ ($a=1,2,\ldots, N^2-1)$, for instance, we get 	\begin{align}
	\delta A_\mu^a&=-\epsilon^\alpha\nabla_\alpha A^a_\mu-A_\alpha^a\nabla_\mu\epsilon^\alpha,\\
	\delta g_{\mu\nu}&=-\nabla_\mu\epsilon_\nu-\nabla_\nu\epsilon_\mu.
	\end{align}
	Inserting these variations into \eqref{invW} and integrating by parts, we obtain the conservation WI
	\begin{align}
	\nabla_\mu\braket{T^{\mu\nu}}+\left(\partial^\mu A^{a\nu}-\partial^\nu A^{a\mu}\right)\braket{J^a_\mu}+A^{a\nu}\,\nabla_\mu\braket{J^{a\mu}}=0.\label{cons1}
	\end{align}
	Similarly, the requirement of invariance under a gauge transformation with a parameter $\theta^a(x)$ gives
	\begin{align}
		\delta A_\mu^a&=\partial_\mu\theta^a+g\,f^{abc}A_\mu^b\theta^c,\\
		\delta g_{\mu\nu}&=0,
	\end{align}
	and the invariance of the generating functional under gauge transformations gives
	\begin{align}
	\nabla_\mu\braket{J^{a\mu}}=gf^{abc}A_\mu^c\braket{J^{b\,\mu}}. 
	\end{align}
Inserting this equation into \eqref{cons1} we obtain the conservation WI's
\begin{subequations}
\begin{align}
\nabla_\mu\braket{T^{\mu\nu}}+F^{a\,\mu\nu}\braket{J^a_\mu}&=0,\\
\nabla_\mu\braket{J^{a\mu}}-gf^{abc}A_\mu^c\braket{J^{b\,\mu}}&=0. 
\end{align}
\end{subequations}
In the Abelian case, which is the case of our interest, diffeomorphism and gauge invariance give the relations
\begin{subequations}
	\begin{align}
	&\nabla_\mu\braket{T^{\mu\nu}}+F_{\mu\nu}\braket{J^{\mu}}=0,\\
	&\nabla_\mu\braket{J^{\mu}}=0. 
\end{align}\label{consWI}
\end{subequations}
By functional differentiations of \eqref{consWI} we derive ordinary WIs for the various correlators involving energy-momentum tensors and conserved currents. In the $\braket{TTJJ}$ case we obtain, after a Fourier transform, the conservation equation
	\begin{align} 
	& {p_1}_{\mu_1} \braket{T^{\mu_1 \nu_1} (p_1) T^{\mu_2 \nu_2} (p_2) J^{\mu_3} (p_3) J^{\mu_4} (p_4)}  \notag\\
	& 
	=\bigg[2\ {p_2}_{\lambda_1} \delta^{\nu_1 (\mu_2} \braket{T^{\nu_2) \lambda_1} (p_1+p_2) J^{\mu_3} (p_3) J^{\mu_4} (p_4)}  - 
	{p_2}^{\nu_1}  \braket{T^{\mu_2 \nu_2} (p_1+p_2) J^{\mu_3} (p_3) J^{\mu_4} (p_4)} \bigg] \notag\\ 
	&
	+ 2 \bigg\{  \bigg[ \delta^{\nu_1 (\mu_2} p_3^{\nu_2)}  \braket{J^{\mu_3} (p_1+p_2+p_3) J^{\mu_4} (p_4)} - \delta^{\nu_1 (\mu_2} \delta^{\nu_2) \mu_3}{p_3}_{\lambda_1 } \braket{J^{\lambda_1} (p_1+p_2+p_3) J^{\mu_4} (p_4)} 
	\notag\\ 
	&\qquad + \, \frac{1}{2}\delta^{\mu_3 \nu_1} p_{3\lambda_1} \braket{J^{\lambda_1}(p_1+p_3) T^{\mu_2 \nu_2}(p_2) J^{\mu_4}(p_4) }- \frac{1}{2}p_3^{\nu_1 } \braket{J^{\mu_3}(p_1+p_3) T^{\mu_2 \nu_2}(p_2) J^{\mu_4}(p_4) } \bigg]+\bigg[ (3 \tor 4) \bigg]\bigg\} ,\label{Cons1}
\end{align}
where the notation $(3\leftrightarrow 4)$ means the exchange of the subscript $3$ with $4$, and the vector current Ward identities
	\begin{equation} 
		p_{i\,\mu_i} \braket{T^{\mu_1 \nu_1} (p_1) T^{\mu_2 \nu_2} (p_2) J^{\mu_3} (p_3)  J^{\mu_4} (p_4)} = 0, \qquad i=3,4.\label{Cons2}
	\end{equation}
	In our conventions, all the momenta, in a given correlator, are incoming.
Furthermore we consider the invariance of the generating functional under Weyl transformations for which the fields transform as in \eqref{www} and
\begin{align}
\delta_\sigma\,A_\mu^a&=0
\end{align}
giving the naive trace Ward identity 
\begin{align}
g_{\mu\nu}\braket{T^{\mu\nu}}=0. \label{traceNaive}
\end{align}
The functional differentiation of \eqref{traceNaive} gives the (non-anomalous) condition
\begin{align}
\delta_{\mu_1 \nu_1} \braket{T^{\mu_1 \nu_1} (p_1) T^{\mu_2 \nu_2} (p_2) J^{\mu_3} (p_3) J^{\mu_4} (p_4) } = -2 \, \braket{T^{\mu_2 \nu_2} (p_1+p_2) J^{\mu_3} (p_3) J^{\mu_4} (p_4) },\label{Cons3}
\end{align}
that is preserved for $d\ne2n$, $n\in\mathbb{N}$. In \appref{AppendixA} we offer more details on the conservation WI's. 

	\section{Decomposition of the correlator}\label{decomp}
As already mentioned, the general form of the $\braket{TTJJ}$ correlator can be constructed by a decomposition into transverse,  longitudinal and trace terms \cite{Bzowski:2013sza}, exploiting its symmetries. 
We start by decomposing the operators $T$ and $J$ in terms of their transverse traceless part and the longitudinal (local) ones 
	\begin{align}
		T^{\mu_i\nu_i}(p_i)&\equiv t^{\mu_i\nu_i}(p_i)+t_{loc}^{\mu_i\nu_i}(p_i),\label{decT}\\
		J^{\mu_i}(p_i)&\equiv j^{\mu_i}(p_i)+j_{loc}^{\mu_i}(p_i),\label{decJ}
	\end{align}
	where
	\begin{align}
		\label{loct}
		t^{\mu_i\nu_i}(p_i)&=\Pi^{\mu_i\nu_i}_{\alpha_i\beta_i}(p_i)\,T^{\alpha_i \beta_i}(p_i), &&t_{loc}^{\mu_i\nu_i}(p_i)=\Sigma^{\mu_i\nu_i}_{\alpha_i\beta_i}(p)\,T^{\alpha_i \beta_i}(p_i),\\
		j^{\mu_i}(p_i)&=\pi^{\mu_i}_{\alpha_i}(p_i)\,J^{\alpha_i }(p_i), &&\hspace{1ex}j_{loc}^{\mu_i}(p_i)=\frac{p_i^{\mu_i}\,p_{i\,\alpha_i}}{p_i^2}\,J^{\alpha_i}(p_i).
	\end{align}
having introduced the transverse-traceless ($\Pi$), transverse $(\pi)$, longitudinal ($\Sigma$) projectors, given respectively by 
\begin{align}
	\label{prozero}
	\pi^{\mu}_{\alpha} & = \delta^{\mu}_{\alpha} - \frac{p^{\mu} p_{\alpha}}{p^2}, \\
	\Pi^{\mu \nu}_{\alpha \beta} & = \frac{1}{2} \left( \pi^{\mu}_{\alpha} \pi^{\nu}_{\beta} + \pi^{\mu}_{\beta} \pi^{\nu}_{\alpha} \right) - \frac{1}{d - 1} \pi^{\mu \nu}\pi_{\alpha \beta}\label{TTproj}, \\
	\Sigma^{\mu_i\nu_i}_{\alpha_i\beta_i}&=\frac{p_{i\,\beta_i}}{p_i^2}\Big[2\delta^{(\nu_i}_{\alpha_i}p_i^{\mu_i)}-\frac{p_{i\alpha_i}}{(d-1)}\left(\delta^{\mu_i\nu_i}+(d-2)\frac{p_i^{\mu_i}p_i^{\nu_i}}{p_i^2}\right)\Big]+\frac{\pi^{\mu_i\nu_i}(p_i)}{(d-1)}\delta_{\alpha_i\beta_i}\equiv\mathcal{I}^{\mu_i\nu_i}_{\alpha_i}p_{i\,\beta_i} +\frac{\pi^{\mu_i\nu_i}(p_i)}{(d-1)}\delta_{\alpha_i\beta_i}\label{Lproj}.
\end{align}
By using the projectors introduced above, the correlator can be written as
\begin{align}
	&\braket{ T^{\mu_1 \nu_1} (p_1) T^{\mu_2 \nu_2} (p_2) J^{\mu_3} (p_3) J^{\mu_4} (p_4) } \notag\\[1ex]
	&\hspace{1cm}=\braket{ t^{\mu_1 \nu_1} (p_1) t^{\mu_2 \nu_2} (p_2) j^{\mu_3} (p_3) j^{\mu_4} (p_4) } +	\braket{ t^{\mu_1 \nu_1} (p_1) t^{\mu_2 \nu_2} (p_2) j^{\mu_3} (p_3) j^{\mu_4} (p_4) } _{loc}\label{decTTJJ}
\end{align}
where
\begin{align}
\delta_{\mu_i\nu_i}\braket{ t^{\mu_1 \nu_1} (p_1) t^{\mu_2 \nu_2} (p_2) j^{\mu_3} (p_3) j^{\mu_4} (p_4) } &=0,\quad i=1,2\,,\label{tracel}\\
p_{\mu_i}\braket{ t^{\mu_1 \nu_1} (p_1) t^{\mu_2 \nu_2} (p_2) j^{\mu_3} (p_3) j^{\mu_4} (p_4) } &=0,\quad i=1,\dots,4\,\label{transverse},
\end{align}
and 
\begin{align}
&\braket{ t^{\mu_1 \nu_1} (p_1) t^{\mu_2 \nu_2} (p_2) j^{\mu_3} (p_3) j^{\mu_4} (p_4) } _{loc}=\braket{ t_{loc}^{\mu_1 \nu_1}T^{\mu_2 \nu_2} J^{\mu_3} J^{\mu_4}  } +\braket{ T^{\mu_1 \nu_1}  t_{loc}^{\mu_2 \nu_2}  J^{\mu_3} J^{\mu_4} } -\braket{ t_{loc}^{\mu_1 \nu_1} t_{loc}^{\mu_2 \nu_2}J^{\mu_3} J^{\mu_4}}\notag\\[1ex]
&= \bigg[\left(\mathcal{I}^{\mu_1\nu_1}_{\alpha_1}p_{1\,\beta_1} +\frac{\pi^{\mu_1\nu_1}(p_1)}{(d-1)}\delta_{\alpha_1\beta_1}\right)\delta^{\mu_2}_{\alpha_2}\delta^{\nu_2}_{\beta_2}+\left(\mathcal{I}^{\mu_2\nu_2}_{\alpha_2}p_{2\,\beta_2} +\frac{\pi^{\mu_2\nu_2}(p_2)}{(d-1)}\delta_{\alpha_2\beta_2}\right)\delta^{\mu_1}_{\alpha_1}\delta^{\nu_1}_{\beta_1}\notag\\
&\hspace{1cm}-\left(\mathcal{I}^{\mu_1\nu_1}_{\alpha_1}p_{1\,\beta_1} +\frac{\pi^{\mu_1\nu_1}(p_1)}{(d-1)}\delta_{\alpha_1\beta_1}\right)\left(\mathcal{I}^{\mu_2\nu_2}_{\alpha_2}p_{2\,\beta_2} +\frac{\pi^{\mu_2\nu_2}(p_2)}{(d-1)}\delta_{\alpha_2\beta_2}\right)\bigg]\,\braket{T^{\alpha_1\beta_1}T^{\alpha_2\beta_2}J^{\mu_3}J^{\mu_4}}.\label{loc}
\end{align}
It is clear now that only the second term in \eqref{decTTJJ}, expressed explicitly in \eqref{loc}, will contain the entire trace and longitudinal contributions, with these two sectors constrained by the conservation WIs \eqref{Cons1}, \eqref{Cons2} and \eqref{Cons3}. Thus, the unknown part of the correlator is contained in its transverse-traceless ($ttjj$), since the remaining longitudinal + trace contributions - the local terms - are related to lower point functions by conservation and trace WIs. Therefore, we can proceed by studying the general decomposition of the transverse-traceless part $\braket{ ttjj }$ into a product of form factors and tensor structures. \\
Due to conditions \eqref{tracel} and \eqref{transverse}, such sector, using the transverse and traceless projectors, can be written in the form
	\begin{equation} 
		\braket{ t^{\mu_1 \nu_1} (p_1) t^{\mu_2 \nu_2} (p_2) j^{\mu_3} (p_3) j^{\mu_4} (p_4) } =\Pi^{\mu_1 \nu_1}_{\alpha_1 \beta_1} (p_1)  \Pi^{\mu_2 \nu_2}_{\alpha_2 \beta_2} (p_2) \pi^{\mu_3}_{\alpha_3} (p_3) \pi^{\mu_4}_{\alpha_4} (p_4) X^{\alpha_1 \beta_1 \alpha_2 \beta_2 \alpha_3\alpha_4},\label{ttjj}
	\end{equation}
where $X^{\alpha_1\dots\alpha_4}$ is a general rank six tensor built out of products of metric tensors and momenta with the appropriate selection of indices. Note that due to the presence of the projectors in \eqref{ttjj}, the terms $\delta^{\alpha_i\beta_i}$, $i=1,2$ or $p_i^{\alpha_i}$, $i=1,\dots,4$ cannot be used as fundamental tensors and vectors to build the $X^{\alpha_1\dots\alpha_4}$ tensor. In addition, the conservation of the total momentum
	\begin{equation} 
		p_1^{\alpha_i} + p_2^{\alpha_i} + p_3^{\alpha_i} + p_4^{\alpha_i} = 0 ,
	\end{equation}
	allows to select for each index $\alpha_i$ a pair of momenta out of the four, to be used in the general construction of $X$. The choice of the independent momenta of the expansion, similarly to the case of 3-point functions discussed in \cite{Bzowski:2013sza}, can be different for each set of contracted tensor indices. We will choose   
	\begin{equation}
		\begin{split}
			&\{(\alpha_1,\beta_1),(\alpha_2,\beta_2)\}\leftrightarrow p_3,p_4,\\
			&\{\alpha_3,\alpha_4\}\leftrightarrow p_1,p_2\,,
		\end{split}\label{choicemom}
	\end{equation}
	as a basis of the expansion, for each pair of indices shown above. Once the decomposition has been performed according to this scheme and the number of form factors identified, one momentum, for instance $p_4$, can be chosen as the dependent one. \\
	This approach is rather economical, since it allows to reduce the number of form factors to a minimum, exploiting the presence of a single $tt$ projector for each external momentum. Regarding the tensor structures built out of metric $\delta$s, the only non vanishing ones appearing in $X^{\alpha_1\dots\alpha_4}$ are
	\begin{align}
	\delta^{\alpha_1\alpha_2},\ \delta^{\alpha_1\alpha_3},\ \delta^{\alpha_1\alpha_4},\ \delta^{\alpha_2\alpha_3},\ \delta^{\alpha_2\alpha_4},\ \delta^{\alpha_3\alpha_4}\label{choicemetr}
	\end{align}
	together with the similar ones obtained by the exchange $\alpha_i\leftrightarrow\beta_i$, $i=1,2$. This strategy has been introduced in \cite{Bzowski:2013sza} for 3-point functions and applied also to the case of 
$4$-point functions in \cite{Coriano:2021nvn}. In the next section, we are going to describe the procedure in order to write explicitly the expression of $X^{\alpha_1\dots\alpha_4}$ in terms of the minimal number of tensor structures and form factors, in general $d$ dimensions \cite{Coriano:2019nkw}.

	\subsection{Orbits of the permutations\label{OrbitsSec}}
$X^{\alpha_1\dots\alpha_4}$ is expressed in terms of tensor structures and form factors using the symmetry of the correlator. The $\braket{TTJJ}$ manifests two types of discrete symmetries related to the permutation group: it must be symmetric under the exchange of the two gravitons ($1 \tor 2$), of the two conserved $J$ currents ($3 \tor 4$), and the combination of both transformations. We label such transformations respectively as $P_{12}$, $P_{34}$ and $P_C=P_{12}P_{34}$.  It is worth mentioning that $P_{12}$ exchanges the pair of indices of the two gravitons and the momenta associated with them, and analogously for the two currents $J$'s.   \\
		The tensorial structures in $X^{\alpha_1\dots\alpha_4}$ will be constructed by using the metric tensors  and the momenta with the choices \eqref{choicemetr} and \eqref{choicemom}. Then, in $X^{\alpha_1\dots\alpha_4}$, there are structures of four different type, depending on the number of metric 
		tensors and momenta used to saturate the number of free indices. We consider the general terms
		\begin{equation} 
			\delta\delta\delta ,  \, \delta\delta p p ,  \, \delta p ppp , \, pppppp,  \label{tensorsectors}
		\end{equation}
		observing that these sectors do not mix when the permutation operator $P_{ij}$ is applied. This property allows us to construct the general symmetric form of each sector separately.\\
As a first step, we determine the orbits of the $P$ operators acting on the tensor structures belonging to each tensorial sector \eqref{tensorsectors}.  This can be achieved by applying all the $P$ transformations to a tensor structure and following the "trajectory" (orbits) in the sector generated by this process.  For instance, in the sector $\delta\delta p p$, we encounter the two orbits
\begin{eqnarray}
	\begin{tikzcd}
		\delta^{\alpha_2 \beta_1} \delta^{\alpha_4 \beta_2} p_3^{\alpha_1} p_1^{\alpha_3}\arrow[r, "P_{12}"] \arrow[rd, "P_C"] & \delta^{\alpha_1 \beta_2} \delta^{\alpha_4 \beta_1} p_3^{\alpha_2} p_2^{\alpha_3} \arrow[d, "P_{34}"]  \\
		\delta^{\alpha_2 \beta_1} \delta^{\alpha_3\beta_2} p_4^{\alpha_1} p_1^{\alpha_4}  \arrow[u, "P_{34}"] & \delta^{\alpha_1 \beta_2} \delta^{\alpha_3\beta_1}  p_4^{\alpha_2} p_2^{\alpha_4} \arrow[l,"P_{12}"]
	\end{tikzcd}\label{orbitex1}
\end{eqnarray}
\begin{eqnarray}
	\begin{tikzcd}
		\delta^{\alpha_1 \alpha_2}  \delta^{\beta_1 \beta_2} p_1^{\alpha_3}p_1^{\alpha_4 }\arrow[rr, "P_{12}", " P_C" '] \arrow["P_{34}"', loop, distance=2em, in=305, out=235] &  & \delta^{\alpha_1 \alpha_2}  \delta^{\beta_1 \beta_2} p_2^{\alpha_3}p_2^{\alpha_4} \arrow["P_{34}"', loop, distance=2em, in=305, out=235].
	\end{tikzcd} \label{orbitex2}
\end{eqnarray}
In this way we decompose every sector \eqref{tensorsectors} into orbits. Every $P$ transformation acts on an orbit irreducibly, i.e. it connects every element on the orbit. The number of orbits for all the sectors equals the number of independent form factors representing the correlator. 
In fact, a representative can be selected for each orbit to which an independent form factor can then be associated. The orbit provides a visual realization of the symmetry properties of the form factors that belong to it. \\
For clarity, let's clarify this statement with an explicit example, using the sector $\delta \delta\delta$.  With the choices \eqref{choicemetr} and \eqref{choicemom} we construct the three possible tensor structures in $X^{\alpha_1\dots\alpha_4}$ as
\begin{equation}
	\delta^{\alpha_1 \alpha_2}  \delta^{\alpha_3\alpha_4}\delta^{\beta_1 \beta_2},\qquad  \delta^{\alpha_1 \alpha_3} \delta^{\alpha_2 \alpha_4}\delta^{\beta_1 \beta_2},\qquad 
	\delta^{\alpha_1 \alpha_4} \delta^{\alpha_2 \alpha_3} \delta^{\beta_1 \beta_2}.
\end{equation}
This sector can be decomposed into two orbits as
\begin{eqnarray}
	\begin{tikzcd}
	\delta^{\alpha_1 \alpha_3} \delta^{\alpha_2 \alpha_4}\delta^{\beta_1 \beta_2}\arrow[rr, "P_{12}", " P_{34}" '] \arrow["P_{C}"', loop, distance=2em, in=305, out=235] &  & \delta^{\alpha_1 \alpha_4} \delta^{\alpha_2 \alpha_3} \delta^{\beta_1 \beta_2} \arrow["P_{C}"', loop, distance=2em, in=305, out=235]
\end{tikzcd}, \qquad \qquad 		
\begin{tikzcd}
\delta^{\alpha_1 \alpha_2}  \delta^{\alpha_3\alpha_4}\delta^{\beta_1 \beta_2}\arrow["{P_{34},P_{12},P_C}"', loop, distance=2em, in=305, out=235] 
\end{tikzcd}
\end{eqnarray}
from which we can conclude that for this sector there exist just two independent form factors related to the representatives of the orbits. Then the general form of $X$ for the sector with three $\delta$s (denoted as zero momentum or "0p") can be directly written as
\begin{align}
X^{\alpha_1\beta_1\alpha_2\beta_2\alpha_3\alpha_4}_{(0p)}=A_1^{(0p)} \delta^{\alpha_1 \alpha_2} \delta^{\alpha_3\alpha_4} \delta^{\beta_1 \beta_2} +A_2^{(0p)} \delta^{\alpha_1 \alpha_3} \delta^{\alpha_2\alpha_4} \delta^{\beta_1 \beta_2}+A_2^{(0p)}(p_1\leftrightarrow p_2) \delta^{\alpha_1 \alpha_4} \delta^{\alpha_2\alpha_3} \delta^{\beta_1 \beta_2},\label{ind3Delta}
\end{align}
where the two independent form factors satisfy the symmetry conditions
\begin{align}
A_1^{(0p)} (p_1\leftrightarrow p_2)&=A_1^{(0p)} (p_3\leftrightarrow p_4)=A_1^{(0p)},\\
A_2^{(0p)} (p_3\leftrightarrow p_4)&=A_2^{(0p)}(p_1\leftrightarrow p_2),\qquad A_2^{(0p)} (p_1\leftrightarrow p_2, p_3\leftrightarrow p_4)=A_2^{(0p)}.
\end{align}
The properties of the orbits within the tensorial sector are directly reflected in the symmetry conditions of the form factors. \\
The method proposed in \cite{Coriano:2019nkw}  is similar, but more involuted, since it starts from the most general tensor structure with a non-minimal number of form factors. By imposing the symmetry conditions under the group of permutations, one obtains the conditions that reduce the number of form factors and the symmetry constraints that they have to satisfy. Indeed, applying the prescription of the example above, we should identify its expression starting from the general ans\"atz
	\begin{equation} 
		X^{\alpha_1\beta_1\alpha_2\beta_2\alpha_3\alpha_4}_{(0p)}=F_1  \delta^{\alpha_1 \alpha_2} \delta^{\beta_1 \beta_2} \delta^{\alpha_3\alpha_4} + 
	F_2  \delta^{\alpha_1 \alpha_3} \delta^{\alpha_2 \alpha_4} \delta^{\beta_1 \beta_2} + F_3\delta^{\alpha_1 \alpha_4} \delta^{\alpha_2 \alpha_3}\delta^{\beta_1 \beta_2}  .
\end{equation}
The invariance of the correlator under the permutation $P_{12}$ reduces the number of independent form factors and gives the symmetry conditions
	\begin{align} 
	F_1&=F_1 (p_1\leftrightarrow p_2), \notag\\
	F_3 &=F_2(p_1\leftrightarrow p_2).
\end{align}
	 The invariance of the correlator under the other symmetry transformations $P_{34}$ and $P_C$, turns into some symmetry conditions on the independent form factors as
	\begin{align} 
		F_1 &= F_1 (p_1\leftrightarrow p_2)= F_1 (p_3\leftrightarrow p_4)=F_1 (p_1\leftrightarrow p_2, p_3\leftrightarrow p_4)\notag\\
	F_2&= F_2 (p_1\leftrightarrow p_2, p_3\leftrightarrow p_4),\qquad F_2(p_1\leftrightarrow p_2) = F_2 (p_3\leftrightarrow p_4) ,
 \end{align}
obtaining again \eqref{ind3Delta}. 
It is now clear that the study of the orbits of the tensor structures under the permutation group, provides directly the answer about the minimal number of independent form factors that are needed in order to describe the general solution of any $4$-point correlator. This procedure can be simply generalized to higher point correlation functions involving operators of any spin. \\
Once we identify and select representative of each orbit for every sector, then the general structure of $X^{\alpha_1\dots\alpha_4}$ can be written down quite easily. In this way, we find that in $d>4$ the general form of $X^{\alpha_1\dots\alpha_4}$ related to $\braket{TTJJ}$ is written in terms of $47$ independent form factors. This number reduces significantly when $d\le4$, as we are going to show in the following sections. \\
The number of tensor structures and independent form factors for the general $d$ dimensional case is listed in \tabref{generalD} and the representative of each orbit are listed in \appref{Orbits}. 
	\begin{table}[h]
		\centering
		\begin{tabular}{|c|c|c|} \hline&&\\[-2ex]
			Sector & $\#$ of tensor structures &$\#$ of orbits \\[1ex] \hline &&\\[-2ex]
			$\delta \delta \delta$ & 3 & 2 \\[1ex]
			$\delta \delta p p$ & 38 & 13 \\[1ex]
			$\delta p p p p$ & 73  & 21 \\[1ex]
			$p p p p p p$ & 36  & 11  \\[1ex]\hline
			Total & 150 &47 \\ \hline
		\end{tabular}
	\caption{ Number of tensor structures and independent form factors in $X^{\alpha_1\dots\alpha_4}$ for the $\braket{TTJJ}$. \label{generalD}}
	\end{table}
\section{Dimensional dependent degeneracies of tensor structures in $d\le5$}

In the previous section we have presented the method to construct the general form of the correlator in order to obtain a minimal number of form factors in the general $d$ dimensional case. In $d\le5$, the structure of $X^{\alpha_1\dots \alpha_4}$ changes, according to the degeneracies of the tensor structures \cite{Coriano:2021nvn,Edgar:2001vv, lovelock_1970, Bzowski:2017poo}. These degeneracies cause a reduction in the number of independent form factors and significantly simplifies the structure of the correlator. In this section, we discuss the dimensional reduction patterns for $d\le5$. 
\subsection{Case of $d=5$}
Following the argument presented in \cite{Edgar:2001vv, lovelock_1970}, for any tensor in $d$-dimensional space there is associated a fundamental tensor identity obtained by antisymmetrizing over $d+1$ indices. In particular let $\mathcal{T}^{A\qquad\ b_1\dots b_l}_{\ \ a_1\dots a_k}=\mathcal{T}^{A\qquad \ [b_1\dots b_l]}_{\ \ [a_1\dots a_k]}$ be a trace-free tensor on all of its indices, where $A$ denotes an arbitrary number of additional lower and/or upper indices.  Then 
\begin{equation}
\mathcal{T}^{A\qquad \ [b_1\dots b_l}_{\ \ [a_1\dots a_k}\ \delta_{a_{k+1}}^{b_{l+1}}\ \dots\  \delta_{a_{k+n]}}^{b_{l+n]}}=0,\label{Lovelock}
\end{equation}
where $n\ge d-k-l+1$ and $n\ge0$. 

In $d=5$, \eqref{Lovelock} admits a tower of tensor identities depending on the values of $k$ and $l$ with the condition $n\ge 6-k-l$ and $n\ge0$ as showed in \tabref{Table2}. 
\begin{table}[h!]
	\centering
	\renewcommand{\arraystretch}{1.2}
\begin{tabular}{|c|c|c|c|c|c|c|c|}
	\hline
	\multicolumn{8}{|c|}{$d=5$} \\ \hline
	\hline
	$k$ & $l$ & $ n$& Tensor rank & $k$ & $l$ & $ n$& Tensor rank\\ \hline \hline
	2 & 0 & \multirow{2}{*}{ $\geq 4$} & \multirow{2}{*}{10} & 5 & 0 & \multirow{3}{*}{ $\geq 1$} & \multirow{3}{*}{7} 
	\\
	1 & 1 & & & 4 & 1 & & \\ \cline{1-4}
	3 & 0 & \multirow{2}{*}{ $\geq 3$} & \multirow{2}{*}{9} & 3 & 2 & & \\ \cline{5-8}
	2 & 1 & & & 6 & 0 & \multirow{4}{*}{ $\geq 0$} & \multirow{4}{*}{6} \\ \cline{1-4}
	4 & 0 & \multirow{3}{*}{ $\geq 2$} & \multirow{3}{*}{8} & 5 & 1 & & \\
	3 & 1 & & & 4 & 2 & & \\ 
	2 & 2 & & & 3 & 3 & & \\ \hline
\end{tabular} \caption{Possible tensor identities depending on the relation $n\ge 6-k-l$. \label{Table2}}
\end{table}

It is worth mentioning that for $l+k\ge6$ we obtain the condition that the tensor itself vanishes for $n=0$, and all the other identities for $n>0$ are trivially satisfied. In order to clarify this point, let us consider the case with $k=3$, $l=3$ for which, in $d=5$, the relation \eqref{Lovelock} becomes
\begin{align}
\mathcal{T}^{A\qquad \ [b_1\,b_2\,b_3}_{\ \ [a_1\,a_2\,a_3}\ \delta_{a_{4}}^{b_{4}}\ \dots\  \delta_{a_{3+n]}}^{b_{3+n]}}=0, \qquad n\ge0.
\end{align}
The first identity, depending on the values of $n$, is obtained for $n=0$
\begin{align}
\mathcal{T}^{A\qquad \ \ b_1\,b_2\,b_3}_{\ \ \,a_1\,a_2\,a_3}=0,\label{n033}
\end{align}
which trivially satisfies all other identities for $n>0$. For example, using the fact that an anti-symmetrization on $n$ indices can be decomposed iteratively on a smaller number of indices, in the case $n=1$ one obtains
\begin{align}
\mathcal{T}^{A\qquad \ [b_1\,b_2\,b_3}_{\ \ [a_1\,a_2\,a_3}\ \delta_{a_{4}]}^{b_{4}]}&=\frac{1}{4}\left(\mathcal{T}^{A\qquad \ [b_1\,b_2\,b_3}_{\ \ [a_1\,a_2\,a_3]}\ \delta_{a_{4}}^{b_{4}]}-\mathcal{T}^{A\qquad \ [b_1\,b_2\,b_3}_{\ \ [a_1\,a_2\,a_4]}\ \delta_{a_{3}}^{b_{4}]} +\mathcal{T}^{A\qquad \ [b_1\,b_2\,b_3}_{\ \ [a_1\,a_4\,a_3]}\ \delta_{a_{2}}^{b_{4}]} -\mathcal{T}^{A\qquad \ [b_1\,b_2\,b_3}_{\ \ [a_4\,a_2\,a_3]}\ \delta_{a_1}^{b_4]}\right)\notag\\[1ex]
&=\frac{1}{4^2}\left(\mathcal{T}^{A\qquad \ [b_1\,b_2\,b_3]}_{\ \ [a_1\,a_2\,a_3]}\ \delta_{a_{4}}^{b_{4}}+\dots+\mathcal{T}^{A\qquad \ [b_4\,b_2\,b_3]}_{\ \ [a_1\,a_2\,a_3]}\ \delta_{a_{4}}^{b_{1}}-\mathcal{T}^{A\qquad \ [b_1\,b_2\,b_3]}_{\ \ [a_1\,a_2\,a_4]}\ \delta_{a_{3}}^{b_{4}}+\dots \right)=0
\end{align}
by means of \eqref{n033}.\\
After a meticulous investigation of all the tensorial identities in $d=5$, we present below only those that affect the form of the tensorial structure of the $TTJJ$. In particular, all the constraints from \tabref{Table2} are equivalent to only one.  To show which one is the constraint for the $TTJJ$ in $d=5$, we consider the case with $k=3$ and $l=3$ that takes the form
\begin{align}
\mathcal{T}^{\ \qquad \ \ b_1\,b_2\,b_3}_{\ a_1\,a_2\,a_3}=0.\label{D5constr}
\end{align}
In order to construct the antisymmetric tensor $\mathcal{T}$, we consider the tensor
\begin{align}
t^{\qquad \ \ \ b_1\,b_2\,b_3}_{\,a_1\,a_2\,a_3}=p_{1\,\raisebox{-0.6ex}{$\scriptstyle[a_1$}}p_{2\,\raisebox{-0.6ex}{$\scriptstyle a_2$}}p_{3\,\raisebox{-0.6ex}{$\scriptstyle a_3]$}}\,p_1^{\,\raisebox{-0.6ex}{$\scriptstyle[b_1$}}p_2^{\,\raisebox{-0.6ex}{$\scriptstyle b_2$}}p_3^{\,\raisebox{-0.6ex}{$\scriptstyle b_3]$}},\label{deft5}
\end{align}
and then the traceless and completely antisymmetric tensor $\mathcal{T}$ is directly written in the form
\begin{align}
\mathcal{T}^{\ \qquad \ \ b_1\,b_2\,b_3}_{\ a_1\,a_2\,a_3}\equiv t^{\ \qquad \ \ \ [a_4\,a_5\,a_6\,}_{\ a_4\,a_5\,a_6}\delta^{b_1}_{a_1}\,\delta^{b_2}_{a_2}\,\delta^{b_3]}_{a_3}.\label{deft6}
\end{align}
This is the only non-trivial tensor that can be constructed using the three independent momenta of a $4$-point function. The use of the metric $\delta_{a_i}^{\,b_i}$ in \eqref{deft5} gives a trivially zero result in five dimensions. In order to find the dimension-dependent equation of the tensor structure of the transverse-traceless part of the $\braket{TTJJ}$, we apply to \eqref{deft6} the $\Pi$ and $\pi$ projectors
\begin{equation}
\Pi^{\mu_1\nu_1\ \,\beta_1}_{\quad\ \,\alpha_1}\ \Pi^{\mu_2\nu_2\ \,\beta_2}_{\quad\ \,\alpha_2}\ \pi^{\mu_3}_{\alpha_3}\,\pi^{\mu_4\,\alpha_4}\,\Big(\mathcal{T}^{\ \qquad \ \ \alpha_1\,\alpha_2\,\alpha_3}_{\ \beta_1\,\beta_2\,\alpha_4}\Big)=0,
\end{equation}
or explicitly
\begin{align}
	&\Pi^{\mu_1\nu_1\ \,\beta_1}_{\quad\ \,\alpha_1}\ \Pi^{\mu_2\nu_2\ \,\beta_2}_{\quad\ \,\alpha_2}\ \pi^{\mu_3}_{\alpha_3}\,\pi^{\mu_4\,\alpha_4}\,\Bigg\{J_{123}\,\delta^{[\alpha_1}_{\beta_1}\delta^{\alpha_2}_{\beta_2}\delta^{\alpha_3]}_{\alpha_4}-3\,J_{\raisebox{-0.6ex}{$\scriptstyle 12;[12$}}\,p_{3]}^{[\alpha_1}\delta^{\alpha_2}_{[\beta_1}\delta^{\alpha_3]}_{\raisebox{-0.2ex}{$\scriptstyle \alpha_4$}}p_{3\raisebox{-0.6ex}{$\scriptstyle \beta_2]$}}\notag\\
	&\hspace{1cm}+6\,J_{\raisebox{-0.6ex}{$\scriptstyle 1;[1$}}\,p_{\raisebox{-0.5ex}{$\scriptstyle 2$}}^{[\sigma}\,p_{3]}^{\alpha_1}\delta^{\alpha_2}_{[\beta_2}\delta^{\alpha_3]}_{\raisebox{-0.2ex}{$\scriptstyle \alpha_4$}}p_{2\raisebox{-0.5ex}{$\scriptstyle \beta_1]$}}\,p_{3\raisebox{-0.5ex}{$\scriptstyle \sigma$}}+6\,J_{\raisebox{-0.6ex}{$\scriptstyle 2;[1$}}\,p_{\raisebox{-0.5ex}{$\scriptstyle 2$}}^{[\sigma}\,p_{3]}^{\alpha_1}\delta^{\alpha_2}_{[\beta_1}\delta^{\alpha_3]}_{\raisebox{-0.2ex}{$\scriptstyle \alpha_4$}}p_{1\raisebox{-0.5ex}{$\scriptstyle \beta_2]$}}\,p_{3\raisebox{-0.5ex}{$\scriptstyle \sigma$}}+4\,p_{3\raisebox{-0.5ex}{$\scriptstyle \sigma$}}\,p_1^{[\mu_3}p_2^{\alpha_1}p_3^{\alpha_2}\delta^{\alpha_3]}_{[\alpha_4}p_{2\raisebox{-0.5ex}{$\scriptstyle \beta_1$}}p_{1\raisebox{-0.5ex}{$\scriptstyle\beta_2]$}}\Bigg\}=0,\label{constr5d}
\end{align}
where 
\begin{align}
	J_{123}&=p_{1\raisebox{-0.5ex}{$\scriptstyle[a_1$}}p_{2\,\raisebox{-0.5ex}{$\scriptstyle a_2$}}p_{3\,\raisebox{-0.5ex}{$\scriptstyle a_3]$}}\,p_1^{[a_1}p_2^{\,a_2}p_3^{a_3]},\qquad J_{12;ij}=p_{1\raisebox{-0.5ex}{$\scriptstyle[a_1$}}p_{2\,\raisebox{-0.5ex}{$\scriptstyle a_2]$}}\,p_i^{[a_1}p_j^{a_2]},\qquad J_{i;j}=(p_i\cdot p_j).
\end{align}
From this analysis, we conclude that in $d=5$ the number of independent form factors is reduced by one according to the constraint \eqref{constr5d} on the tensor structures.  

\subsection{Case of $d=4$}
Analogously to the previous case, in $d=4$ one has to consider tensor identities derived from \eqref{Lovelock}. However, in this case, we are going to show that there is an even  more efficient way to identify the independent tensorial structures.  
Indeed, in $d=4$, the metric $\delta$ is not an independent tensor and one can construct a different basis in which it can be re-expressed. For this purpose, we construct a new orthogonal four-vector $n^\mu$ using the completely antisymmetric tensor $\epsilon^{\mu_1\dots\mu_4}$ and three of the four external momenta in the form
\begin{equation} 
n^{\mu_1}= \epsilon^{\mu_1 \mu_2 \mu_3 \mu_4}\, {p_1}_{\mu_2}\, {p_2}_{\mu_3}\, {p_3}_{\mu_4} . \label{ndef}
\end{equation}
Notice that this reduction is possible if one has in $d$ dimensions at least $d-1$ independent external momenta in a correlation function. 
The vector \eqref{ndef} is obviously transverse with respect to $p_1 , p_2 , p_3$, i.e. $n\cdot p_i=0$.  Having defined such a vector, we use this new basis, that we call the $n$-$p$ basis, to construct all the tensorial structures that we use to define the correlation function. 
In this new basis, the metric tensor is expressed as 
\begin{equation}
	 \label{DepDeltaIm} \delta^{\mu \nu} = \sum_{i=1}^4 (Z^{-1})_{ji} \, P_i^\mu P_j^\nu, 
\end{equation}
where $Z^{-1}$ is the inverse of the Gram matrix $Z=[P_i \cdot P_j]_{i,j=1}^d $ and  $P_j^\mu\in\{p_1^\mu,p_2^\mu,p_3^\mu,n^\mu\}$. In particular, the Gram matrix in this case is written as
\begin{align}
Z=\left(\begin{matrix}
p_1^2&p_1\cdot p_2&p_1\cdot p_3&0\\
p_1\cdot p_2&p_2^2&p_2\cdot p_3&0\\
p_1\cdot p_3&p_2\cdot p_3&p_3^2&0\\
0&0&0&n^2
\end{matrix}\right)
\end{align}
from which we obtain the expression of $\delta$ as
\begin{equation} 
	\label{DepDelta} 
	\delta^{\mu \nu} = \sum_{i,j=1}^3 \Big[(p_{i+1} \cdot p_{j-1} ) (p_{i-1} \cdot p_{j+1} ) - (p_{i-1} \cdot p_{j-1} ) (p_{i+1} \cdot p_{j+1} ) \Big] \frac{p_i^\mu p_j^\nu}{n^2} + \frac{n^\mu n^\nu}{n^2} , 
\end{equation}
where the indices $i,j$ are labelled mod-3 and 
\begin{equation}
\label{nsquare} n^2 = -p_1^2 p_2^2 p_3^2 + p_1^2 \, (p_2 \cdot p_3 )^2+p_2^2 \, (p_1 \cdot p_2 )^2 +
p_3^2 \, (p_1 \cdot p_2 )^2 - 2 \, (p_1 \cdot p_2 )\, (p_2 \cdot p_3) \, (p_1 \cdot p_3) . 
\end{equation}
With the expression for the metric given in \eqref{DepDelta}, and  using the constraint $\delta^{\alpha_i\beta_i}\,\Pi^{\mu_i\nu_1}_{\alpha_i\beta_i}=0$, we find 
\begin{align}
	\Pi_{\alpha_i \beta_i}^{\mu_i \nu_i}(p_i) \, n^{\alpha_i} n^{\beta_i} &= \Pi_{\alpha_i \beta_i}^{\mu_i \nu_i}(p_i)\bigg[
	\Big(p_{i-1}^2 \, p_i^2 - (p_i \cdot p_{i-1})^2 \Big)\, p_{i+1}^{\alpha_i} \, p_{i+1}^{\beta_i} +
	\Big(p_{i+1}^2\,  p_i^2 - (p_i \cdot p_{i+1})^2 \Big) \, p_{i-1}^{\alpha_i} \, p_{i-1}^{\beta_i} \notag\\ 
	& +2 \Big((p_i \cdot p_{i+1}) \, (p_i \cdot p_{i-1}) - p_i^2(p_{i-1} \cdot p_{i+1}) \Big) \, p_{i-1}^{\alpha_i} \, p_{i+1}^{\beta_i } \bigg],\qquad i=1,2
\end{align}
with the indices labelled  mod-3. With the conventions for independent momenta and indices made in \eqref{choicemom}, the constraint can be explicitly rewritten as
\begin{align}
	\Pi_{\alpha_i \beta_i}^{\mu_i \nu_i}(p_i) \, n^{\alpha_i} n^{\beta_i} &= \Pi_{\alpha_i \beta_i}^{\mu_i \nu_i}(p_i)\bigg[p_3^{\alpha_i}p_3^{\beta_i}\bigg(p_i^2(p_i+p_4)^2-(p_i\cdot (p_i+p_4))^2\bigg)\notag\\
	&\hspace{-2cm}+p_4^{\alpha_i}p_4^{\beta_i}\bigg(p_i^2\,p_3^2-(p_i\cdot p_3)^2\bigg)+2p_3^{\alpha_i}p_4^{\beta_i}\bigg(\big(p_i\cdot (p_i+p_4)\big)(p_i\cdot p_3)-p_i^2\,\big((p_i+p_4)\cdot p_3\big)\bigg)\bigg],\quad i=1,2\,.\label{constraintsNN}
\end{align}
The previous constraints, occurring when at least two $n$ vectors are contracted with a transverse traceless projector, allow us to write the decomposition of the transverse-traceless part \eqref{ttjj} in $d=4$ in terms of just three sectors, i.e. 
\begin{equation}
 \label{tensorsectorsNP1} \, nnnnpp, \, nnpppp, \, pppppp .
\end{equation}
It is worth to point out that in this new basis the term $n^{\alpha_1}n^{\beta_1}n^{\alpha_2}n^{\beta_2}n^{\alpha_3}n^{\alpha_4}$ cannot appear in the decomposition of $X^{\alpha_1\dots\alpha_4}$, since  because of the constraints \eqref{constraintsNN} it would be rewritten as $p_i^{\alpha_1}p_j^{\beta_1}p_k^{\alpha_2}p_l^{\beta_2}n^{\alpha_3}n^{\alpha_4}$ with $i,j,k,l=3,4$. \\
Furthermore, in this new basis we notice a drastic reduction of the number of independent tensorial structures of \eqref{tensorsectorsNP1}. By analysing the orbits of the tensor structures under the permutation group, we select the representatives of each of them (see \appref{Orbits}) and thus determine the number of independent form factors in $d=4$. The result is considerably simplified respect to the general case. We conclude that in this specific case the transverse traceless part of the $\braket{TTJJ}$ is parametrised by 34 independent form factors as summarized in \tabref{4D}.
	\begin{table}[h!]
	\centering
	\begin{tabular}{|c|c|c|} \hline&&\\[-2ex]
		Sector & $\#$ of tensor structures &$\#$ of orbits \\[1ex] \hline &&\\[-2ex]
		$n n n n p p$ & 4 & 2 \\[1ex]
		$n n p p p p$ &  73& 21 \\[1ex]
		$p p p p p p$ &  36 & 11 \\[1ex]\hline
			Total &113 &34 \\ \hline
	\end{tabular}
	\caption{ Number of tensor structures and independent form factors in $d=4$ with the $n$-$p$ basis. \label{4D}}
\end{table}

\subsection{Case of $d=3$}
In the case of $d=3$ the decomposition of the $\braket{ttjj}$ reduces significantly. Indeed,  also in this case the $\delta$ is not an independent tensor but it can be written in terms of the external  momenta. 
Furthermore, contrary to the case of $d=4$, there is no need to introduce an additional orthogonal vector.  The three independent momenta $p_1$, $p_2$ and $p_3$ are sufficient to span the new basis. The three dimensional version of \eqref{DepDeltaIm} and \eqref{DepDelta} is immediate and we have
\begin{align} 
	\delta^{(3)\,\mu \nu} &=\sum_{i=1}^3 (Z^{-1})_{ji} \, p_i^\mu p_j^\nu =  \frac{1}{J}\sum_{i,j=1}^3 \Big[(p_{i-1} \cdot p_{j-1} ) (p_{i+1} \cdot p_{j+1} )-(p_{i+1} \cdot p_{j-1} ) (p_{i-1} \cdot p_{j+1} )  \Big]_{\text{mod} 3} p_i^\mu p_j^\nu ,\label{delta3d}
\end{align}
where $J$ is the determinant of the Gram matrix given as
\begin{equation}
J=p_1^2 p_2^2 p_3^2 - p_1^2 \, (p_2 \cdot p_3 )-p_2^2 \, (p_1 \cdot p_2 ) -
p_3^2 \, (p_1 \cdot p_2 ) + 2 \, (p_1 \cdot p_2 )\, (p_2 \cdot p_3) \, (p_1 \cdot p_3).
\end{equation}
The form of the $\delta$ in three dimensions in \eqref{delta3d} allows us to conclude that the only structure present in \eqref{tensorsectors} is formed just by momenta, and we have to study a single tensorial sector made of six momenta.  Also in this case we have some constraints coming from the contraction of \eqref{delta3d} with a transverse-traceless projector. Indeed, the property $\delta^{\alpha_i \beta_i} \Pi_{\alpha_i \beta_i}^{\mu_i \nu_i} = 0$ gives
\begin{equation} 
	\Pi^{\mu_i\nu_i}_{\alpha_i\beta_i}(p_i)\, p_3^{\alpha_i}p_4^{\alpha_i}=	\Pi^{\mu_i\nu_i}_{\alpha_i\beta_i}(p_i)\,\frac{1}{J_i}\Big[\Big(p_i^2\,p_3^2-(p_i\cdot p_3)^2\Big)p_4^{\alpha_i}p_4^{\beta_i}+\Big(p_i^2(p_i+p_4)^2-\big(p_i\cdot(p_i+p_4)\big)^2\Big)p_3^{\alpha_i}p_3^{\beta_i}\Big],\quad i=1,2
\end{equation}
where 
\begin{equation}
J_i=2\Big[p_i^2\,\big(p_3\cdot(p_i+p_4)\big)-(p_i\cdot p_3)\,\big(p_i\cdot(p_i+p_4)\big)\Big],\quad i=1,2\,.
\end{equation}
By using these constraints we find that at $d=3$ the number of tensor structures is smaller respect to the general case and that of $d=4$. Following the procedure described in the previous sections, we find that the number of independent form factors in this case is reduced to five, as shown in \tabref{3D}. 
\begin{table}[h!]
	\centering
	\begin{tabular}{|c|c|c|} \hline&&\\[-2ex]
		Sector & $\#$ of tensor structures &$\#$ of orbits \\[1ex] \hline &&\\[-2ex]
		$p p p p p p$ & 16 & 5 \\[1ex]\hline
	\end{tabular}
	\caption{ Number of tensor structures and independent form factors in $d=3$. \label{3D}}
\end{table}

\subsection{Summary }
We are ready to summarize the results obtained in the previous sections. \\
By studying the orbits of the tensor structures under the group of permutations, we have identified the minimal tensor structures and form factors needed to construct the transverse-traceless part of the correlator. \\
This study has been first performed in the case of $d\ge6$. For $d\le5$, because of degeneracies in the tensor structures, one needs to take into account identity \eqref{Lovelock}. In particular, for $d=5$, we find one independent constraint that reduces the number of form factors by one. In $d=4$, we have shown how to rewrite the metric $\delta_{\mu\nu}$ in the basis $n$-$p$ by introducing an orthogonal vector $n$ given by \eqref{ndef}. This approach can also be performed for $d=3$, showing that the independent tensor structures are considerably reduced, as is the number of independent form factors. \\
It is worth mentioning that when it is possible to define a new independent tensor base, 
the identities \eqref{Lovelock} are automatically satisfied. This property has been checked in the case of the $TTJ$. The generalization of this approach to any $3$- or $4$- point correlator will be presented elsewhere. The number of minimal form factors for specific dimensions is summarised in \tabref{FormFacTTJJ}. 

\begin{table}[h!]
	\centering
	\begin{tabular}{|c|c|c|c|c|} \hline&&&&\\[-2ex]
		Correlator & $d\ge6$ &$d=5$&$d=4$ & $d=3$ \\[1ex] \hline &&&&\\[-2ex]
		$\braket{TTJJ}$ & 47 & 46 & 34 &5\\[1ex]\hline
	\end{tabular}
	\caption{Number of independent form factors needed to characterize the $\braket{TTJJ}$ depending on the dimensions $d$. \label{FormFacTTJJ}}
\end{table}
\section{Divergences and Renormalization}\label{DivAndRen}
In the previous sections, we have shown how to decompose the correlator in terms of a minimal number of form factors and tensorial structures. In particular, we have seen that this number depends on the specific spacetime dimensions $d$. In particular, one has to consider the $d$-dependent degeneracies in order to identify such minimal decomposition, which may be divergent. \\
Working in DR with the minimal subtraction (MS) scheme, the tensor degeneracy identities in the $n$-$p$ basis should be carefully analyzed in the $\varepsilon\to0$ limit, with $\varepsilon\equiv d-4$. 
This is clear if we consider the case where $d = 4$, where the new vector n is defined using the Levi Civita tensor in four dimensions. This tensor does not allow extending the basis $n-p$ outside the four dimensions, which is necessary in DR. The procedure we then use is to consider the general decomposition in $d$ dimensions, extract the divergences as poles in $1/\varepsilon$ around the target dimension, then renormalize the correlator with an appropriate counterterm, and once we have the finished result for $\varepsilon\to 0$ , then consider the change of basis to obtain a minimal decomposition.\\
Let us now consider the possible cases in which divergences may occur, and make use of the scaling behaviour of the entire correlator. The scale invariance of the correlator is expressed through the dilatation Ward identity which for a specific form factor 
$A_j^{(d\,;\,N_p)}$ takes the form
\begin{equation}
\left(d-2-N_p-\sum_{i=1}^3\,p_i^\mu\frac{\partial}{\partial p_i^\mu}\right)\,A_j^{(d\,;\,N_p)}=0,
\end{equation}
where $N_p$ is the number of momenta multiplying the form factor in the decomposition. This equation characterizes the scaling behaviour of $A^{(d),N_p}$ and allows to identify quite easily those among all which will be manifestly divergent in the UV regime. In general, their behaviour is summarized in \tabref{TabDg}. 
	\begin{table}[h]
\begin{center}
	\begin{tabular}{  c | c | c | c | c  }
		Form Factor &$A_j^{(d\ge5,N_p=6)}$&$A_j^{(d,N_p=4)}$&$A_j^{(d,N_p=2)}$&$A_j^{(d,N_p=0)}$\\ [0.5ex]
		\hline Degree & $d-8$ & $d-6$ & $d-4$ &$d-2$		
		\\ 	\hline
		UV divergent in $d\ge8$  & \cmark & \cmark & \cmark &\cmark \\
		\hline
		UV divergent in $d=6,7$  & \xmark & \cmark & \cmark &\cmark \\
		\hline
		UV divergent in $d=4,5$  & \xmark & \xmark & \cmark &\cmark \\
	\end{tabular}
\end{center}\caption{\label{TabDg}UV scaling behaviour of $A^{(d;N_p)}_j$ depending on the dimensions.}
\end{table}

In all the cases, whenever divergences are present, they will show up as single poles in the regulator $\varepsilon$. The procedure of renormalization, obtained by the inclusion of the counterterm, will remove these divergences and will generate an anomaly. In the next section, we are going to discuss the appearance of the conformal anomaly after the renormalization procedure, for the $d=4$ case. 
	\section{Perturbative analysis in the conformal case}
	In this section we direct our attention towards an important aspect of our analysis, which will allow us to extend the result of the $TJJ$ correlator presented in \cite{Coriano:2018bbe,Coriano:2018zdo} to the $TTJJ$. The two correlators are connected via the hierarchical structures of the trace and conservation WIs and the renormalization procedure.\\
We use the free field theory realization of the $TTJJ$ in order to study the structure of the anomalous Ward identities once the conformal symmetry is broken after its renormalization.  After the renormalization, the vertex will separate into a renormalized part of the form \eqref{decTTJJ} and an anomaly part, following the same pattern of the $TJJ$. This anomaly part will be compared directly with the result coming from the variation of the anomaly effective action. 
		\begin{figure}[ht]
		\centering
		\subfigure[]{\includegraphics[scale=0.4]{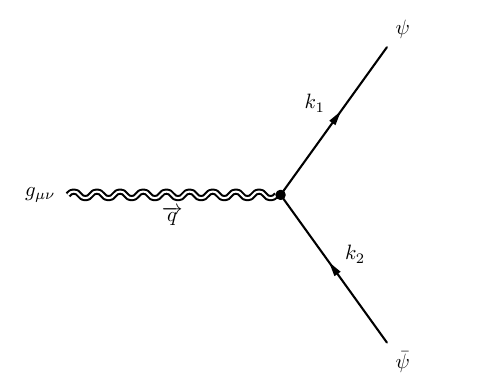}}\hspace{0.2cm}
		\subfigure[]{\includegraphics[scale=0.4]{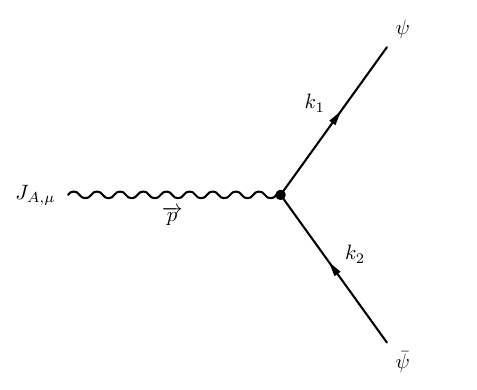}}\hspace{0.5cm}
		\subfigure[]{\includegraphics[scale=0.4]{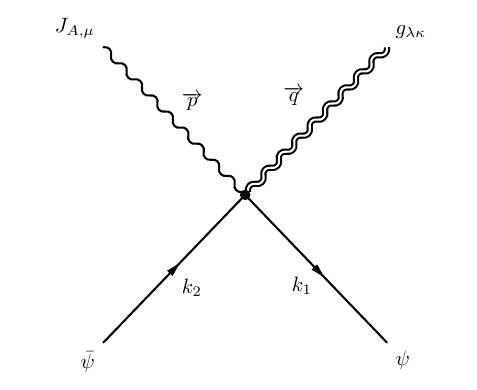}}\hspace{0.5cm}
		\subfigure[]{\includegraphics[scale=0.4]{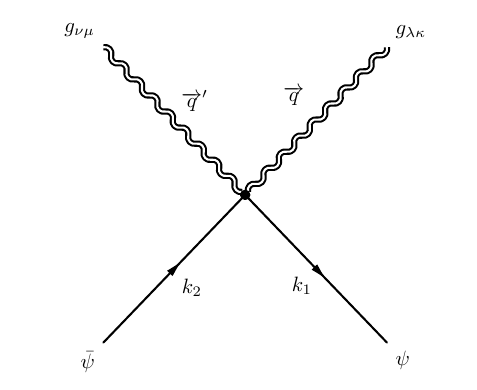}}\hspace{0.5cm}
		\subfigure[]{\includegraphics[scale=0.4]{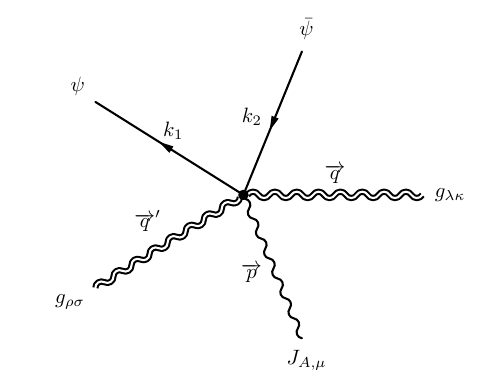}}\hspace{0.5cm}
		\caption{Vertices with fermions for the construction of the $TTJJ$ correlator.\label{vertici}}
	\end{figure}
	We start our analysis by defining our conventions for the perturbative sectors. 
	The fundamental classical action $\sm_0$ can be defined as a sum of two sectors, the scalar and the fermion sectors, which will be considered separately. It can  be defined in the form 
\beq
\sm_0=\sum_l S_{l\, scalar}+ 	\sum_{l'}S_{l', fermion}
\eeq 
where $l=1,2\ldots N_s$ and $l'=1,2\ldots N_f$ enumerate the conformally coupled scalars and fermions in the total action. We will consider the case of a single scalar and a single fermion, correcting the results for their multiplicities at the end.  \\	
The  action for the fermion field in a gravitational background is 
	\begin{align}
	S_{fermion}&=\int\,d^dx\,V\,\left[\frac{i}{2}\,V^{\mu}_a\,\bar\psi\,\gamma^a\,\left(D_\mu\psi\right)-\frac{i}{2}\,V^{\mu}_a\,\left(D_\mu^\dagger\bar\psi\right)\,\gamma^a\,\psi\right],
	\end{align}
	where $V^\mu_a$ is the vielbein and $V$ its determinant, $D_\mu$ is the covariant derivative defined as
	\begin{align}
	D_\mu\psi&=\left(\nabla_\mu+i\,e\,A_\mu\right)\psi=\left(\partial_\mu+i\,e\,A_\mu+\frac{1}{2}\,\omega_{\mu\,ab}\,\Sigma^{ab}\right)\psi,\\
	D_\mu\bar\psi&=\left(\nabla_\mu-i\,e\,A_\mu\right)\bar\psi=\left(\partial_\mu-i\,e\,A_\mu-\frac{1}{2}\,\omega_{\mu\,ab}\,\Sigma^{ab}\right)\bar\psi,
	\end{align}
	where $\Sigma^{ab}$ are the generators of the Lorentz group in the case of a spin 1/2-field, and
	\begin{align}
	\omega_{\mu\,ab}\equiv\,V^{\nu}_a\left(\partial_\mu\,V_{\nu\,b}-\Gamma^\lambda_{\mu\nu}\,V_{\lambda\,b}\right),
	\end{align} 
	being the spin connection in the holonomic (metric) definition, with the antisymmetric property $\omega_{\mu\,ab}=-\omega_{\mu\,ba}$. 
	The Latin and Greek indices are related to the (locally) flat basis and the curved background respectively. Using the explicit expression of the generators of the Lorentz group one can re-expresses the action as follows
	\begin{align}
	S_{fermion}&=\int\,d^dx\,V\left[\frac{i}{2}\,\,\bar\psi\,V^{\mu}_a\gamma^a\,\left(\partial_\mu\psi\right)-\frac{i}{2}\,\left(\partial_\mu\bar\psi\right)V^{\mu}_a\,\gamma^a\,\psi-e\,\bar\psi V^{\mu}_a\gamma^a\,A_\mu\,\psi-\frac{i}{4}\,\omega_{\mu\,ab}\,V^{\mu}_c\,\bar\psi\,\gamma^{abc}\psi\,\right],\label{FermAct}
	\end{align}
	where $\gamma^{abc}$ is a completely antisymmetric product of gamma matrices defined as
	\begin{equation}
	\gamma^{abc}\equiv\frac{1}{3!}\left(\gamma^a\gamma^b\gamma^c+\text{antisymmetric}\right).
	\end{equation}
	We recall the property
	\begin{align}
	\gamma^c\,\Sigma^{ab}+\Sigma^{ab}\,\gamma^c=-\gamma^{abc}
	\end{align}
	where $\Sigma^{ab}=-\frac{1}{4}[\gamma^a,\gamma^b]$. 
	Taking a first variation of the action with respect to the metric one can construct the energy momentum tensor as
	\begin{align}
	T^{\mu\nu}=-\frac{i}{2}\Bigg[\,\bar\psi\,\gamma^{(\mu}\nabla^{\nu)}\psi-\nabla^{(\mu}\bar\psi\,\gamma^{\nu)}\psi-g^{\mu\nu}\left(\bar\psi\,\gamma^\lambda\nabla_\lambda\psi-\nabla_\lambda\bar\psi\,\gamma^\lambda\psi\right)\Bigg]-e\,\bar\psi\,\left(g^{\mu\nu}\gamma^\lambda\,A_{\lambda}-\gamma^{(\mu}A^{\nu)}\right)\psi
	.\label{T}
	\end{align}
	The computation of the vertices can be done by taking functional derivatives of the action with respect to the metric and the gauge field and Fourier transforming to momentum space. They are given in \figref{vertici} and their explicit expressions have been collected in \appref{AppendixB}. 
	\begin{figure}[ht]
	\centering
	\subfigure[]{\includegraphics[scale=0.4]{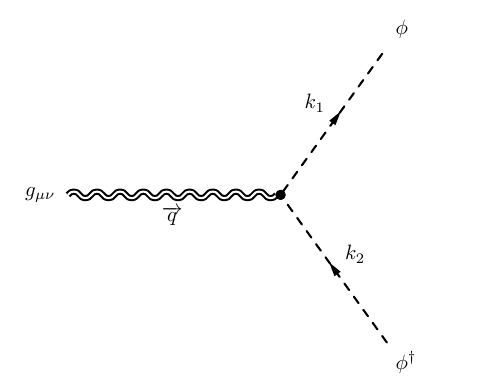}}\hspace{0.2cm}
	\subfigure[]{\includegraphics[scale=0.4]{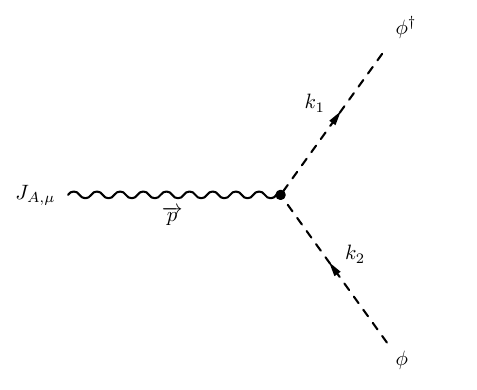}}\hspace{0.5cm}
	\subfigure[]{\includegraphics[scale=0.4]{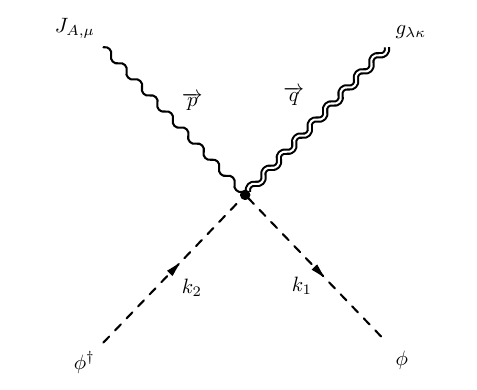}}\hspace{0.5cm}
	\subfigure[]{\includegraphics[scale=0.4]{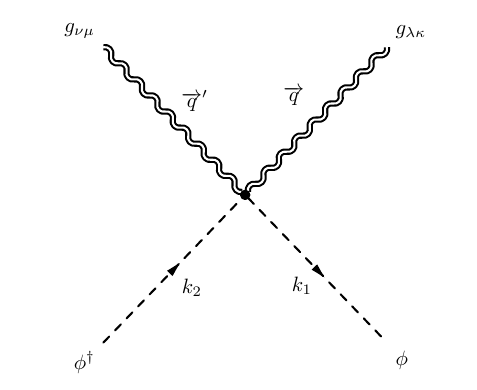}}\hspace{0.5cm}
	\subfigure[]{\includegraphics[scale=0.4]{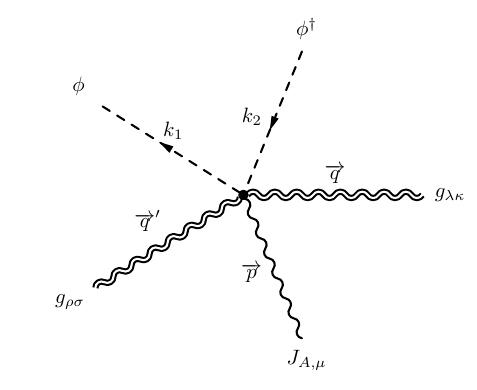}}\hspace{0.5cm}
	\caption{Vertices with scalars for the construction of the $TTJJ$ correlator.\label{vertici2}}
\end{figure}
	The action of a scalar coupled to a gauge field in a curved  background is defined by the functional 
	\begin{equation}
	S_{scalar}=\int d^dx\,\sqrt{-g}\left(\left|D_\mu\phi\right|^2+\frac{(d-2)}{4(d-1)}R|\phi|^2\right),\label{ScalAct}
	\end{equation}
	where $R$ is the scalar curvature and $\phi$ denotes a complex scalar field and $D_\mu\phi=\nabla_\mu\phi+ie\,A_\mu\phi$, the covariant derivative for the coupling to the gauge field $A_\mu$. The vertices for the scalar interactions depicted in \figref{vertici2} are listed in \appref{AppendixB}. It is worth mentioning that we have not considered the vertex with two gauge fields/two scalars and two gravitons because it contributes as a massless tadpole at one-loop, and vanishes in DR.
	
	\subsection{Feynman Diagrams}
	In order to find all the Feynman diagrams that will contribute to the correlation function, we start from the definition of the energy momentum tensor as given in \eqref{anomx0}.
The $TTJJ$ correlator around flat space is extracted by taking four derivatives of the bare effective action $\sm_B$ with respect to the metric and the gauge field, evaluated when the sources are turned off 
	\begin{align}
	\braket{T^{\mu_1\nu_1}(x_1)\,T^{\mu_2\nu_2}(x_2)\,J^{\mu_3}(x_3)\,J^{\mu_4}(x_4)}=\,4\,\frac{\delta^4\,\mathcal{S}_B}{\delta g_{\mu_1\nu_1}(x_1)\,\delta g_{\mu_2\nu_2}(x_2)\,\delta A_{\mu_3}(x_3)\,\delta A_{\mu_4}(x_4)}\Bigg|_{g=\delta,\,A=0}\,.
	\end{align}
We will discuss the Euclidean case, with background metric $\delta_{\mu\nu}$. The analytic continuation of our results to Minkowski space are pretty straightforward, since the basic master integrals appearing in the computations are the scalar self-energy, the triangle and the box diagram. Having denoted with $S_0$ the conformal invariant classical action in 
$\sm_B$, we have 	

		\begin{align}
	&\braket{T^{\mu_1\nu_1}(x_1)\,T^{\mu_2\nu_2}(x_2)\,J^{\mu_3}(x_3)\,J^{\mu_4}(x_4)}=\,4\,\bigg\{\,\left\langle\frac{\delta S_0}{\delta g_1}\ \frac{\delta S_0}{\delta g_2}\ \frac{\delta S_0}{\delta A_3}\ \frac{\delta S_0}{\delta A_4}\right\rangle-\left\langle\frac{\delta^2 S_0}{\delta g_1\,\delta g_2}\ \frac{\delta S_0}{\delta A_3}\ \frac{\delta S_0}{\delta A_4}\right\rangle\notag\\[1.5ex]
	&
	\hspace{0.5cm}-\left\langle\frac{\delta^2 S_0}{\delta g_1\,\delta A_3}\ \frac{\delta S_0}{\delta g_2}\ \frac{\delta S_0}{\delta A_4}\right\rangle
	-\left\langle\frac{\delta^2 S_0}{\delta g_1\,\delta A_4}\ \frac{\delta S_0}{\delta g_2}\ \frac{\delta S_0}{\delta A_3}\right\rangle
	-\left\langle\frac{\delta S_0}{\delta g_1}\ \frac{\delta^2 S_0}{\delta g_2\,\delta A_3}\ \frac{\delta S_0}{\delta A_4}\right\rangle
	-\left\langle\frac{\delta S_0}{\delta g_1}\ \frac{\delta^2 S_0}{\delta g_2\,\delta A_4}\ \frac{\delta S_0}{\delta A_3}\right\rangle\notag\\[1.5ex]
	&
	\hspace{1cm}+\left\langle\frac{\delta^2 S_0}{\delta g_1\,\delta A_3}\ \frac{\delta^2 S_0}{\delta g_2\,\delta A_4}\right\rangle 
	+\left\langle\frac{\delta^2 S_0}{\delta g_1\,\delta A_4}\ \frac{\delta^2 S_0}{\delta g_2\,\delta A_3}\right\rangle
	+\left\langle\frac{\delta^3 S_0}{\delta g_1\,\delta g_2\,\delta A_3}\ \frac{\delta S_0}{\delta A_4}\right\rangle
	+\left\langle\frac{\delta^3 S_0}{\delta g_1\,\delta g_2\,\delta A_4}\ \frac{\delta S_0}{\delta A_3}\right\rangle\Bigg\},\label{topology}
	\end{align}
	where for sake of simplicity we have indicated for $g_i=g_{\mu_i\nu_i}(x_i)$, $i=1,\,2$ and for $A_j=A_{\mu_j}(x_j)$, $j=3,\,4$. In \eqref{topology} the angle brackets denote the vacuum expectation value and all the terms are referred to a particular topology for the Feynman diagrams. In particular, we have the box diagram topology, the first term in \eqref{topology}, then the triangle diagrams topology expressed by the second term and the second line of \eqref{topology}, and finally the bubble diagrams written in the last line of the same equation. As we have already mentioned, we discard the term 
	\begin{equation}
	\left\langle\frac{\delta^4 S_0}{\delta g_1\,\delta g_2\,\delta A_4\delta A_3}\right\rangle\label{tadpole}
	\end{equation}
	in \eqref{topology}, because it generates a tadpole diagram at one-loop. 
	\begin{figure}[t]
	\centering
	\subfigure[]{\includegraphics[scale=0.65]{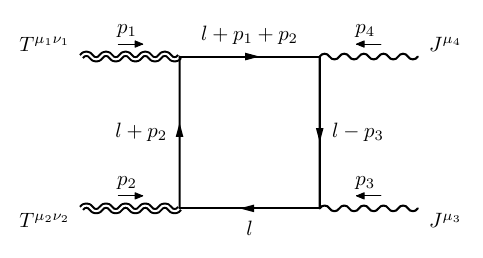}}\hspace{5ex}
	\subfigure[]{\includegraphics[scale=0.65]{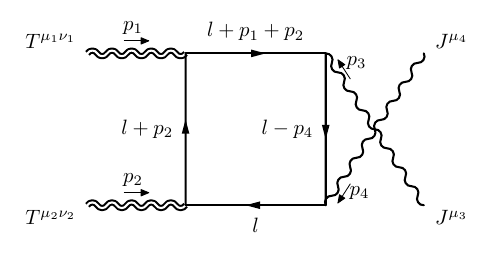}}\hspace{5ex}
	\subfigure[]{	\raisebox{-2.5ex}{\includegraphics[scale=0.65]{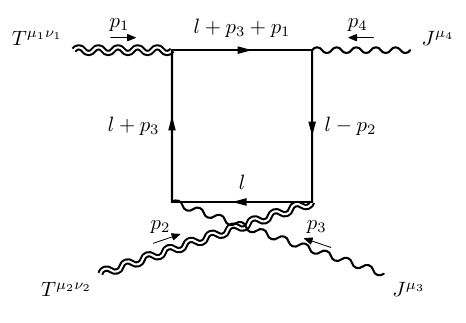}}}
	\caption{Examples of box diagrams that contribute to the $TTJJ$ correlator.\label{BoxDiag}}
	\end{figure} 
For the fermionic case, examples of diagrams are shown in \figref{BoxDiag}. For instance, the first in \figref{BoxDiag} (a), is summarized by the expression
	\begin{align}
	&V^{\mu_1\nu_1\mu_2\nu_2\mu_3\mu_4}_{1,fermion}(p_1,p_2,p_3,p_4)=\notag\\
	&=-\int\frac{d^dl}{(2\pi)^d}\,\frac{\Tr\left[V_{T\bar\psi\psi}^{\mu_1\nu_1}(l+p_2,l+p_1+p_2)\,\left(\slashed{l}+\slashed{p}_2\right)\,V_{T\bar \psi\psi}^{\mu_2\nu_2}(l,l+p_2)\left(\slashed{l}\right)\,V_{J \bar \psi \psi}^{\mu_3}\,\left(\slashed{l}-\slashed{p}_3\right)\,V_{J \bar \psi \psi}^{\mu_4}\,\left(\slashed{l}+\slashed{p}_1+\slashed{p}_2\right)\right]}{l^2\,\left(l+p_2\right)^2\,\left(l+p_2+p_1\right)^2\,\left(l-p_3\right)^2},
	\end{align}	

	\begin{figure}[t]
	\centering
	\includegraphics[scale=0.65]{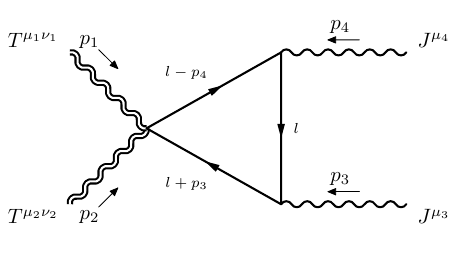}\hspace{5ex}
	\includegraphics[scale=0.65]{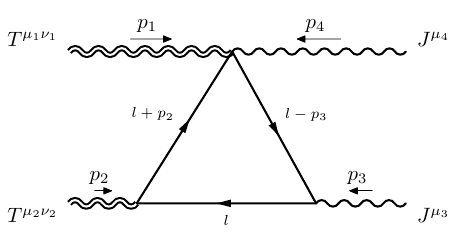}\hspace{5ex}
	\includegraphics[scale=0.65]{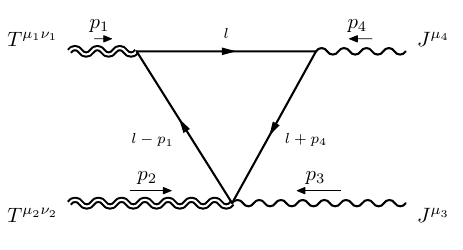}
	\caption{Examples of triangle diagrams of the triangle topology.\label{TriangleDiag}}
\end{figure} 

\begin{figure}[h]
	\centering
	\includegraphics[scale=0.65]{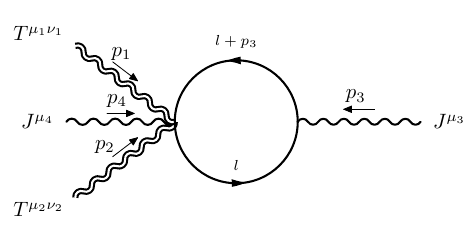}\hspace{8ex}
	\includegraphics[scale=0.65]{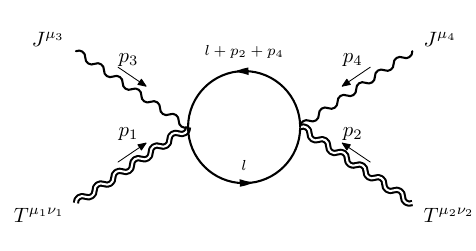}
	\caption{Examples of bubble diagrams of bubble topology.\label{BubbleDiag}}
\end{figure}

A few additional examples of triangle diagrams contributing to the $TTJJ$ can be found in \figref{TriangleDiag}. The first diagram, for instance, is expressed as
\begin{align}
	V^{\mu_1\nu_1\mu_2\nu_2\mu_3\mu_4}_{2,fermion}(p_1,p_2,p_3,p_4)=-\int\frac{d^dl}{(2\pi)^d}\,\frac{\Tr\left[
		V_{TT \bar \psi \psi}^{\mu_1 \nu_1 \mu_2 \nu_2} (l+p_3, l-p_4)\,  \left( \slashed l + \slashed p_3\right) \, V_{J \bar \psi \psi}^{\mu_3} \,\big(\slashed{l}\big)\,V_{J \bar \psi \psi}^{\mu_4}\,\left(  \slashed{l} - \slashed p_4 \right)\right] }{l^2(l+p_3)^2(l-p_4)^2}.
\end{align}	
Other topologies are given by the bubble contributions. Some examples of these are shown in \figref{BubbleDiag}. For instance, for the first diagram we have
		\begin{align}
V_{3,fermion}^{\mu_1\nu_1\mu_2\nu_2\mu_3\mu_4}(p_1,p_2,p_3,p_4)=-\int\frac{d^dl}{(2\pi)^d}\,\frac{\Tr\left[
	V_{TTJ \bar \psi \psi}^{\mu_1 \nu_1 \mu_2 \nu_2 \mu_4} (l+p_3, l)\,  \left( \slashed l+ \slashed p_3\right) \, V_{J \bar \psi \psi}^{\mu_3} (l,l+p_3)\slashed l \right]}{l^2(l+p_3)^2}.
	\end{align}

\begin{figure}[h!]
	\centering
	\includegraphics[scale=0.45]{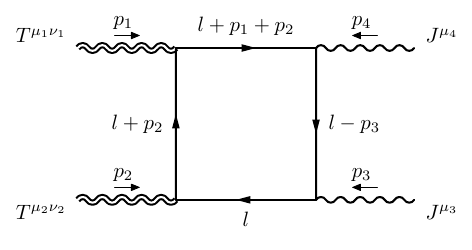}\hspace{5ex}
	\includegraphics[scale=0.45]{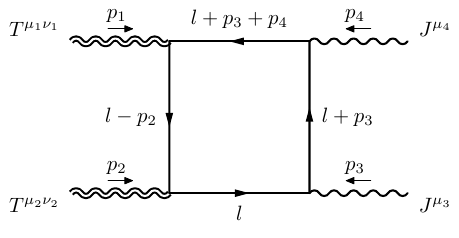}\hspace{5ex}
	\includegraphics[scale=0.45]{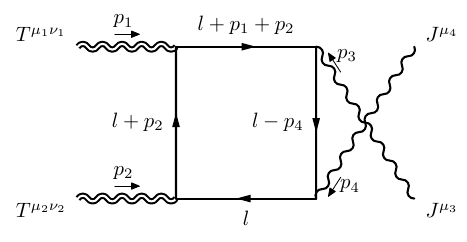}\hspace{5ex}
	\includegraphics[scale=0.45]{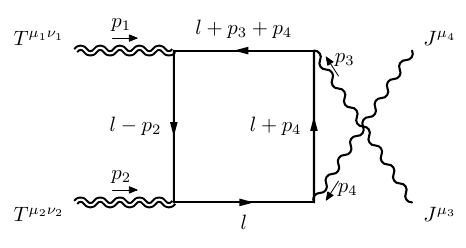}\\[1ex]
	\raisebox{-2ex}{\includegraphics[scale=0.45]{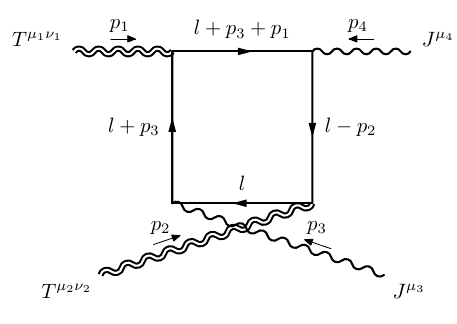}}\hspace{5ex}
	\raisebox{-2ex}{\includegraphics[scale=0.45]{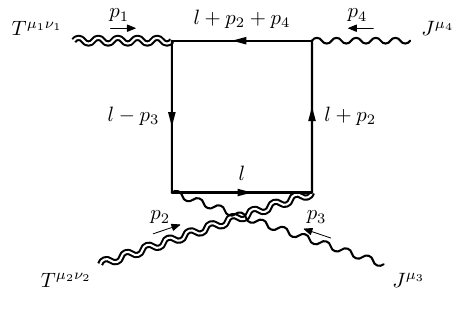}}\hspace{5ex}
	\includegraphics[scale=0.45]{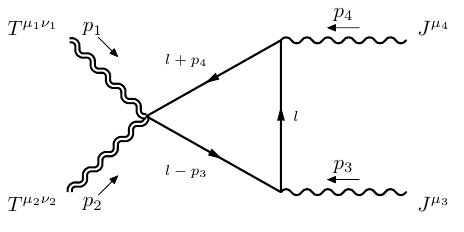}\hspace{5ex}
	\includegraphics[scale=0.45]{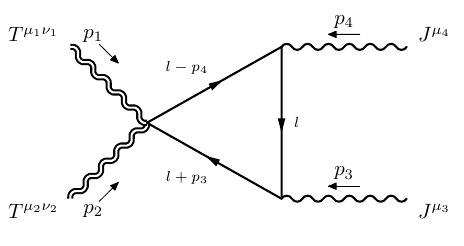}\\[1ex]
	\includegraphics[scale=0.45]{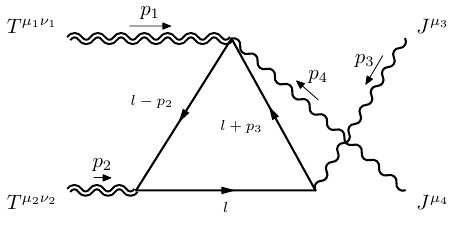}\hspace{5ex}
	\includegraphics[scale=0.45]{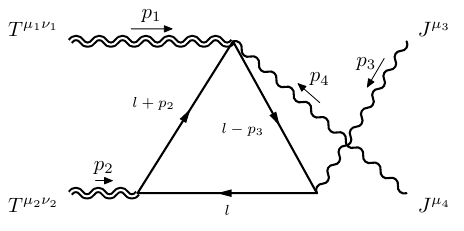}\hspace{5ex}
	\includegraphics[scale=0.45]{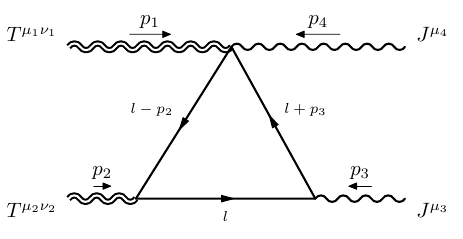}\hspace{5ex}
	\includegraphics[scale=0.45]{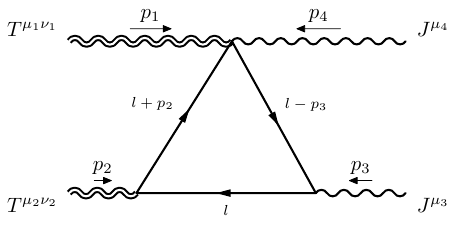}\\[1ex]
	\includegraphics[scale=0.45]{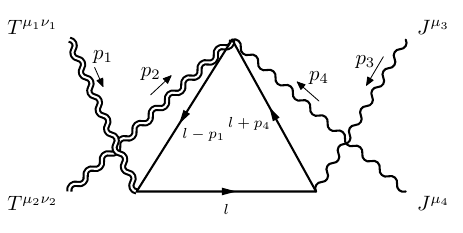}\hspace{5ex}
	\includegraphics[scale=0.45]{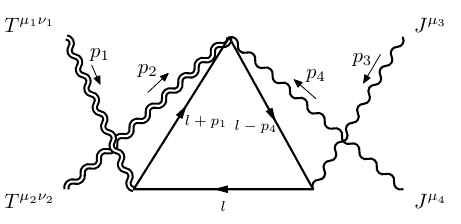}\hspace{5ex}
	\includegraphics[scale=0.45]{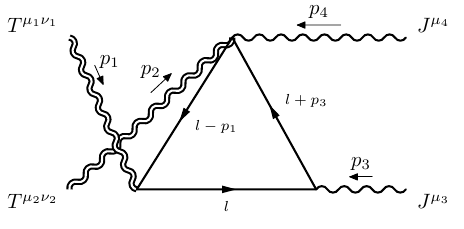}\hspace{5ex}
	\includegraphics[scale=0.45]{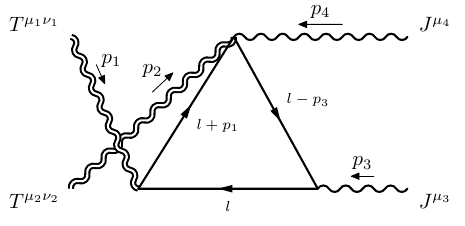}\\[1.5ex]
	\hspace{3ex}\includegraphics[scale=0.45]{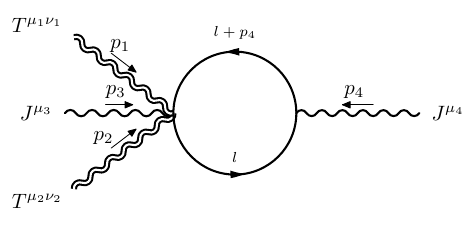}\hspace{4ex}
	\includegraphics[scale=0.45]{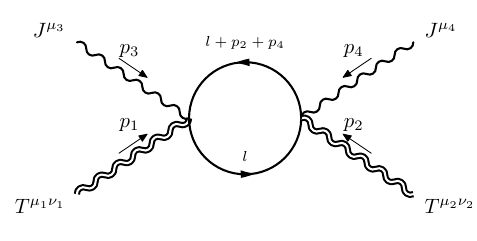}\hspace{5ex}
	\includegraphics[scale=0.45]{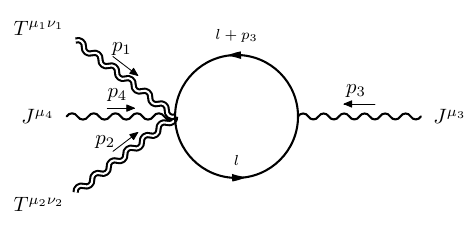}\hspace{5ex}
	\includegraphics[scale=0.45]{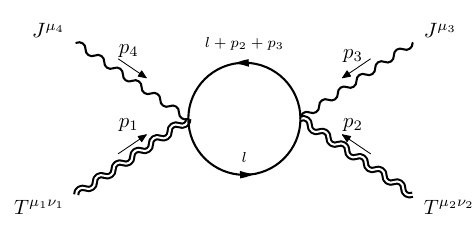}\hspace{5ex}
	\caption{Diagrams with fermions for the $TTJJ$ correlator.\label{DiagramF}}
\end{figure}

\begin{figure}
	\centering
	\includegraphics[scale=0.45]{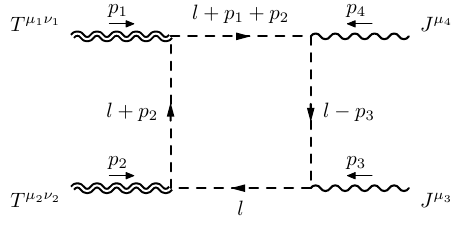}\hspace{5ex}
	\includegraphics[scale=0.45]{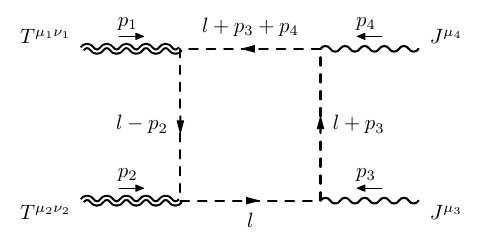}\hspace{5ex}
	\includegraphics[scale=0.45]{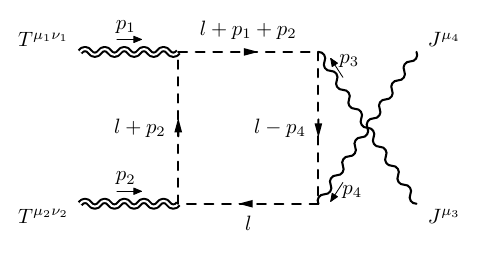}\hspace{5ex}
	\includegraphics[scale=0.45]{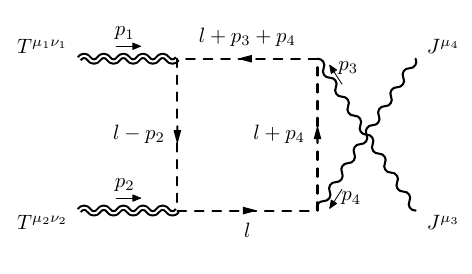}\hspace{5ex}\\[2ex]
	\raisebox{-1ex}{\includegraphics[scale=0.45]{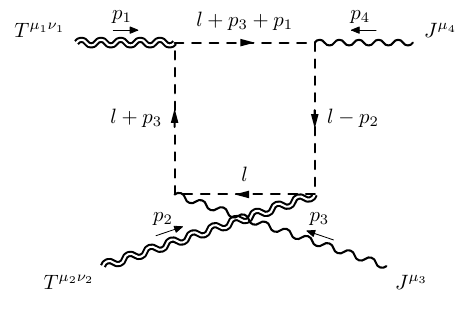}}\hspace{5ex}
	\raisebox{-1ex}{\includegraphics[scale=0.45]{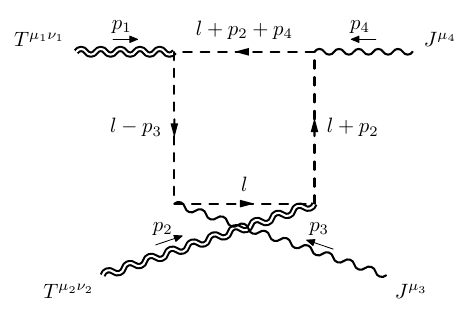}}\hspace{5ex}
	\raisebox{1ex}{\includegraphics[scale=0.45]{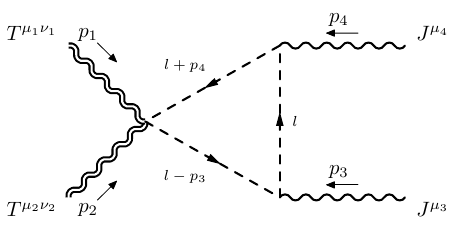}}\hspace{5ex}
	\includegraphics[scale=0.45]{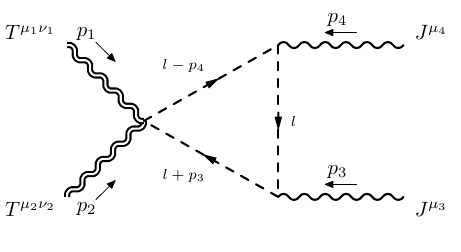}\hspace{5ex}\\[1ex]
	\includegraphics[scale=0.45]{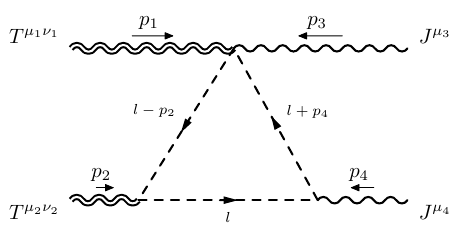}\hspace{5ex}
	\includegraphics[scale=0.45]{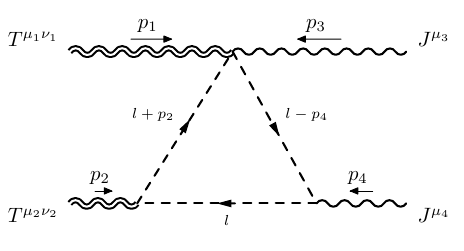}\hspace{5ex}
	\includegraphics[scale=0.45]{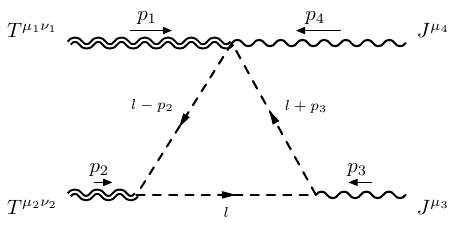}\hspace{5ex}
	\includegraphics[scale=0.45]{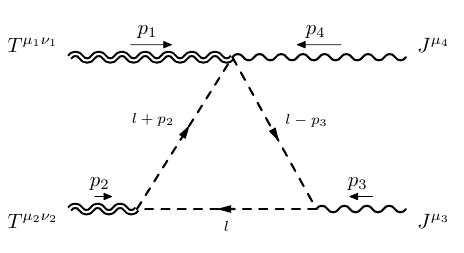}\hspace{5ex}\\[1ex]
	\includegraphics[scale=0.45]{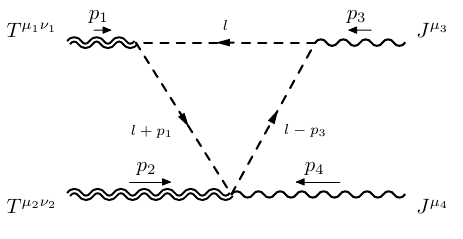}\hspace{5ex}
	\includegraphics[scale=0.45]{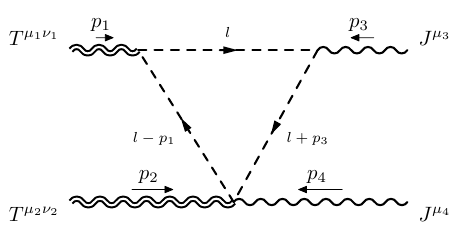}\hspace{5ex}
	\includegraphics[scale=0.45]{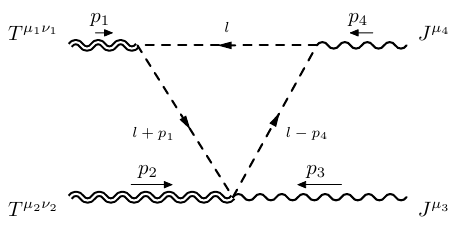}\hspace{5ex}
	\includegraphics[scale=0.45]{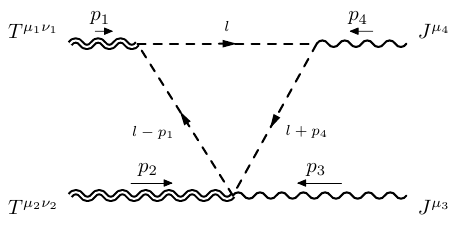}\hspace{5ex}\\[1ex]
	\includegraphics[scale=0.45]{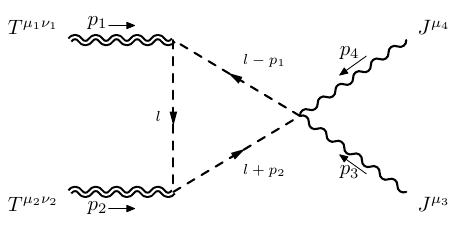}\hspace{5ex}
	\includegraphics[scale=0.45]{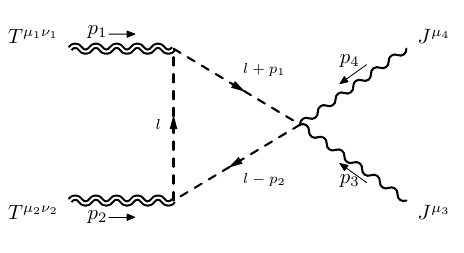}\hspace{5ex}
	\includegraphics[scale=0.45]{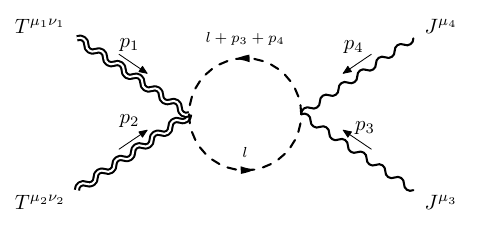}\hspace{5ex}
	\includegraphics[scale=0.45]{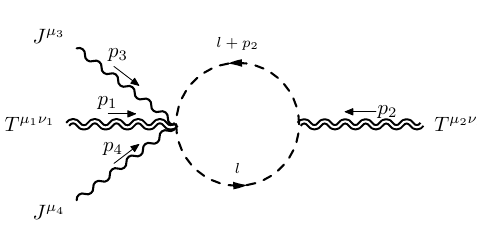}\hspace{5ex}\\[1ex]
	\includegraphics[scale=0.45]{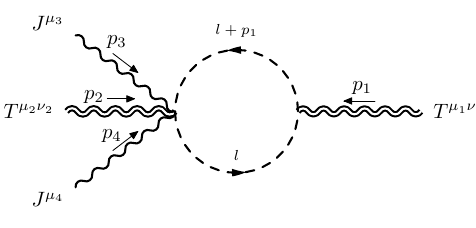}\hspace{5ex}
	\includegraphics[scale=0.45]{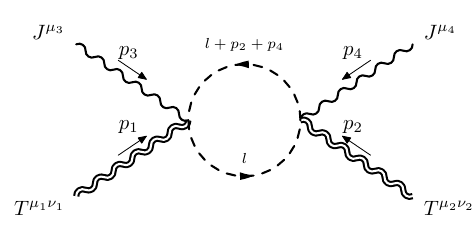}\hspace{5ex}
	\includegraphics[scale=0.45]{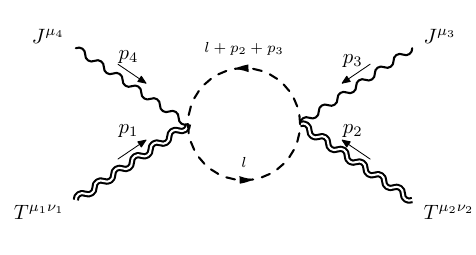}\hspace{5ex}
	\includegraphics[scale=0.45]{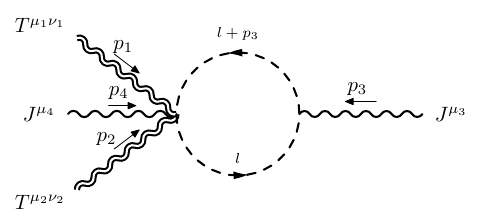}\hspace{5ex}\\[1ex]
	\includegraphics[scale=0.45]{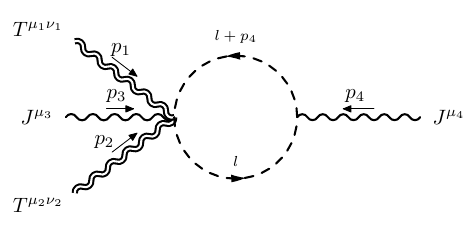}\hspace{5ex}
	\caption{Diagrams with scalars that contribute to the $TTJJ$ correlator.\label{DiagramS}}
\end{figure}

All the diagrams computed in the fermionic and scalar cases are given in \figref{DiagramF} and \figref{DiagramS}. 
The perturbative realization of the $TTJJ$ will be written down as the sum of these amplitudes, formally given by the expression
\begin{equation}
\braket{T^{\mu_1\nu_1}(p_1)T^{\mu_2\nu_2}(p_2)J^{\mu_3}(p_3)J^{\mu_4}(p_4)}=4\,\sum_{i\in\mathcal{G}} S_i\,V_i^{\mu_1\nu_1\mu_2\nu_2\mu_3\mu_4}(p_1,p_2,p_3,p_4),
\label{ssum}
\end{equation}
where $\mathcal{G}$ is the set of all the diagrams listed in \figref{DiagramF} and \figref{DiagramS}, and $S_i$ is the symmetry factor of each diagram. 
This perturbative realisation of the $TTJJ$ in $d$ dimensions satisfies the conservation and trace Ward  identities \eqref{Cons1}, \eqref{Cons2}, \eqref{Cons3}. However, around $d=4$, some loop integrals diverge and the renormalisation procedure of this correlator will lead to the appearance  of an anomalous part (the trace anomaly). Scale-breaking contributions will be naturally associated with the renormalization procedure, and they are not accounted for by the trace anomaly condition, as we will discuss in the next sections. \\
From \eqref{ssum}, by using the transverse $\pi$ and transverse traceless $\Pi$ projectors defined in \eqref{Lproj} and \eqref{TTproj}, we obtain the perturbative realization of the transverse traceless part 
\begin{equation}
	\braket{t^{\mu_1\nu_1}(p_1)t^{\mu_2\nu_2}(p_2)t^{\mu_3}(p_3)t^{\mu_4}(p_4)}=\Pi^{\mu_1\nu_1}_{\alpha_1\beta_1}(p_1)\Pi^{\mu_2\nu_2}_{\alpha_2\beta_2}(p_2)\pi^{\mu_3}_{\alpha_3}(p_3)\pi^{\mu_4}_{\alpha_4}(p_4)\left[4\,\sum_{i\in\mathcal{G}} S_i\,V_i^{\alpha_1\beta_1\alpha_2\beta_2\alpha_3\alpha_4}(p_1,p_2,p_3,p_4)\right],
\end{equation}
and then identify the form factors described in \secref{decomp}. 
The computation of all the diagrams has been performed explicitly and the expressioon of the renormalized CWI's satisfied by the form factors of the $TTJJ$ will be discussed separately for $d=4$, since they are rather involved. 

\subsection{Divergences in $d=4$ and renormalization}

	We have computed the explicit expressions of the form factors in \secref{DivAndRen} and classified the divergent ones, that coincide with the list given in \tabref{TabDg}. As already mentioned, their expressions are finite at $d=3$. Here, in this section, we focus on the analysis of the structure of their  divergences in DR at $d=4$. 	
	At $d=4$, from \tabref{TabDg}, we expect that the form factors multiplying the tensorial structures with $(2\delta,2p)$ and $(3\delta)$ are divergent. We actually find, form the perturbative calculations, that the divergent part of the transverse traceless component is written, after some manipulation, as
	\begin{align}
	&\braket{t^{\mu_1\nu_1}t^{\mu_2\nu_2}j^{\mu_3}j^{\mu_4}}^{div}=\,\Pi^{(4-\varepsilon)\,\mu_1\nu_1}_{\alpha_1\beta_1}\Pi^{(4-\varepsilon)\,\mu_2\nu_2}_{\alpha_2\beta_2}\pi^{\mu_3}_{\alpha_3}\pi^{\mu_4}_{\alpha_4}\ \left(\frac{e^2\,(4N_f+N_s)}{12\pi^2\varepsilon}\right)\,\bigg\{  \delta^{\alpha_1 \alpha_2}\delta^{\beta_1 \beta_2} p_3^{\alpha_4}p_4^{\alpha_3}\notag\\[1ex]
	&-4\,\delta^{\alpha_1 \alpha_2} \delta^{\alpha_3 (\beta_2}p_4^{\beta_1)}p_3^{\alpha_4} +4 \delta^{\alpha_1 \alpha_2} \delta^{\alpha_3 \alpha_4}p_3^{(\beta_1} p_4^{\beta_2)}-4 \delta^{\alpha_1 \alpha_2} \delta^{\alpha_4 (\beta_2}p_3^{\beta_1)} p_4^{\alpha_3} 
	+2 \delta^{\alpha_2 \alpha_3} \delta^{\alpha_4 \beta_2}p_4^{\beta_1} p_3^{\alpha_1}+2\delta^{\alpha_1 \alpha_3} \delta^{\alpha_4 \beta_1}p_4^{\beta_2} p_3^{\alpha_2}\notag\\[1ex]
	&\quad-2 p_4^{\beta_1}p_3^{\beta_2} \delta^{\alpha_1 \alpha_3} \delta^{\alpha_2 \alpha_4}-2 p_4^{\beta_2} p_3^{\beta_1} \delta^{\alpha_1 \alpha_4} \delta^{\alpha_2 \alpha_3}+(s-p_3^2+p_4^2)\left[2\delta^{\alpha_1(\alpha_4}\delta^{\alpha_3)\alpha_2}\delta^{\beta_1\beta_2}-\frac{1}{2}\delta^{\alpha_1\alpha_2}\delta^{\beta_1\beta_2}\delta^{\alpha_3\alpha_4}\right]\bigg\},\label{ttjjDiv}
	\end{align}
where $N_f$ and $N_s$ indicate the the number of fermion and scalar families respectively, that are arbitrary. The projectors $\Pi$ are expanded around $d=4$ as
\begin{align}
\Pi^{(4-\varepsilon)\,\mu_1\nu_1}_{\alpha_1\beta_1}=\Pi^{(4)\,\mu_1\nu_1}_{\alpha_1\beta_1}-\frac{\varepsilon}{9}\,\pi^{\mu_1\nu_1}\pi_{\alpha_1\beta_1}+O(\varepsilon^2)
\end{align} 
with $\Pi^{(4)}$ the transverse traceless projectors defined in \eqref{TTproj} with $d=4$. 

The counterterm vertex is 
\begin{equation}
\braket{T^{\mu_1\nu_1}T^{\mu_2\nu_2}J^{\mu_3}J^{\mu_4}}_{count}=-\frac{1}{\varepsilon}\sum_{I=f,s}\,N_I\,\beta_c(I)\,V_{F^2}^{\mu_1\nu_1\mu_2\nu_2\mu_3\mu_4}(p_1,p_2,p_3,p_4),
\end{equation}
where
\begin{align}
&V_{F^2}^{\mu_1\nu_1\mu_2\nu_2\mu_3\mu_4}(p_1,p_2,p_3,p_4)=4\int\,d^dx\prod_{k=1}^4d^dx_k\left(\frac{\delta^4\left(\sqrt{-g}\,F^2\right)(x)}{\delta g_{\mu_1\nu_1}(x_1)\delta g_{\mu_2\nu_2}(x_2)\delta A_{\mu_3}(x_3)\delta A_{\mu_4}(x_4)}\right)_{g\to\delta}\,e^{i\sum_j^4p_jx_j}
\end{align}

From this counterterm we can extract its transverse traceless part that can be written as
\begin{align}
	&\braket{t^{\mu_1\nu_1}t^{\mu_2\nu_2}j^{\mu_3}j^{\mu_4}}_{count}=\,\Pi^{(4-\varepsilon)\,\mu_1\nu_1}_{\alpha_1\beta_1}\Pi^{(4-\varepsilon)\,\mu_2\nu_2}_{\alpha_2\beta_2}\pi^{\mu_3}_{\alpha_3}\pi^{\mu_4}_{\alpha_4}\ \left(-\sum_{I=s,f}\frac{8}{\varepsilon}\,\beta_c(I)\,N_I\right)\,\bigg\{  \delta^{\alpha_1 \alpha_2}\delta^{\beta_1 \beta_2} p_3^{\alpha_4}p_4^{\alpha_3} \notag\\[1ex]
	&-4\,\delta^{\alpha_1 \alpha_2} \delta^{\alpha_3 (\beta_2}p_4^{\beta_1)}p_3^{\alpha_4} +4 \delta^{\alpha_1 \alpha_2} \delta^{\alpha_3 \alpha_4}p_3^{(\beta_1} p_4^{\beta_2)}-4 \delta^{\alpha_1 \alpha_2} \delta^{\alpha_4 (\beta_2}p_3^{\beta_1)} p_4^{\alpha_3} 
	+2 \delta^{\alpha_2 \alpha_3} \delta^{\alpha_4 \beta_2}p_4^{\beta_1} p_3^{\alpha_1}+2\delta^{\alpha_1 \alpha_3} \delta^{\alpha_4 \beta_1}p_4^{\beta_2} p_3^{\alpha_2}\notag\\[1ex]
	&\hspace{1.5cm}-2 p_4^{\beta_1}p_3^{\beta_2} \delta^{\alpha_1 \alpha_3} \delta^{\alpha_2 \alpha_4}-2 p_4^{\beta_2} p_3^{\beta_1} \delta^{\alpha_1 \alpha_4} \delta^{\alpha_2 \alpha_3}+(s-p_3^2+p_4^2)\left[2\delta^{\alpha_1(\alpha_4}\delta^{\alpha_3)\alpha_2}\delta^{\beta_1\beta_2}-\frac{1}{2}\delta^{\alpha_1\alpha_2}\delta^{\beta_1\beta_2}\delta^{\alpha_3\alpha_4}\right]\bigg\},\label{ttjjCount}
\end{align}
where  $s=(p_1+p_2)^2$. From \eqref{ttjjDiv} and \eqref{ttjjCount} we notice that the divergences are removed with the choices
\begin{equation}
\beta_c(s)=\frac{e^2}{96\,\pi^2},\qquad \beta_c(f)=\frac{e^2}{24\,\pi^2}.
\end{equation}
It is worth mentioning, as expected, that these are exactly the same choices that renormalize the $2$-point function $\braket{JJ}$ and $3$-point function $\braket{TTJ}$ as well as all the other $n$-point functions involving two conserved currents.  

Furthermore, the counterterm contribution satisfies the constraints
\begin{align}
	&p_{1\mu_1}\braket{T^{\mu_1\nu_1}(p_1)T^{\mu_2\nu_2}(p_2)J^{\mu_3}(p_3)J^{\mu_4}(p_4)}_{count}=\notag\\
	& 
	=\bigg[2\ {p_2}_{\lambda_1} \delta^{\nu_1 (\mu_2} \braket{T^{\nu_2) \lambda_1} (p_1+p_2) J^{\mu_3} (p_3) J^{\mu_4} (p_4)}_{count}  - 
	{p_2}^{\nu_1}  \braket{T^{\mu_2 \nu_2} (p_1+p_2) J^{\mu_3} (p_3) J^{\mu_4} (p_4)}_{count} \bigg] \notag\\ 
	&
	+ 2 \bigg\{  \bigg[ \delta^{\nu_1 (\mu_2} p_3^{\nu_2)}  \braket{J^{\mu_3} (p_1+p_2+p_3) J^{\mu_4} (p_4)}_{count} - \delta^{\nu_1 (\mu_2} \delta^{\nu_2) \mu_3}{p_3}_{\lambda_1 } \braket{J^{\lambda_1} (p_1+p_2+p_3) J^{\mu_4} (p_4)} _{count}
	\notag\\ 
	& + \, \frac{1}{2}\delta^{\mu_3 \nu_1} p_{3\lambda_1} \braket{J^{\lambda_1}(p_1+p_3) T^{\mu_2 \nu_2}(p_2) J^{\mu_4}(p_4) }_{count}- \frac{1}{2}p_3^{\nu_1 } \braket{J^{\mu_3}(p_1+p_3) T^{\mu_2 \nu_2}(p_2) J^{\mu_4}(p_4) }_{count} \bigg]+\bigg[ (3 \tor 4) \bigg]\bigg\} ,
\end{align}
where 
\begin{align}
\braket{J^{\mu_3}(p_3)J^{\mu_4}(p_4)}_{count}&=\left(-\sum_{I=s,f}\frac{1}{\varepsilon}\,\beta_c(I)\,N_I\right)\int d^dx\,d^dx_3\,d^dx_4\left(\frac{\delta^2\,F^2(x)}{\delta A_{\mu_3}(x_3)\delta A_{\mu_4}(x_4)}\right)\,e^{ip_3x_3+ip_4x_4}\notag\\
&=\left(-\sum_{I=s,f}\frac{4}{\varepsilon}\,\beta_c(I)\,N_I\right)\bigg[\delta^{\mu_3\mu_4}(p_3\cdot p_4)-p_4^{\mu_3}p_3^{\mu_4}\bigg]
\end{align}
is the counterterm 2-point function of two photons when $p_3=-p_4$. It is the counterterm of the photon self-energy at one-loop with intermediate scalars and fermions, as clear from the sum over 
$s$ and $f$ present in the equation above. 

It is worth mentioning that this counterterm renormalizes the $2$-point function $\braket{JJ}$, perturbatively expressed as
\begin{align}
\braket{J^{\mu_3}(-p_4)J^{\mu_4}(p_4)}=\frac{e^2}{(4\pi)^2}\frac{2(d-2)N_f+N_s}{(d-1)}\bigg[\delta^{\mu_3\mu_4}p_4^2-p_4^{\mu_3}p_4^{\mu_4}\bigg]\,B_0(p_4^2),
\end{align}
with $B_0(p_4^2)$ denoting the scalar (bubble) 2-point function, where the divergent part is extracted in DR as
\begin{align}
\braket{J^{\mu_3}(-p_4)J^{\mu_4}(p_4)}_{div}=\frac{e^2}{24\pi^2}\frac{4N_f+N_s}{\varepsilon}\bigg[\delta^{\mu_3\mu_4}p_4^2-p_4^{\mu_3}p_4^{\mu_4}\bigg].
\end{align}

Coming to the the trace Ward identities, we also have 
\begin{align}
	&\delta_{\mu_1\nu_1}\braket{T^{\mu_1\nu_1}(p_1)T^{\mu_2\nu_2}(p_2)J^{\mu_3}(p_3)J^{\mu_4}(p_4)}^{count}=\notag\\[1ex]
	&=2\big[\sqrt{g}\,F^2\big]^{\mu_2\nu_2\mu_3\mu_4}(p_2,p_3,p_4)-2 \, \braket{T^{\mu_2 \nu_2} (p_1+p_2) J^{\mu_3} (p_3) J^{\mu_4} (p_4) }_{count}
\end{align}

At this point, having identified the anomalous conservation and trace WIs satisfied by the correlator, its renormalized expression is given by  
\begin{align}
\braket{ T^{\mu_1 \nu_1} (p_1)J^{\mu_3} (p_3) J^{\mu_4} (p_4) }_{Ren}&=\braket{ T^{\mu_1 \nu_1} (p_1)J^{\mu_3} (p_3) J^{\mu_4} (p_4) }_{fin}+\braket{ T^{\mu_1 \nu_1} (p_1)J^{\mu_3} (p_3) J^{\mu_4} (p_4) }_{anomaly}
\end{align}
where the anomaly contribution is given by 
\begin{align}
\braket{ T^{\mu_1 \nu_1} (p_1)J^{\mu_3} (p_3) J^{\mu_4} (p_4) }_{anomaly}=\sum_{I=s,f}\beta_c(I)\,\frac{\pi^{\mu_1\nu_1}(p_1)}{3}\Big[F^2\Big]^{\mu_3\mu_4}(p_3,p_4).\label{anomTJJ}
\end{align}
These direct computations are performed with no reference to the reconstruction procedure 
into transverse/traceless, trace and longitudinal sectors introduced in the previous sections, that allows the identification of a minimal number of form factors. 


\subsection{The reconstruction of the renormalized correlator from the 
transverse-traceless sector\label{reconstructionAnomaly}}
Once we have removed the divergences from the transverse traceless part, we are able to reconstruct the entire correlator. The transverse traceless part, after the renormalization, acquires an anomalous contributions, with an anomalous dilatation WI satisfied by the corresponding form factors. The longitudinal part, instead, is affected by the presence of the trace anomaly. In this section, we are going to discuss this point in more detail.  \\
The renormalized correlator is given by

\begin{align}
&\braket{ T^{\mu_1 \nu_1} (p_1) T^{\mu_2 \nu_2} (p_2) J^{\mu_3} (p_3) J^{\mu_4} (p_4) }_{Ren}=\notag\\ &=\Big(\braket{ T^{\mu_1 \nu_1} (p_1) T^{\mu_2 \nu_2} (p_2) J^{\mu_3} (p_3) J^{\mu_4} (p_4) }+\braket{ T^{\mu_1 \nu_1} (p_1) T^{\mu_2 \nu_2} (p_2) J^{\mu_3} (p_3) J^{\mu_4} (p_4) }_{count} \Big)_{d\to4}\notag\\
&=\braket{ T^{\mu_1 \nu_1} (p_1) T^{\mu_2 \nu_2} (p_2) J^{\mu_3} (p_3) J^{\mu_4} (p_4) }_{fin}+\braket{ T^{\mu_1 \nu_1} (p_1) T^{\mu_2 \nu_2} (p_2) J^{\mu_3} (p_3) J^{\mu_4} (p_4) }_{anomaly},
\end{align}
where the bare correlator and the counterterm are re-expressed in terms of a finite renormalized correlator not contributing to the trace Ward identity, and a second part which accounts for the trace anomaly.
In order to show this, we consider the first term in the longitudinal part of the correlator in \eqref{loc}, for which the divergent contribution is given by
\begin{align}
&\braket{ t_{loc}^{\mu_1 \nu_1}T^{\mu_2 \nu_2} J^{\mu_3} J^{\mu_4}}_{div} =\left(\mathcal{I}^{\mu_1\nu_1}_{\alpha_1}p_{1\,\beta_1} +\frac{\pi^{\mu_1\nu_1}(p_1)}{(d-1)}\delta_{\alpha_1\beta_1}\right)\braket{T^{\alpha_1\beta_1}T^{\mu_2 \nu_2} J^{\mu_3} J^{\mu_4}}_{div}\notag\\
&=\mathcal{I}^{\mu_1\nu_1}_{\alpha_1}\bigg\{\bigg[2\ {p_2}_{\lambda_1} \delta^{\alpha_1 (\mu_2} \braket{T^{\nu_2) \lambda_1} (p_1+p_2) J^{\mu_3} (p_3) J^{\mu_4} (p_4)}_{div}  - 
{p_2}^{\alpha_1}  \braket{T^{\mu_2 \nu_2} (p_1+p_2) J^{\mu_3} (p_3) J^{\mu_4} (p_4)} _{div}\bigg] \notag\\ 
&
+ 2 \bigg[  \bigg( \delta^{\alpha_1 (\mu_2} p_3^{\nu_2)}  \braket{J^{\mu_3} (p_1+p_2+p_3) J^{\mu_4} (p_4)}_{div} - \delta^{\alpha_1 (\mu_2} \delta^{\nu_2) \mu_3}{p_3}_{\lambda_1 } \braket{J^{\lambda_1} (p_1+p_2+p_3) J^{\mu_4} (p_4)}_{div}
\notag\\ 
& + \, \frac{1}{2}\delta^{\mu_3 \alpha_1} p_{3\lambda_1} \braket{J^{\lambda_1}(p_1+p_3) T^{\mu_2 \nu_2}(p_2) J^{\mu_4}(p_4) }_{div}- \frac{1}{2}p_3^{\alpha_1 } \braket{J^{\mu_3}(p_1+p_3) T^{\mu_2 \nu_2}(p_2) J^{\mu_4}(p_4) }_{div}\bigg)+\bigg((3 \tor 4) \bigg)\bigg]
\bigg\}\notag\\
&-2\,\frac{\pi^{\mu_1\nu_1}(p_1)}{(d-1)}\braket{T^{\mu_2 \nu_2} (p_1+p_2) J^{\mu_3} (p_3) J^{\mu_4}(p_4) }_{div} .
\end{align}
By adding the counterterm part, we are able to renormalize this expression, and taking the limit $d\to4$ we find
\begin{align}
&\braket{ t_{loc}^{\mu_1 \nu_1}T^{\mu_2 \nu_2} J^{\mu_3} J^{\mu_4}}_{Ren}=\bigg(\braket{ t_{loc}^{\mu_1 \nu_1}T^{\mu_2 \nu_2} J^{\mu_3} J^{\mu_4}}_{div}+\braket{ t_{loc}^{\mu_1 \nu_1}T^{\mu_2 \nu_2} J^{\mu_3} J^{\mu_4}}_{count}\bigg)_{d\to4}\notag\\
&=\braket{ t_{loc}^{\mu_1 \nu_1}T^{\mu_2 \nu_2} J^{\mu_3} J^{\mu_4}}_{fin}^{(d=4)}+\braket{ t_{loc}^{\mu_1 \nu_1}T^{\mu_2 \nu_2} J^{\mu_3} J^{\mu_4}}_{anom}^{(d=4)},
\end{align}
where the anomalous part is explicitly given by
\begin{align}
	&\braket{ t_{loc}^{\mu_1 \nu_1}T^{\mu_2 \nu_2} J^{\mu_3} J^{\mu_4}}_{anom}^{(d=4)}=\notag\\
	&=\bigg\{\mathcal{I}^{(d=4)\,\mu_1\nu_1}_{\alpha_1}\bigg[2\ {p_2}_{\lambda_1} \delta^{\alpha_1 (\mu_2} \braket{T^{\nu_2) \lambda_1} (p_1+p_2) J^{\mu_3} (p_3) J^{\mu_4} (p_4)}_{anom}  - 
	{p_2}^{\alpha_1}  \braket{T^{\mu_2 \nu_2} (p_1+p_2) J^{\mu_3} (p_3) J^{\mu_4} (p_4)} _{anom} \notag\\ 
	&
	+ \bigg( \delta^{\mu_3 \alpha_1} p_{3\lambda_1} \braket{J^{\lambda_1}(p_1+p_3) T^{\mu_2 \nu_2}(p_2) J^{\mu_4}(p_4) }_{anom}- p_3^{\alpha_1 } \braket{J^{\mu_3}(p_1+p_3) T^{\mu_2 \nu_2}(p_2) J^{\mu_4}(p_4) }_{anom}\bigg)\notag\\
	&+ \bigg( \delta^{\mu_4 \alpha_1} p_{4\lambda_1} \braket{J^{\lambda_1}(p_1+p_4) T^{\mu_2 \nu_2}(p_2) J^{\mu_3}(p_3) }_{anom}- p_4^{\alpha_1 } \braket{J^{\mu_4}(p_1+p_4) T^{\mu_2 \nu_2}(p_2) J^{\mu_3}(p_3) }_{anom}\bigg)\bigg]\notag\\
	&-2\,\frac{\pi^{\mu_1\nu_1}(p_1)}{3}\braket{T^{\mu_2 \nu_2} (p_1+p_2) J^{\mu_3} (p_3) J^{\mu_4}(p_4) }_{anom} +2\frac{\pi^{\mu_1\nu_1}(p_1)}{3}\big[\sqrt{g}\,F^2\big]^{\mu_2\nu_2\mu_3\mu_4}(p_2,p_3,p_4)\bigg\},
\end{align}
and where the $\braket{TJJ}_{anom}$ 3-point function is given in \eqref{anomTJJ}. We make this contribution explicit by using \eqref{anomTJJ} to obtain
\begin{align}
	&\braket{ t_{loc}^{\mu_1 \nu_1}T^{\mu_2 \nu_2} J^{\mu_3} J^{\mu_4}}_{anom}^{(d=4)}=\notag\\
&=\beta_C\,\bigg\{\mathcal{I}^{(d=4)\,\mu_1\nu_1}_{\alpha_1}\bigg[2\ {p_2}_{\lambda_1}  \frac{\delta^{\alpha_1 (\mu_2}\pi^{\nu_2) \lambda_1} (p_1+p_2)}{3}\left[F^2\right]^{\mu_3\mu_4}(p_3,p_4) - 
{p_2}^{\alpha_1}  \frac{\pi^{\mu_2 \nu_2} (p_1+p_2) }{3}\left[F^2\right]^{\mu_3\mu_4}(p_3,p_4) \notag\\ 
&
+ \frac{\pi^{\mu_2\nu_2}(p_2)}{3}\bigg( \delta^{\mu_3 \alpha_1} p_{3\lambda_1} \left[F^2\right]^{\lambda_1\mu_4}(p_1+p_3,p_4)- p_3^{\alpha_1 } \left[F^2\right]^{\mu_3\mu_4}(p_1+p_3,p_4)\bigg)\notag\\
&+ \frac{\pi^{\mu_2\nu_2}(p_2)}{3}\bigg( \delta^{\mu_4 \alpha_1} p_{4\lambda_1} \left[F^2\right]^{\lambda_1\mu_3}(p_1+p_4,p_3)- p_4^{\alpha_1 } \left[F^2\right]^{\mu_3\mu_4}(p_3,p_1+p_4)\bigg)\bigg]\notag\\
&-2\,\frac{\pi^{\mu_1\nu_1}(p_1)}{3}\frac{\pi^{\mu_2\nu_2}(p_1+p_2)}{3} \left[F^2\right]^{\mu_3\mu_4}(p_3,p_4)+2\frac{\pi^{\mu_1\nu_1}(p_1)}{3}\big[\sqrt{g}\,F^2\big]^{\mu_2\nu_2\mu_3\mu_4}(p_2,p_3,p_4)\bigg\}\label{tlocTJJ},
\end{align}
where we have defined 
\begin{equation}
\beta_C\equiv\left(\sum_{I=s,f}\beta_c(I)\right).
\end{equation}
Then, the second term in \eqref{loc} is obtained from $\braket{ T^{\mu_1 \nu_1}t_{loc}^{\mu_2 \nu_2} J^{\mu_3} J^{\mu_4}}_{anom}^{(d=4)}$ with the replacement $(1\leftrightarrow2)$. Similarly, one writes the last term $\braket{t_{loc}t_{loc}JJ}_{anom}$, as presented in \appref{moreReconstr}, in order to write the entire anomalous contribution given as
 \begin{align}
\braket{ T^{\mu_1 \nu_1}T^{\mu_2 \nu_2} J^{\mu_3} J^{\mu_4}}_{anom}^{(d=4)}&=\braket{ t_{loc}^{\mu_1 \nu_1}T^{\mu_2 \nu_2} J^{\mu_3} J^{\mu_4}}_{anom}^{(d=4)}+\braket{T^{\mu_1 \nu_1}t_{loc}^{\mu_2 \nu_2} J^{\mu_3} J^{\mu_4}}_{anom}^{(d=4)}-\braket{ t_{loc}^{\mu_1 \nu_1}t_{loc}^{\mu_2 \nu_2} J^{\mu_3} J^{\mu_4}}_{anom}^{(d=4)}\notag\\
&=\braket{T^{\mu_1 \nu_1}T^{\mu_2 \nu_2} J^{\mu_3} J^{\mu_4}}_{0-residue}+\braket{ T^{\mu_1 \nu_1}T^{\mu_2 \nu_2} J^{\mu_3} J^{\mu_4}}_{pole}\label{resultFinal}.
 \end{align}
In this expression we have defined 
\begin{align}
\braket{ T^{\mu_1 \nu_1}T^{\mu_2 \nu_2} J^{\mu_3} J^{\mu_4}}_{pole}&=\beta_C\ \bigg\{2\frac{\pi^{\mu_1\nu_1}(p_1)}{3}\left(\big[\sqrt{g}\,F^2\big]^{\mu_2\nu_2\mu_3\mu_4}(p_2,p_3,p_4)-\frac{\pi^{\mu_2\nu_2}(p_1+p_2)}{3} \left[F^2\right]^{\mu_3\mu_4}(p_3,p_4)\right)\notag\\
&+2\frac{\pi^{\mu_2\nu_2}(p_2)}{3}\left(\big[\sqrt{g}\,F^2\big]^{\mu_1\nu_1\mu_3\mu_4}(p_1,p_3,p_4)-\frac{\pi^{\mu_1\nu_1}(p_1+p_2)}{3} \left[F^2\right]^{\mu_3\mu_4}(p_3,p_4)\right)\notag\\
&+2\,\frac{\pi^{\mu_1\nu_1}(p_1)}{3}\frac{\pi^{\mu_2\nu_2}(p_2)}{3}\,\Big[F^2\Big]^{\mu_3\mu_4}(p_3,p_4)\bigg\}\label{pole}
\end{align}
with the properties
\begin{align}
\delta_{\mu_i\nu_i}\braket{ T^{\mu_1 \nu_1}T^{\mu_2 \nu_2} J^{\mu_3} J^{\mu_4}}_{pole}&=2\beta_C\,\left(\big[\sqrt{g}\,F^2\big]^{\mu_j\nu_j\mu_3\mu_4}(p_j,p_3,p_4)-\frac{\pi^{\mu_j\nu_j}(p_1+p_2)}{3} \left[F^2\right]^{\mu_3\mu_4}(p_3,p_4)\right),\\
p_{\mu_i}\braket{ T^{\mu_1 \nu_1}T^{\mu_2 \nu_2} J^{\mu_3} J^{\mu_4}}_{pole}&=\frac{2\beta_C\,\pi^{\mu_j\nu_j}(p_j)}{3}\,\,p_{i\mu_i}\left(\big[\sqrt{g}\,F^2\big]^{\mu_i\nu_i\mu_3\mu_4}(p_i,p_3,p_4)-\frac{\pi^{\mu_i\nu_i}(p_1+p_2)}{3} \left[F^2\right]^{\mu_3\mu_4}(p_3,p_4)\right),
\end{align}
where $i\ne j\in\{1,2\}$, and the $0$-residue term, instead, is written explicitly as
\begin{align}
	&\braket{T^{\mu_1 \nu_1}T^{\mu_2 \nu_2} J^{\mu_3} J^{\mu_4}}_{0-residue}=\notag\\
	&=\beta_C\,\bigg\{\mathcal{I}^{(d=4)\,\mu_1\nu_1}_{\alpha_1}\Pi^{\mu_2\nu_2}_{\alpha_2\beta_2}\left(\frac{2}{3}\ {p_2}_{\lambda_1}  \delta^{\alpha_1 \alpha_2}\pi^{\beta_2\lambda_1} (p_1+p_2)- 
	\frac{1}{3}\,{p_2}^{\alpha_1}  \pi^{\alpha_2 \beta_2} (p_1+p_2) \right)\left[F^2\right]^{\mu_3\mu_4}(p_3,p_4)\notag\\
	&\hspace{1.5cm}+\mathcal{I}^{(d=4)\,\mu_2\nu_2}_{\alpha_2}\Pi^{\mu_1\nu_1}_{\alpha_1\beta_1}\left(\frac{2}{3}\ {p_1}_{\lambda_2}  \,\delta^{\alpha_1 \alpha_2}\pi^{\beta_1\lambda_2} (p_1+p_2)- 
	\frac{1}{3}{p_1}^{\alpha_2}  \,\pi^{\alpha_1 \beta_1} (p_1+p_2) \right)\left[F^2\right]^{\mu_3\mu_4}(p_3,p_4)\notag\\	
	&\hspace{1.5cm}+\mathcal{I}^{(d=4)\,\mu_1\nu_1}_{\alpha_1}\mathcal{I}^{(d=4)\,\mu_2\nu_2}_{\alpha_2}\ \delta^{\alpha_1\alpha_2}\,p_{2\lambda_1}\,p_{2\lambda_2}\frac{\pi^{\lambda_1\lambda_2}(p_1+p_2)}{3}\,\Big[F^2\Big]^{\mu_3\mu_4}(p_3,p_4)\bigg\},\label{0-residue}
\end{align}
with the property
\begin{align}
&\delta_{\mu_i\nu_i}\braket{ T^{\mu_1 \nu_1}T^{\mu_2 \nu_2} J^{\mu_3} J^{\mu_4}}_{0-residue}=0,\\
&p_{1\mu_1}\braket{ T^{\mu_1 \nu_1}T^{\mu_2 \nu_2} J^{\mu_3} J^{\mu_4}}_{0-residue}=\notag\\
&=\beta_C\,\bigg\{\Pi^{\mu_2\nu_2}_{\alpha_2\beta_2}\left(\frac{2}{3}\ {p_2}_{\lambda_1}  \delta^{\nu_1 \alpha_2}\pi^{\beta_2\lambda_1} (p_1+p_2)- 
\frac{1}{3}\,{p_2}^{\nu_1}  \pi^{\alpha_2 \beta_2} (p_1+p_2) \right)\left[F^2\right]^{\mu_3\mu_4}(p_3,p_4)\notag\\
&\hspace{1.5cm}+\ \delta^{\nu_1\alpha_2}\,p_{2\lambda_1}\,p_{1\lambda_2}\frac{\pi^{\lambda_1\lambda_2}(p_1+p_2)}{3}\,\Big[F^2\Big]^{\mu_3\mu_4}(p_3,p_4)\bigg\},\notag\\
&p_{1\mu_1}p_{2\mu_2}\braket{ T^{\mu_1 \nu_1}T^{\mu_2 \nu_2} J^{\mu_3} J^{\mu_4}}_{0-residue}=\beta_C\,\bigg\{\delta^{\nu_1\alpha_2}\,p_{2\lambda_1}\,p_{2\lambda_2}\frac{\pi^{\lambda_1\lambda_2}(p_1+p_2)}{3}\,\Big[F^2\Big]^{\mu_3\mu_4}(p_3,p_4)\bigg\}.\label{0residue}
\end{align}
These relations show that the anomaly part of the action satisfies the conservation Ward identity
\begin{align}
\label{int}
	&p_{1\mu_1}\braket{T^{\mu_1\nu_1}(p_1)T^{\mu_2\nu_2}(p_2)J^{\mu_3}(p_3)J^{\mu_4}(p_4)}_{anom}=\notag\\
	& 
	=\bigg[2\ {p_2}_{\lambda_1} \delta^{\nu_1 (\mu_2} \braket{T^{\nu_2) \lambda_1} (p_1+p_2) J^{\mu_3} (p_3) J^{\mu_4} (p_4)}_{anom}  - 
	{p_2}^{\nu_1}  \braket{T^{\mu_2 \nu_2} (p_1+p_2) J^{\mu_3} (p_3) J^{\mu_4} (p_4)}_{anom} \bigg] \notag\\ 
	&
	+  \bigg\{  \bigg[  \, \delta^{\mu_3 \nu_1} p_{3\lambda_1} \braket{J^{\lambda_1}(p_1+p_3) T^{\mu_2 \nu_2}(p_2) J^{\mu_4}(p_4) }_{anom}- p_3^{\nu_1 } \braket{J^{\mu_3}(p_1+p_3) T^{\mu_2 \nu_2}(p_2) J^{\mu_4}(p_4) }_{anom} \bigg]+\bigg[ (3 \tor 4) \bigg]\bigg\} ,
\end{align}
and the anomalous trace Ward identity
\begin{align}
&\delta_{\mu_1\nu_1}\braket{T^{\mu_1\nu_1}(p_1)T^{\mu_2\nu_2}(p_2)J^{\mu_3}(p_3)J^{\mu_4}(p_4)}_{anom}=\delta_{\mu_1\nu_1}\braket{ T^{\mu_1 \nu_1}(p_1)T^{\mu_2 \nu_2}(p_2) J^{\mu_3}(p_3) J^{\mu_4}(p_4)}_{pole}\notag\\
&=2\beta_C\,\left(\big[\sqrt{g}\,F^2\big]^{\mu_2\nu_2\mu_3\mu_4}(p_j,p_3,p_4)-\frac{\pi^{\mu_2\nu_2}(p_1+p_2)}{3} \left[F^2\right]^{\mu_3\mu_4}(p_3,p_4)\right).
\end{align}
Therefore, the anomaly part of the correlator satisfies the conservation WI by itself. A direct, explicit computation of the diagrams allows to check these relations for $d=4$. The explicit anomalous conformal Ward Identities satisfied by the single form factors of the decompositions will be discussed elsewhere.

\section{4-point functions from anomaly induced actions}
In this section we turn our attention towards a study of the same correlator using the formalism of the anomaly induced actions. Actions of these type are variational solutions of the anomaly constraint. \\
We denote by $\mathcal{S}_A[g]$ the functional identified by the integration of the trace anomaly 
\begin{equation}
2\,g_{\mu\nu}\,\frac{\delta\,\,\mathcal{S}_A[g]}{\delta g_{\mu\nu}(x)}=\sqrt{-g}\bigg[b\,C^2+b'\,E+\beta_C\,F^2\bigg].
\end{equation}
This functional is obtained by a Weyl rescaling of the metric $g_{\mu\nu}(x)=e^{2\phi(x)}\,\bar{g}_{\mu\nu}(x)$, to give 
\begin{align}
	\mathcal{S}_A[g]=\mathcal{S}[\bar g]+\Delta\,\mathcal{S}[\sigma,\bar g],
\end{align}
such that its conformal variation
\begin{equation}
2\,g_{\mu\nu}\,\frac{\delta\,\mathcal{S}_A[g]}{\delta g_{\mu\nu}(x)}=\frac{\delta\,\mathcal{S}_A[g]}{\delta \phi(x)}\Bigg|_{g=e^{2\phi}\bar{g}}=\frac{\delta\,\big(\Delta \mathcal{S}[\phi,\bar g]\big)}{\delta \phi(x)}=\sqrt{-g}\bigg[b\,C^2+b'\,E+\beta_C\,F^2\bigg]\Bigg|_{g=e^{2\phi}\bar{g}}\label{Weylconstr}
\end{equation}
is the anomaly. The explicit expression for $\Delta \mathcal{S}$ can be obtained after the integration of \eqref{Weylconstr} along a path $\tilde{\phi}(x,\lambda)=\lambda\,\phi(x)$ with $0\ge\lambda\ge1$ as
\begin{align}
\Delta\mathcal{S}[\phi,\bar g]&=\int\,d^4x\int_0^1\,d\lambda\,\frac{\delta\,\big(\Delta \mathcal{S}[\phi,\bar g]\big)}{\delta \phi(x)}\Bigg|_{\phi=\tilde{\phi}(x,\lambda)}\frac{\partial\,\tilde{\phi}(x,\lambda)}{\partial\lambda}\notag\\
&=\int\,d^4x\int_0^1\,d\lambda\,\bigg[\sqrt{-g}\bigg(b\,C^2+b'\,E+\beta_C\,F^2\bigg)\bigg]_{g=e^{2\tilde\phi(x,\lambda)}\bar{g}}\,\phi(x).
\end{align}
After the integration $\Delta \mathcal{S}$ is written as
\begin{align}
	\Delta \mathcal{S}[\phi,\bar g]&=b'\,\int\,d^4x\,\sqrt{-\bar g}\bigg[2\,\phi\,\bar\Delta_4\phi+\bigg(\bar E-\frac{2}{3}\bar \square\bar R\bigg)\phi\bigg]+\int d^4x\,\sqrt{-\bar g}\bigg[b\,\bar C^2+\beta_C\,\bar F^2\bigg]\phi,\label{deltaW}
\end{align}
up to terms which are $\sigma$ independent, \textit{i.e.} conformally invariant, and hence do not contribute to the variation \eqref{Weylconstr}. In \eqref{deltaW} we have defined the fourth order operator $\Delta_4$ as
\begin{equation}
	\Delta_4=\nabla_{\mu}\left(\nabla^\mu\nabla^\nu+2R^{\mu\nu}-\frac{2}{3}R\,g^{\mu\nu}\right)\nabla_\nu=\square^2+2R^{\mu\nu}\,\nabla_{\mu}\nabla_\nu-\frac{2}{3}R\,\square+\frac{1}{3}\,\nabla^\mu R\,\nabla_{\mu}.\label{Delta4}
\end{equation}
In order to obtain a nonlocal form of the anomaly induced action, one can impose a condition $\chi(g)=0$,
such that the relation $\chi(g\,e^{-2\phi})=0$ is satisfied. From this condition one solves for $\phi$ in terms of a function of the metric, mainly $\phi=\Sigma(g)$, such that
\begin{equation}
	\bar{g}_{\mu\nu}=e^{-2\,\Sigma(g)}\,g_{\mu\nu}, \quad g_{\alpha\beta}\frac{\delta}{\delta g_{\alpha\beta}}\,\bar{g}_{\mu\nu}[g]=0.
\end{equation}
Thus the conformal decomposition is expressed as 
\begin{align}
	\mathcal{S}[g]&=\bar{\mathcal{S}}[g]+\mathcal{S}_A[g,\Sigma(g)]\\
	\mathcal{S}_A[g,\Sigma(g)]&=\,\int\,d^4x\,\sqrt{- g}\bigg[b'\bigg( E-\frac{2}{3} \square R\bigg)\Sigma+\bigg(b\, C^2+\beta_C\, F^2\bigg)\Sigma-2 b'\,\Sigma\,\Delta_4\Sigma\bigg]\label{WA}
\end{align}
where $\bar{\mathcal{S}}[g]$ and $\mathcal{S}_A[g,\Sigma(g)]$ are its conformal-invariant and anomalous parts respectively.
There are two distinct exactly solvable conformal choices in $4D$, in which the gauge parameter $\Sigma(g)$ can be calculated in a closed form, as a functional of $g$. They are discussed in \cite{Barvinsky:1995it}. One is due to Fradkin and Vilkovisky (FV) \cite{Fradkin:1978yw}
\begin{align}
	\chi_{FV}(g)=R(g)
\end{align}
with 
\begin{align}
	\Sigma_{FV}=-\ln\left(1+\frac{1}{6}(\square-R/6)^{-1}\,R\right)\label{FVgauge}
\end{align}
where $(\square-R/6)^{-1}$ is the inverse of the operator $\square-R/6$, playing the role of the corresponding Faddeev-Popov operator. Another choice is due to Riegert \cite{Riegert:1984kt} with
\begin{align}
	\chi_{R}(g)=E(g)-\frac{2}{3}\square\,R(g),
\end{align}
for which
\begin{align}
	\Sigma_{R}=\frac{1}{4}\,(\Delta_4)^{-1}\left(E-\frac{2}{3}\square\,R\right),\label{Rgauge}
\end{align}
with $(\Delta_4)^{-1}$ the inverse of the operator in \eqref{Delta4}. 

\subsection{Variation of the anomaly effective action}

In general, the two choices of $\Sigma$ should be equivalent, giving the same result when the action is varied with respect to the metric. In order to see if this is the case, we expand $\mathcal{S}_A[g,\Sigma(g)]$ around flat space by the  simultaneous expansions of the metric $g_{\mu\nu}$, the gauge field $A_\mu$, and the local factor $\Sigma$
	\begin{subequations}
		\begin{align}
		g_{\mu \nu} & = g_{\mu \nu}^{(0)} +g_{\mu \nu}^{(1)}+g_{\mu \nu}^{(2)}+\dots\equiv  \eta_{\mu \nu} + h_{\mu \nu} + h^{(2)}_{\mu \nu} + \dots \\
		A_\mu &= A^{(0)}_\mu + A^{(1)}_\mu + A^{(2)}_\mu + \dots\\
		\Sigma&=\Sigma^{(0)}+ \Sigma^{(1)}+ \Sigma^{(2)}+ \dots.
	\end{align}
	\end{subequations}
The local effective action from \eqref{WA} is then written at the first order in $h_{\mu\nu}$ as
\begin{align}
\mathcal{S}^{(1)}_A[g,\Sigma(g)]&=\beta_C\,\int\,d^4x\,F^2\,\Sigma^{(1)}\label{1st}
\end{align}
neglecting the terms that are pure gravitational. We are assuming that $\Sigma^{(0)}=0$ as we will prove it in the next sections for the two choices of the parameter $\Sigma$. The first order in the fluctuation $h_{\mu\nu}$ of the expansion is related to the structure of the $TJJ$ anomaly contribution. In order to extract the contribution to the $TTJJ$ we need the second order in $h_{\mu\nu}$, given as
\begin{align}
	\mathcal{S}^{(2)}_A[g,\Sigma(g)]&=\beta_C\int\,d^4x\bigg\{\,F^2\,\Sigma^{(2)}+(\sqrt{-g}F^2)^{(1)}\,\Sigma^{(1)}\bigg\}.\label{2nd}
\end{align}
\subsection{The Riegert Action}
We consider first the Riegert choice, for which
\begin{equation} 
	\sqrt{-g}\,\Delta_4\,\Sigma_R = {1 \over 4}\Big(E - {2 \over 3}\,\square R \Big),\label{Riegert}
\end{equation}
where we have used the relation on the Green function of the $\Delta_4$. 
Expanding \eqref{Riegert} and taking order by order in the power of the fluctuations we determine the relations
\begin{align} 
	\Sigma_R^{(0)} &= 0, \notag\\
	\Sigma_R^{(1)} &= -{1 \over 6} \, {1 \over \square_0} \, R^{(1)}, \notag\\
	\Sigma_R^{(2)} &=  {1 \over 4}\, {1 \over \square_0^2}\, E^{(2)} +{1 \over 6}\, {1 \over \square_0^2} \, (\sqrt{-g} \Delta_4)^{(1)} \, {1 \over \square_0}\, R^{(1)} - {1 \over 6} \, {1 \over \square_0^2} \, \square_1\, R^{(1)} - {1 \over 6} \, {1 \over \square_0} \, R^{(2)},
\end{align}
Substituting this expression into \eqref{1st} we obtain the contribution to the $TJJ$ as
\begin{align}
	\mathcal{S}_A^{(1)} [g , \Sigma_R (g)]&=\beta_C\,\int\,d^4x\,F^2\,\Sigma_R^{(1)}=-\frac{\beta_C}{6}\int\,d^4x\,F^2\,\frac{1}{\square_0} \, R^{(1)},
\end{align}
while the second order term is expressed as
\begin{align} 
	\mathcal{S}_A^{(2)} [g , \Sigma_R (g)] &= 
	\beta_C \,\int d^4 x \, \Big( (\sqrt{-g} \, F^2)^{(1)} \, \Sigma_R^{(1)} + F^2 \, \Sigma_R^{(2)} \Big) \notag\\
	&=
	\beta_C  \int d^4 x \, \Big(-{1 \over 6} \, (\sqrt{-g} \, F^2)^{(1)} \, {1 \over \square_0} \, R^{(1)}  
	+ {1 \over 4} \, F^2 {1 \over \square_0^2} \, E^{(2)}
	+ {1 \over 6} \, F^2 {1 \over \square_0^2} \, (\sqrt{-g} \Delta_4)^{(1)} \, {1 \over \square_0} \, R^{(1)} \notag\\ &
	\qquad - {1 \over 6} \, F^2 {1 \over \square_0^2} \, \square_1 \, R^{(1)} 
	- \, {1 \over 6} \, F^2 \, {1 \over \square_0} \, R^{(2)} \Big),
\end{align}
and by using the expansion of the Paneitz operator $\Delta_4$
\begin{equation} ( \sqrt{-g} \Delta_4 )^{(1)} = (\sqrt{-g} \, \square^2)^{(1)} + 2 \, \partial_\mu \, \Big( R^{\mu \nu} - {1 \over 3} g^{\mu \nu}\,R \Big)^{(1)} \partial_\nu  
\end{equation}
with
\begin{align} 
	(\sqrt{-g} \, \square^2)^{(1)} &=
	(\sqrt{-g} \, \square)^{(1)} \square_0  + \square_0\,\square_1
\end{align}
and integrating by parts we obtain
\begin{align} 
	\mathcal{S}_A^{(2)} [g , \Sigma_R (g)] &=
	-\frac{\beta_C}{2}\,\int d^4 x \int d^4 x' \, \Bigg\{\,
	{1 \over 3}\, (\sqrt{-g} \, F^2)_x^{(1)} \left( {1 \over \square_0} \right)_{xx'}R^{(1)}_{x'}  
	- {1 \over 2}\, F^2_x \left( {1 \over \square_0^2} \right)_{xx'} E^{(2)}_{x'}
	\notag\\ &\hspace{-2cm}
	- {1\over 3}\int d^4 x'' \left[2\, F^2_x\left( {1 \over \square_0^2} \right)_{xx'} \partial_\mu \Big( R^{\mu \nu} 
	- {1 \over 3} \eta^{\mu \nu} R \Big)^{(1)}_{x'} \partial_\nu \left( {1 \over \square_0} \right)_{x' x''} R^{(1)}_{x''} 
	+\, F^2_x \left( {1 \over \square_0} \right)_{xx'}( \square_1)_{x'} \left( {1 \over \square_0} \right)_{x'x''}R^{(1)}_{x''} \right]\notag\\
	&
	+ {1 \over 3}\, F^2_x \left( {1 \over \square_0} \right)_{xx'} R^{(2)}_{x'} + {1 \over 3}\, F^2_x \left( {1 \over \square_0} \right)_{xx'} \big(\sqrt{-g}\big)^{(1)}R^{(1)}_{x'}\Bigg\}
\end{align}
In the expression above, the three terms affected by the Green function of $\Box^2$  identify logarithmic contributions which can be correctly defined only by the inclusion of a 
scale $\mu_F$, in order to define their Fourier transform to momentum space. Indeed they can be correctly  transformed to momentum space only in  the presence of such cutoff. 
Obviously, one would expect such cutoff to disappear from the final expression, in such a way that the correlator satisfies, in its anomaly part, the same WIs valid for the anomaly part of its perturbative realization.

In momentum space, we obtain 
\begin{align}  
	&\braket{T^{\mu_1\nu_1}(p_1)T^{\mu_2\nu_2}(p_2)J^{\mu_3}(p_3)J^{\mu_4}(p_4)}_{anom}^{R}= \notag\\
	&=
	2\beta_C  \Bigg\{\,
	{1 \over 3}\, \big[\sqrt{-g} \, F^2\big]^{\mu_1\nu_1\mu_3\mu_4}(p_1,p_3,p_4)\pi^{\mu_2\nu_2}(p_2) +	{1 \over 3}\, \big[\sqrt{-g} \, F^2\big]^{\mu_2\nu_2\mu_3\mu_4}(p_2,p_3,p_4) \pi^{\mu_1\nu_1}(p_1)\notag\\
	&- {1 \over 2}\, \big[F^2\big]^{\mu_3\mu_4}(p_3,p_4) G(p_1+p_2) [E]^{\mu_1\nu_1\mu_2\nu_2}(p_1,p_2)+ {1 \over 3}\, \big[F^2\big]^{\mu_3\mu_4}(p_3,p_4)\,{1 \over (p_1+p_2)^2} \, \big[R\big]^{\mu_1\nu_1\mu_2\nu_2}(p_1,p_2)
	\notag\\ 
	&
	- {2\over 3} \,\big[ F^2\big]^{\mu_3\mu_4}(p_3,p_4)G(p_1+p_2)\,(p_1+p_2)_{\mu}\, \Big[ R^{\mu \nu} 
	- {1 \over 3} \eta^{\mu \nu} R \Big]^{\mu_1\nu_1}(p_1) \ p_{2\,\nu}\ \pi^{\mu_2\nu_2}(p_2)\notag\\
	&
	- {2\over 3} \,\big[ F^2\big]^{\mu_3\mu_4}(p_3,p_4)G(p_1+p_2)\,(p_1+p_2)_{\mu}\, \Big[ R^{\mu \nu} 
	- {1 \over 3} \eta^{\mu \nu} R \Big]^{\mu_2\nu_2}(p_2) \ p_{1\,\nu}\ \pi^{\mu_1\nu_1}(p_1)\notag\\
	&-\frac{1}{3}\big[F^2\big]^{\mu_3\mu_4}(p_3,p_4) {1 \over (p_1+p_2)^2}\bigg[\big[\square_1\big]^{\mu_1\nu_1}(p_1,p_2) \pi^{\mu_2\nu_2}(p_2)+\big[\square_1\big]^{\mu_2\nu_2}(p_2,p_1) \pi^{\mu_1\nu_1}(p_1)\bigg]\notag\\
	&+ {1 \over 3}\, \big[ F^2\big]^{\mu_3\mu_4}(p_3,p_4) {1 \over (p_1+p_2)^2}\left[\frac{p_2^2}{2}\delta^{\mu_1\nu_1}\,\pi^{\mu_2\nu_2}(p_2)+\frac{p_1^2}{2}\delta^{\mu_2\nu_2}\,\pi^{\mu_1\nu_1}(p_1)\right]\Bigg\}\ \delta^{(4)}\left(\sum_i^4\,p_i\right)
\end{align}
where $G(p)=c \log( p^2/\mu_F^2)$ as pointed out also in \cite{Stergiou:2022qqj,Brust:2016gjy,Nesterov:2010jh} and we have used the relations in \appref{FunctionalVariations}. 

One can easily figure out that the $\mu_F$ dependence in the sum logarithmic correlators does not cancel. As we are going to discuss next, a similar problem emerges in the case of the FV gauge. This, as we are now going to clarify, indicates an inconsistency of the expansion around the flat spacetime limit. \\
Indeed, a close look at \eqref{int}, the hierarchical equation of the conservation WI, relates on the lhs a 4-point function which remains logarithmic even after an ordinary differentiation, to 3-point functions on the rhs which are not, and are correctly described by the two anomaly-induced actions that we have considered. \\
In the case of the Riegert action, therefore, this shortcoming appears to be related to $1/\Box^2$ terms. \\
This could provide an indication that these types of action do not allow a consistent expansion starting from the quartic order.  

\subsection{Fradkin-Vilkovisky Gauge}
We consider then the Fradkin Vilkoviski choice for which 
\begin{equation} \Sigma_{FV} = - \log \left( 1 + {1 \over 6} G_{\square - R/6} \, R \right) = \sum_{k=1}^\infty (-1)^{k} {1 \over k\,2^{k} \, 3^k}  \left( G_{\square - R/6} \, R \right)^k .\label{exVil}
\end{equation}
In order to obtain the first and second order contributions in $h_{\mu\nu}$ with this choice we need the relations of expansion of the Green function of the conformal Laplacian. These are obtained considering
\begin{equation}
	\sqrt{-g}\left(\square-\frac{R}{6}\right)G_{\square - R/6}=1,
\end{equation}
and expanding and collecting the terms order by order we obtain
\begin{align} 
	G^{(0)}_{\square - R/6} &= {1 \over \square_0} \notag\\
	G^{(1)}_{\square - R/6} &= - {1 \over \square_0} \, [\sqrt{-g}\,\square]^{(1)}\, {1 \over \square_0} + {1 \over 6}\, {1 \over \square_0}\, R^{(1)}\, {1 \over \square_0} .
\end{align}
With this results we use \eqref{exVil} to get
\begin{equation}
\begin{split} 
	&\Sigma_{FV}^{(0)} = 0,\hspace{2cm}\Sigma_{FV}^{(1)} = -{1 \over 6}\, {1 \over \square_0}\, R^{(1)}, \\
	&\Sigma_{FV}^{(2)} = -{1 \over 6}\, {1 \over \square_0}\, R^{(2)} + {1 \over 6} \,{1 \over \square_0} \, [\sqrt{-g}\,\square]^{(1)} \, {1 \over \square_0}\,R^{(1)} - {1 \over 2^3 3^2} \, {1 \over \square} \, R^{(1)} \, {1 \over \square} \, R^{(1)}.\label{FV}
\end{split}
\end{equation}
It is worth noticing that the two choices are equivalent at the first order, indeed
\begin{align}
	\Sigma_R^{(1)}=\Sigma_{FV}^{(1)}
\end{align}
and then the anomalous contribution to the three point function $TJJ$ is the same by using one choice or the other. Indeed, we obtain
\begin{equation}
	\mathcal{S}_A^{(1)} [g , \Sigma_{FV} (g)]=\mathcal{S}_A^{(1)} [g , \Sigma_{R} (g)]=-\frac{\beta_C}{6}\int\,d^4x\,F^2\,\frac{1}{\square_0} \, R^{(1)},
\end{equation}
showing the equivalence of the two choices at the first order. At the second order, by inserting \eqref{FV} in the expansion \eqref{Riegert}, we have
\begin{align} 
	&\mathcal{S}_A^{(2)} [g , \Sigma_{FV} (g)] =
	-{\beta_C \over 6 } \int d^4 x \int d^4 x' \, \bigg\{ 
	(\sqrt{-g}  F^2)_x^{(1)} \left( {1 \over \square_0} \right)_{xx'} R^{(1)}_{x'}
	+ F^2_x \left( {1 \over \square_0} \right)_{xx'} R^{(2)}_{x'}
	\notag\\ 
	&\hspace{1cm}+ \int d^4 x'' \, \bigg[\, {1 \over 12} \, F^2_x \left( {1 \over \square_0} \right)_{xx'} R^{(1)}_{x'} \, \left( {1 \over \square_0} \right)_{x'x''} R^{(1)}_{x''} 
	- \, F^2_x \left( {1 \over \square_0} \right)_{xx'} [\sqrt{-g}\,\square]^{(1)}_{x'} \, \left( 1 \over \square_0 \right)_{x'x''} R^{(1)}_{x''} \, \bigg] \bigg\},\label{ExpFV}
\end{align}
noticing that at the second order the expansions with the two different choices of $\Sigma$ give distinguished results, and they differ for terms like
\begin{equation}
	\mathcal{S}_A^{(2)} [g , \Sigma_{FV} (g)]-\mathcal{S}_A^{(2)} [g , \Sigma_R (g)]  \sim {1 \over 2}\, F^2_x \left( {1 \over \square_0^2} \right)_{xx'} E^{(2)}_{x'}+\dots
\end{equation}
We compute now the contribution of the anomaly effecgive action with the Fradkin Vilkovisky gauge, to the $TTJJ$ in momentum space. From \eqref{ExpFV} and by using the variations in \appref{FunctionalVariations} we have
\begin{align} 
&\braket{T^{\mu_1\nu_1}(p_1)T^{\mu_2\nu_2}(p_2)J^{\mu_3}(p_3)J^{\mu_4}(p_4)}_{anom}^{FV}=\notag\\
&=\frac{2}{3}\beta_C\, \Bigg\{ \big[(\sqrt{-g} F^2)\big]^{\mu_2 \nu_2 \mu_3 \mu_4 }(p_2, p_3,p_4) \pi^{\mu_1\nu_1}(p_1) +\big[(\sqrt{-g} F^2)\big]^{\mu_1 \nu_1 \mu_3 \mu_4 }(p_1, p_3,p_4) \pi^{\mu_2\nu_2}(p_2) \notag\\ 
&+\big[F^2\big]^{\mu_3\mu_4}(p_3,p_4)\frac{1}{(p_1+p_2)^2}\bigg(\big[R\big]^{\mu_1\nu_1\mu_2\nu_2}(p_1,p_2)+\frac{1}{12}(p_1^2+p_2^2)\pi^{\mu_1\nu_1}(p_1)\pi^{\mu_2\nu_2}(p_2)\bigg)\notag\\
&-\big[F^2\big]^{\mu_3\mu_4}(p_3,p_4)\frac{1}{(p_1+p_2)^2}\bigg(\big[\sqrt{-g}\,\square\big]^{\mu_1\nu_1}(p_1,p_2)\pi^{\mu_2\nu_2}(p_2)+\big[\sqrt{-g}\,\square\big]^{\mu_2\nu_2}(p_2,p_1)\pi^{\mu_1\nu_1}(p_1)\bigg)\Bigg\}.
\end{align}
One can check by explicit computation in momentum space - in this case the expression is transformable by the Fourier integral - that the trace Ward identity is violated. Indeed we have
\begin{align}
&\delta_{\mu_1\nu_1}\delta_{\mu_2\nu_2}\braket{T^{\mu_1\nu_1}(p_1)T^{\mu_2\nu_2}(p_2)J^{\mu_3}(p_3)J^{\mu_4}(p_4)}_{anom}^{FV}=\notag\\
&=\frac{2\,\beta_C}{3} \Bigg\{\big[F^2\big]^{\mu_3\mu_4}(p_3,p_4)\frac{1}{(p_1+p_2)^2}\bigg(9(p_1\cdot p_2)+\frac{3}{4}(p_1^2+p_2^2)\bigg)-3\big[F^2\big]^{\mu_3\mu_4}(p_3,p_4)\Bigg\}.
\end{align}

This is, obviously, at variance with the trace Ward identity satisfied by the anomaly contribution coming from the perturbative expansion discussed in the previous sections.  In the case of the FV action, such Ward identities are expected to be expressed only in terms of hierarchical equations involving  only the 3- and 4- point functions identified by the same action. \\
This result raises some important issues concerning the consistency of the flat spacetime limit of such actions. Notice that the FV action also introduces a conformal decomposition related to a logarithmic gauge choice. The corresponding expansion, in FV, is also essentially deprived of a physical scale, although it is possible to formulate an explicit expansion of such gauge conditions in terms of the dimensionless combination  $R\Box^{-1}$.

\section{New proposal}
In this section we propose the solution of the problem discussed so far. The result we are going to discuss comes from the explicit calculation of the anomalous part of the correlator $\braket{TTJJ}$ performed in \secref{reconstructionAnomaly}. In particular we write the correct expansion at the second order in curvature of an effective action that reproduces \eqref{resultFinal}. We find that at the second order in the metric fluctuation, the desired expansion of the anomaly effective action should be
\begin{align}
	\mathcal{S}_{anom}^{(2)}&=-\frac{\beta_C}{6} \int d^4 x \int d^4 x' \, \bigg\{\,
	 (\sqrt{-g} \, F^2)_x^{(1)} \left( {1 \over \square_0} \right)_{xx'}R^{(1)}_{x'}  
	+ \, F^2_x \left( {1 \over \square_0} \right)_{xx'} R^{(2)}_{x'} 
	\notag\\ &\hspace{-2cm}
	+\int d^4 x'' \bigg[\, F^2_x \left( {1 \over \square_0} \right)_{xx'}( \square_1)_{x'} \left( {1 \over \square_0} \right)_{x'x''}R^{(1)}_{x''} -\frac{1}{6}\, F^2_x \left( {1 \over \square_0} \right)_{xx'}\,R^{(1)}_{x'}\, \left( {1 \over \square_0} \right)_{x'x''}R^{(1)}_{x''}  \notag\\
	&+\frac{1}{3}R^{(1)}_x\left( {1 \over \square_0} \right)_{xx'}\,F^2_{x'}\left( {1 \over \square_0} \right)_{x'x''}R^{(1)}_{x''}\bigg]\bigg\}.
\end{align}
The contribution to the anomaly part of the $\braket{TTJJ}$ of this expansion in momentum space reads as
\begin{align}
&\braket{T^{\mu_1\nu_1}(p_1)T^{\mu_2\nu_2}(p_2)J^{\mu_3}(p_3)J^{\mu_4}(p_4)}_{anom}=\notag\\
&=\frac{2\beta_C}{3} \bigg\{\,
(\sqrt{-g} \, F^2)^{\mu_1\nu_1\mu_3\mu_4}(p_1,p_3,p_4) \,{1 \over p_2^2}\,[R]^{\mu_2\nu_2}(p_2)+(\sqrt{-g} \, F^2)^{\mu_2\nu_2\mu_3\mu_4}(p_2,p_3,p_4) \,{1 \over p_1^2}\,[R]^{\mu_1\nu_1}(p_1)\notag\\
&+ \, [F^2]^{\mu_3\mu_4}(p_3,p_4) \,{1 \over (p_1+p_2)^2} [R]^{\mu_1\nu_1\mu_2\nu_2}(p_1,p_2)+\frac{2}{3}[R]^{\mu_1\nu_1}(p_1){1 \over p_1^2} \ [F^2]^{\mu_3\mu_4}(p_3,p_4)\ {1 \over p_2^2}\  [R]^{\mu_2\nu_2}(p_2)
\notag\\ 
&
+[F^2]^{\mu_3\mu_4}(p_3,p_4) \,{1 \over (p_1+p_2)^2} \left[[\square_1]^{\mu_1\nu_1}(p_1,p_2) {1 \over p_2^2} [R]^{\mu_2\nu_2}(p_2)+[\square_1]^{\mu_2\nu_2}(p_2,p_1) {1 \over p_1^2} [R]^{\mu_1\nu_1}(p_1)\right] \notag\\
&-\frac{1}{6}\, [F^2]^{\mu_3\mu_4}(p_3,p_4) \ {1 \over (p_1+p_2)^2}\ \bigg[ [R]^{\mu_1\nu_1}(p_1)\   {1 \over p_1^2} \ [R]^{\mu_2\nu_2}(p_2)+[R]^{\mu_2\nu_2}(p_2)\   {1 \over p_2^2} \ [R]^{\mu_1\nu_1}(p_1)\bigg]\bigg\}\delta^{(4)}\left(\sum_{i=1}^4\,p_i\right)
\end{align}
Using the expression of the functional variations in momentum space given in \appref{FunctionalVariations}, we write
\begin{align}
	&\braket{T^{\mu_1\nu_1}(p_1)T^{\mu_2\nu_2}(p_2)J^{\mu_3}(p_3)J^{\mu_4}(p_4)}_{anom}\notag\\
	&=\braket{T^{\mu_1 \nu_1}T^{\mu_2 \nu_2} J^{\mu_3} J^{\mu_4}}_{0-residue}+\braket{ T^{\mu_1 \nu_1}T^{\mu_2 \nu_2} J^{\mu_3} J^{\mu_4}}_{pole}+\braket{ T^{\mu_1 \nu_1}T^{\mu_2 \nu_2} J^{\mu_3} J^{\mu_4}}_{inv}
\end{align}
where the first two terms are exactly those in \eqref{pole} and \eqref{0residue}, and the last term is a Weyl invariant term written as
\begin{align}
&\braket{ T^{\mu_1 \nu_1}T^{\mu_2 \nu_2} J^{\mu_3} J^{\mu_4}}_{inv}=\notag\\
&=\frac{1}{3}\beta_C\,\Pi^{\mu_1\nu_1}_{\alpha_1\beta_1}(p_1)\Pi^{\mu_2\nu_2}_{\alpha_2\beta_2}(p_2)\left\{\frac{\delta^{\alpha_1\alpha_2}}{(p_1+p_2)^2}\left[2 p_1^{\beta_2}p_2^{\beta_1}-\delta^{\beta_1\beta_2}\left((p_1+p_2)^2-p_1\cdot p_2\right)\right][F^2]^{\mu_3\mu_4}(p_3,p_4)\right\},
\end{align}
and it does not contribute to the trace anomaly part. In principle this Weyl invariant contribution can be removed by an inclusion of a Weyl invariant term in the effective action. It is worth mentioning that the expression derived satisfies correctly the conservation and anomalous trace Ward identities.

\section{Conclusions}  
In this work, we have performed an explicit computation of the $TTJJ$ correlator using a free field theory realization. We have identified the general structure of the corresponding form factors in its tensorial decomposition. In particular, we have implemented a method that simplifies their number, exploiting the momentum dependent degeneracies in combination with the classification of their orbits. The approach we presented is general and will be applied to similar correlators in future works. \\
We have used this correlator to investigate the structure of the anomalous CWIs satisfied by it and compared the results with the analogous prediction for its anomaly part, as identified by the anomaly-induced actions. Such actions are expected to reproduce the anomaly contribution to all orders in the external gravitational field. Indeed, this has been verified in the case of  $3$-point functions for two special choices of the conformal decomposition of the background metric that generate two different types of anomaly actions.\\
Indeed we have shown that, for 3-point functions, the two actions of Riegert and FV type are both in agreement with the perturbative analysis, once that we have isolated from the perturbative correlator its anomaly contribution.\\
As explained in the previous sections, this separation is uniquely defined since the anomaly part of a correlator is identified by the condition that $1)$ it reproduces the anomaly contribution in the hierarchy and $2)$ it is separately conserved. In other words, such part satisfies anomalous CWIs while the non anomalous part satisfies CWIs 
which are ordinary (non anomalous), and both parts are separately conserved. \\
As we move to 4-point functions, our analysis shows that the conservation WI for the anomaly part should be modified by an additional traceless contribution (that we have called the "$0$-residue" part). 
In other words, at the level of $4$-point functions, the anomaly part of the correlator is not just identified by the sequence of pole-like contributions, which are part of the trace WIs, but also by a traceless part. This allows, again, to separate the correlator into two parts, separately conserved and satisfying, again separately, anomalous CWIs and ordinary CWIs, respectively, as for 3-point functions. \\
These analyses demonstrate the important role of free field theory realizations in extracting such information, which is not a priori predictable from general considerations in this class of theories. This point has been first noticed in a previous analysis of the anomaly contribution to the $TTTT$ ($4T$) \cite{Coriano:2021nvn}.\\
 In the case of the $4T$ correlator, that we hope to investigate in the future by similar methods, such (0-residue) part could be identified by the $\epsilon$ expansion of the counterterms 
$\frac{1}{\epsilon}V_E$ and $\frac{1}{\epsilon}V_{C^2}$, without the need to determine the entire structure of the correlator. Indeed, one only needs to perform a consistent decomposition of the counterterms up to $O(1)$ in $\epsilon$ to achieve the goal. \\
We recall that the anomaly, in DR, comes for the expansion of $V_{E/C^2/F^2}$ up to 
$O(\epsilon)$, which in flat space is polynomial in the momenta. The application of the transverse-traceless/longitudinal/trace decomposition, introduced in \cite{Bzowski:2013sza}, has allowed extracting such 0-residue term from the finite part of the counterterms, indicating that such terms are essential in the definition of an anomaly action defined directly from the perturbative expansion.  
Therefore, the anomalous CWIs and the conservation WI are both satisfied only if we add to the pole-term contributions such $0$-residue term. The perturbative analysis of the $TTJJ$ that we have discussed is similar to the case of the $4T$ discussed in previous work, and exhibits the same features.   \\
On the other end, our analysis shows that the anomaly-induced actions that we have investigated fail - at the level of 4-point functions -  to generate conserved anomaly parts, while they succeed in 3-point functions. At the same time, as we have shown, they do not satisfy the hierarchical  CWIs that we expect. \\
The result is puzzling - but rather interesting - since it may bring under even closer scrutiny this class of actions, originally introduced as formal solutions of a variational problem.\\
We also remark that conservation WIs are associated with the diffeomorphism invariance, and one indeed expects that the conservation of the stress energy tensor should hold. Indeed it does, at least for the correlators of rank $2$ and $3$.

\subsection{Possible resolutions}

Our conclusions, given these findings, are open-ended and call for further investigations of such correlation functions. One crucial issue that needs to be addressed is if such actions can be consistently expanded around a flat spacetime - in the absence of any physical 
scale in their expressions - beyond a specific order, without encountering the puzzling behaviour that we have identified in our analysis. 
The need to introduce an extra scale $\mu_F$ in the expression of the 
$TTJJ$ vertex using Riegert's conformal decomposition, is an indication that the action cannot be correctly Fourier transformed to momentum space. This is in clear contradiction with the result we have obtained for the anomaly part of the perturbative $TTJJ$ and corresponding WIs, which are, obviously, free of any logarithmic term. In Riegert's action, this point is, in a way, expected, given that the $\Delta_4$ operator is quartic and it has been noticed before \cite{Erdmenger:1996yc}.\\
It is nevertheless remarkable that the action successfully predicts the behaviour of the $TTT$, as shown in \cite{Coriano:2017mux}. 
On the other hand, the FV action does not suffer from such shortcomings since it can be transformed in momentum space. However, the gauge choice $(\Sigma_{FV})$ used for its definition is formally very involved, being defined as a logarithmic expansion of a curvature dependent Green function. Even though no scale is present in the logarithm, one may ask if the expansion is justified in the flat spacetime limit, and $R\,\square^{-1}$ is a reasonable variable that can appear in an expansion. The issues that we have identified could be related to such intrinsic limitations. The answer to this question may be found by extending these types of analysis to a curved space, to begin with, such as to a maximally symmetric space or to other spaces where dimensionful constants are naturally present.\\
Finally, one possibility is to proceed with a modification of  such anomaly induced action by the inclusion of Weyl invariant terms. These extra terms can be identified, by investigating more closely the mismatch between the perturbative analysis and the predictions of such anomaly-induced actions, as done in this work for the $TTJJ$.\\
In all these analyses, it is obvious that the safest way to deal with these problems is to proceed with perturbative tests, that provide a safe reference for any further investigation of these issues. In principle, the strategy we have presented can be extended any n-point function.
These important aspects are currently being investigated.

\vspace{1cm}

 \centerline{\bf Acknowledgements}
The work of C. C. and R. T. is funded by the European Union, Next Generation EU, PNRR project "National Centre for HPC, Big Data and Quantum Computing", project code CN00000013 and by INFN iniziativa specifica QFT-HEP.
  M. M. M. is supported by the European Research Council (ERC) under the European Union as Horizon 2020 research and innovation program (grant agreement No818066) and by Deutsche Forschungsgemeinschaft (DFG, German Research Foundation) under Germany's Excellence Strategy EXC-2181/1 - 390900948 (the Heidelberg STRUCTURES Cluster of Excellence). C.C. thanks  Manuel Asorey, Andrei Barvinsky, John Donoghue, Michael Duff, Stefano Liberati, Jos\`e Lemos, Emil Mottola, Roberto Percacci and Ilya Shapiro, for discussions on some of the points addressed in this work and all the participants to the workshop "Quantum Field Theory and Black Hole Tests of Quantum Gravity" at SISSA Trieste.  He thanks Stefano Liberati, Emil Mottola and Enrico Barausse for organising the workshop. M. M. M. thanks Razvan Gurau, Luca Lionni and Davide Lettera for discussions. 
  Finally, we thank Mario Cret\`i and Stefano Lionetti for discussions.   
\newpage
	\appendix

\section{Ward identities}\label{AppendixA}
\subsection{Diffeomorphism invariance}
In this section, we derive the conservation and trace Ward identities for the $TJJ$ and $TTJJ$. 
We start from the invariance under diffeomorphism, for which the energy momentum tensor has to satisfy 
\begin{equation} 
	\nabla_{\mu_1} \braket{T^{\mu_1 \nu_1}} + F^{\ \,\nu_1}_{\mu_1 } \braket{J^{\mu_1}} = 0,
\end{equation}
that can be re-written as
\begin{equation} 
	2 \, \partial_{\mu_1} \dfun{\mathcal{S}}{g_{\mu_1 \nu_1}} + 2\, \Gamma^{\nu_1}_{\mu_1 \lambda} \dfun{\mathcal{S} }{g_{\mu_1 \lambda}} + F_{\mu_1}^{ \ \ \nu_1}\, \dfun{\mathcal{S}}{A_{\mu_1}} = 0  .
\end{equation}

Taking functional variations with respect to the gauge fields we find
\begin{align}
		2 \, \partial_{\mu_1} \left(\frac{\delta^3\mathcal{S}}{\delta g_{\mu_1 \nu_1}\delta A_{\mu_3}\delta A_{\mu_4}}\right) + 2\, \Gamma^{\nu_1}_{\mu_1 \lambda} \left(\frac{\delta^3\mathcal{S} }{\delta g_{\mu_1 \lambda}\delta A_{\mu_3}\delta A_{\mu_4}}\right) + \frac{\delta F_{\mu_1}^{ \ \ \nu_1}}{\delta A_{\mu_3}} \frac{\delta^2\mathcal{S}}{\delta A_{\mu_1}\delta A_{\mu_4}} + \frac{\delta F_{\mu_1}^{ \ \ \nu_1}}{\delta A_{\mu_4}}\frac{\delta^2\mathcal{S}}{\delta A_{\mu_1}\delta A_{\mu_3}}= 0,\label{TJJwa}
\end{align}
where we have discarded the terms that in the limit $A\to0$ does not contribute. Taking one more functional derivative with respect to the metric and neglecting terms that vanish in the limit $g\to\delta$ we have
\begin{align}
	&2 \, \partial_{\mu_1} \left(\frac{\delta^4\mathcal{S}}{\delta g_{\mu_1 \nu_1}\delta g_{\mu_2 \nu_2}\delta A_{\mu_3}\delta A_{\mu_4}}\right) + 2\,\left(\frac{\delta \Gamma^{\nu_1}_{\mu_1 \lambda}}{\delta g_{\mu_2\nu_2}}\right) \left(\frac{\delta^3\mathcal{S} }{\delta g_{\mu_1 \lambda}\delta A_{\mu_3}\delta A_{\mu_4}}\right) +\left( \frac{\delta^2 F_{\mu_1}^{ \ \ \nu_1}}{\delta A_{\mu_3}\delta g_{\mu_2\nu_2}}  \right)\left(\frac{\delta^2\mathcal{S}}{\delta A_{\mu_1}\delta A_{\mu_4}} \right)\notag\\
	&+ \left(\frac{\delta^2 F_{\mu_1}^{ \ \ \nu_1}}{\delta A_{\mu_4}\delta g_{\mu_2\nu_2}} \right)\left( \frac{\delta^2\mathcal{S}}{\delta A_{\mu_1}\delta A_{\mu_3}}\right)+\left(\frac{\delta F_{\mu_1}^{ \ \ \nu_1}}{\delta A_{\mu_3}} \right)\left( \frac{\delta^3\mathcal{S}}{\delta A_{\mu_1}\delta A_{\mu_4}\delta g_{\mu_2\nu_2}}\right)+\left( \frac{\delta F_{\mu_1}^{ \ \ \nu_1}}{\delta A_{\mu_4}} \right)\left( \frac{\delta^3\mathcal{S}}{\delta A_{\mu_1}\delta A_{\mu_3}\delta g_{\mu_2\nu_2}}\right)= 0.\label{TTJJwa}
\end{align}
Then by using the explicit expressions of the functional derivatives written as
\begin{align}
\frac{\delta F_{\mu_1}^{ \ \ \nu_1}(x_1)}{\delta A_{\mu_i}(x_i)}\bigg|_{g=\delta,A=0}&=\big(\delta^{\nu_1\mu_i}\partial_{\mu_1}-\delta^{\mu_i}_{\mu_1}\partial^{\nu_1}\big)\delta_{x_1x_i}\\[1ex]
\frac{\delta^2 F_{\mu_1}^{ \ \ \nu_1}(x_1)}{\delta A_{\mu_i}(x_i)\delta g_{\mu_2\nu_2}(x_2)}\bigg|_{g=\delta,A=0}&=-\delta_{x_1x_2}\delta^{\nu_1(\mu_2}\delta^{\nu_2)\lambda}\big(\delta^{\mu_i}_{\lambda}\partial_{\mu_1}-\delta^{\mu_i}_{\mu_1}\partial_{\lambda}\big)\delta_{x_1x_i}\\[1ex]
\dfun{\Gamma^{\nu_1}_{\mu_1 \lambda} (x_1)}{g_{\mu_2 \nu_2} (x_2)}\bigg |_{g=\delta}&=  {1\over 2} \bigg(\delta^{\nu_1 (\mu_2} \delta^{\nu_2)}_{\mu_1} \partial_{\lambda} +
\delta^{\nu_1 (\mu_2} \delta^{\nu_2)}_{\lambda} \partial_{\mu_1} - 
\delta^{(\mu_2}_{\lambda} \delta^{\nu_2)}_{\mu_1} \partial^{\nu_1} \bigg) \delta_{x_1, x_2}
\end{align}
with $\delta_{x_i\,x_j}=\delta^{(d)}(x_i-x_j)$, we rewrite \eqref{TJJwa}  to obtain the conservation Ward identities for the $\braket{TJJ}$ as
\begin{align}
	 \partial_{\mu_1}\braket{T^{\mu_1\nu_1}(x_1)J^{\mu_3}(x_3)J^{\mu_4}(x_4)}=&- \big[\big(\delta^{\nu_1\mu_3}\partial_{\mu_1}-\delta^{\mu_3}_{\mu_1}\partial^{\nu_1}\big)\delta_{x_1x_3}\big]\braket{J^{\mu_1}(x_1)J^{\mu_4}(x_4)} \notag\\
	 &-\big[ \big(\delta^{\nu_1\mu_4}\partial_{\mu_1}-\delta^{\mu_4}_{\mu_1}\partial^{\nu_1}\big)\delta_{x_1x_4}\big]\braket{J^{\mu_1}(x_1)J^{\mu_3}(x_3)},
\end{align}
and from \eqref{TTJJwa} the conservation Ward identities for the $\braket{TTJJ}$ reads as
\begin{align}
	&\, \partial_{\mu_1} \braket{T^{\mu_1 \nu_1}(x_1)T^{\mu_2 \nu_2}(x_2)J^{\mu_3}(x_3)J^{\mu_4}(x_4)} =- \big[\big(2\delta^{\nu_1 (\mu_2} \delta^{\nu_2)}_{\mu_1} \partial_{\lambda} - 
	\delta^{(\mu_2}_{\lambda} \delta^{\nu_2)}_{\mu_1} \partial^{\nu_1} \big) \delta_{x_1, x_2}\big]\braket{T^{\mu_1 \lambda}(x_1)J^{\mu_3}(x_3)J^{\mu_4}(x_4)} \notag\\
	&+\bigg\{2\delta_{x_1x_2}\big[\big(\delta^{\nu_1(\mu_2}\delta^{\nu_2)\mu_4}\partial_{\mu_1}-\delta^{\mu_4}_{\mu_1}\delta^{\nu_1(\mu_2}\partial^{\nu_2)}\big)\delta_{x_1x_4}\big]\braket{J^{\mu_1}(x_1)J^{\mu_3}(x_3)}\notag\\
	&\hspace{1cm}-\big[\big(\delta^{\nu_1\mu_4}\partial_{\mu_1}-\delta^{\mu_4}_{\mu_1}\partial^{\nu_1}\big)\delta_{x_1x_4}\big]\braket{J^{\mu_1}(x_1)J^{\mu_3}(x_3)T^{\mu_2\nu_2}(x_2) }\bigg\}+\big\{3\leftrightarrow 4\big\}.
\end{align}
In momentum space, these conservation Ward identities are written as
\begin{align} 
	&{p_1}_{\mu_1} \braket {T^{\mu_1 \nu_1} (p_1) J^{\mu_3} (p_3) J^{\mu_4} (p_4)} =\bigg[\left( \delta^{\mu_3 \nu_1} {p_3}_{\mu_1} - \delta^{\mu_3}_{\mu_1} {p_3}^{\nu_1} \right) \braket{J^{\mu_1} (p_1+p_3) J^{\mu_4} (p_4)}\bigg]+ \big[3 \tor 4\big],\\[1ex]
	&{p_1}_{\mu_1} \braket{T^{\mu_1 \nu_1} (p_1) T^{\mu_2 \nu_2} (p_2) J^{\mu_3} (p_3) J^{\mu_4} (p_4)} 
	 =\left( 2\delta^{\nu_1 (\mu_2} \delta^{\nu_2)}_{\mu_1} {p_2}_{\lambda_1} - 
	\delta^{(\mu_2}_{\lambda_1} \delta^{\nu_2)}_{\mu_1} {p_2}^{\nu_1} \right) \braket{T^{\mu_1 \lambda_1} (p_1+p_2) J^{\mu_3} (p_3) J^{\mu_4} (p_4)}  \notag\\[0.7ex]
	&\hspace{2cm}
	+\Big\{2 \left( \delta^{\mu_3}_{\mu_1} \delta^{\nu_1 (\mu_2} p_3^{\nu_2)}  - \delta^{\nu_1 (\mu_2} \delta^{\nu_2) \mu_3}{p_3}_{\mu_1 } \right)  \braket{J^{\mu_1} (p_1+p_2+p_3) J^{\mu_4} (p_4)} 
	\notag\\ 
	&\hspace{4cm}+\big(\delta^{\nu_1\mu_3}p_{3\mu_1}-\delta^{\mu_3}_{\mu_1}p_3^{\nu_1}\big)\braket{J^{\mu_1}(p_1+p_3)J^{\mu_4}(p_4)T^{\mu_2\nu_2}(p_2)}\Big\}+ \big\{3 \tor 4\big\}.
\end{align}

\subsection{Gauge invariance}

The invariance under gauge transformation implies the conservation of the current
\begin{equation}
	 \nabla_\mu \braket{J^\mu} = 0. 
 \end{equation}
By taking functional derivatives with respect to the metric and the gauge fields and considering the flat spacetime limit, we find the conservation Ward Identities for the $3$- and $4$- point functions
\begin{align} 
	&\partial_{\mu_3} \braket{T^{\mu_1 \nu_1} (x_1) J^{\mu_3} (x_3)  J^{\mu_4} (x_4)} = 0 ,\\[1ex]
	&\partial_{\mu_3} \braket{T^{\mu_1 \nu_1} (x_1) T^{\mu_2 \nu_2} (x_2) J^{\mu_3} (xx_3)  J^{\mu_4} (x_4)} = 0 ,
\end{align}
that in momentum space become
\begin{align} 
	&p_{3\mu_3} \braket{T^{\mu_1 \nu_1} (x_1) J^{\mu_3} (x_3)  J^{\mu_4} (x_4)} = 0 ,\\[1ex]
	&p_{3\mu_3} \braket{T^{\mu_1 \nu_1} (x_1) T^{\mu_2 \nu_2} (x_2) J^{\mu_3} (xx_3)  J^{\mu_4} (x_4)} = 0,
\end{align}
and similar equations for the contraction of $p_{4\mu_4}$. 

\subsection{Weyl invariance}

The invariance under Weyl transformation implies that the energy momentum tensor has zero trace, mainly
\begin{equation} 
	g_{ \mu \nu} \braket{T^{\mu \nu}} = 0,
\end{equation}
in general $d$ dimensions.  Taking functional derivatives of the previous equation we have the trace Ward identities 
 \begin{align} 
 	&\delta_{\mu_1 \nu_1} \braket{T^{\mu_1 \nu_1} (p_1)  J^{\mu_3} (p_3) J^{\mu_4} (p_4) } =  0,\\[1ex]
 	&\delta_{\mu_1 \nu_1} \braket{T^{\mu_1 \nu_1} (p_1) T^{\mu_2 \nu_2} (p_2) J^{\mu_3} (p_3) J^{\mu_4} (p_4) } = -2 \braket{T^{\mu_2 \nu_2} (p_1+p_2) J^{\mu_3} (p_3) J^{\mu_4} (p_4) }
  \end{align}

In $d=4$, due to the quantum effect, the trace of the energy momentum tensor acquires a non-zero trace contribution
\begin{align}
	g_{ \mu \nu} \braket{T^{\mu \nu}}_g =\bigg[b\,C^2+b'\,E+\beta_C\,F^2\bigg],
\end{align}
leading to the anomalous trace Ward identities for the correlation functions
\begin{align}
&\delta_{\mu_1 \nu_1} \braket{T^{\mu_1 \nu_1} (p_1)  J^{\mu_3} (p_3) J^{\mu_4} (p_4) } =  \beta_C\,\Big[F^2\Big]^{\mu_3\mu_4}(p_3,p_4),\\[1.5ex]
&\delta_{\mu_1 \nu_1} \braket{T^{\mu_1 \nu_1} (p_1) T^{\mu_2 \nu_2} (p_2) J^{\mu_3} (p_3) J^{\mu_4} (p_4) } =2\beta_C\,\Big[\sqrt{-g}F^2\Big]^{\mu_2\nu_2\mu_3\mu_4}(p_2,p_3,p_4) \notag\\
&\hspace{8cm}-2 \braket{T^{\mu_2 \nu_2} (p_1+p_2) J^{\mu_3} (p_3) J^{\mu_4} (p_4) },
\end{align}
where the functional derivatives of the field strength $F^2$ with respect tot he gauge field are expressed in \eqref{Fmu3mu4} and \eqref{gFmu3mu4}. 

\section{Orbits representatives of the permutations\label{Orbits}}

\subsection{General $d$ dimensional case}
In this section we list the representative of the orbits for each sector of tensor structures used to construct the transverse traceless part in general $d$ dimensions. In particular for the $\delta\delta\delta$ sector we have chosen the two representative
\begin{equation} 
\delta^{\mu_1 \mu_2} \delta^{\nu_1 \nu_2} \delta^{\mu_3 \mu_4}, \qquad
\delta^{\mu_1 \mu_2} \delta^{\mu_3 \nu_1} \delta^{\mu_4 \nu_2}.
\end{equation}
Moving forward, for the $\delta \delta pp$ sector we find the following representatives
\begin{equation}
\begin{split}
{p_1}^{\mu_3} {p_1}^{\mu_4} \delta^{\mu_1\mu_2} \delta^{\nu_1 \nu_2},  \qquad
{p_1}^{\mu_3} {p_2}^{\mu_4} \delta^{\mu_1\mu_2} \delta^{\nu_1\nu_2},  \qquad
{p_3}^{\mu_1} {p_1}^{\mu_3} \delta^{\mu_2\nu_1} \delta^{\mu_4\nu_2},  \qquad 
{p_3}^{\mu_2} {p_1}^{\mu_3} \delta^{\mu_1\nu_2} \delta^{\mu_4\nu_1},  \\
{p_3}^{\mu_1} {p_1}^{\mu_4} \delta^{\mu_2\nu_1} \delta^{\mu_3\nu_2},   \qquad  
{p_3}^{\mu_2} {p_1}^{\mu_4} \delta^{\mu_1\nu_2} \delta^{\mu_3\nu_1},  \qquad   
{p_3}^{\mu_1} {p_3}^{\nu_1} \delta^{\mu_2\mu_3} \delta^{\mu_4\nu_2},  \qquad   
{p_3}^{\mu_1} {p_3}^{\mu_2} \delta^{\mu_3\mu_4} \delta^{\nu_1\nu_2},    \\    
{p_3}^{\mu_1} {p_3}^{\mu_2} \delta^{\mu_3\nu_1} \delta^{\mu_4\nu_2},  \qquad  
{p_3}^{\mu_1} {p_4}^{\nu_1} \delta^{\mu_2\mu_3} \delta^{\mu_4\nu_2}, \qquad   
{p_3}^{\mu_1} {p_4}^{\mu_2} \delta^{\mu_3 \mu_4} \delta^{\nu_1\nu_2},\\
{p_3}^{\mu_1} {p_4}^{\mu_2} \delta^{\mu_3\nu_1} \delta^{\mu_4\nu_2}, \qquad
{p_3}^{\mu_1} {p_4}^{\mu_2} \delta^{\mu_3\nu_2} \delta^{\mu_4\nu_1},
\end{split}
\end{equation}
and for the $\delta pp pp$ sector we have
\begin{equation}
\begin{split}
{p_1}^{\mu_3} {p_1}^{\mu_4} {p_3}^{\mu_1} {p_3}^{\mu_2} \delta^{\nu_1\nu_2}, \qquad   
{p_1}^{\mu_3} {p_1}^{\mu_4} {p_3}^{\mu_1} {p_4}^{\mu_2} \delta^{\nu_1\nu_2}, \qquad   
{p_1}^{\mu_3} {p_2}^{\mu_4} {p_3}^{\mu_1} {p_3}^{\mu_2} \delta^{\nu_1\nu_2},   \\    
{p_1}^{\mu_3} {p_2}^{\mu_4} {p_3}^{\mu_1} {p_4}^{\mu_2} \delta^{\nu_1\nu_2},\qquad   
{p_1}^{\mu_3} {p_2}^{\mu_4} {p_3}^{\mu_2} {p_4}^{\mu_1} \delta^{\nu_1\nu_2},\qquad   
{p_1}^{\mu_3} {p_3}^{\mu_1} {p_3}^{\mu_2} {p_3}^{\nu_1} \delta^{\mu_4\nu_2},  \\  
{p_1}^{\mu_3} {p_3}^{\mu_1} {p_3}^{\mu_2} {p_3}^{\nu_2} \delta^{\mu_4\nu_1},\qquad   
{p_1}^{\mu_4} {p_3}^{\mu_1} {p_3}^{\mu_2} {p_3}^{\nu_1} \delta^{\mu_3\nu_2},\qquad   
{p_1}^{\mu_4} {p_3}^{\mu_1} {p_3}^{\mu_2} {p_3}^{\nu_2} \delta^{\mu_3\nu_1},\\      
{p_1}^{\mu_3} {p_3}^{\mu_1} {p_3}^{\nu_1} {p_4}^{\mu_2} \delta^{\mu_4\nu_2},\qquad   
{p_1}^{\mu_3} {p_3}^{\mu_1} {p_3}^{\mu_2} {p_4}^{\nu_1} \delta^{\mu_4\nu_2},\qquad   
{p_1}^{\mu_3} {p_3}^{\mu_1} {p_3}^{\mu_2} {p_4}^{\nu_2} \delta^{\mu_4\nu_1},\\      
{p_1}^{\mu_3} {p_3}^{\mu_2} {p_3}^{\nu_2} {p_4}^{\mu_1} \delta^{\mu_4\nu_1},\qquad   
{p_1}^{\mu_4} {p_3}^{\mu_1} {p_3}^{\nu_1} {p_4}^{\mu_2} \delta^{\mu_3\nu_2},\qquad   
{p_1}^{\mu_4} {p_3}^{\mu_1} {p_3}^{\mu_2} {p_4}^{\nu_1} \delta^{\mu_3\nu_2},\\     
{p_1}^{\mu_4} {p_3}^{\mu_1} {p_3}^{\mu_2} {p_4}^{\nu_2} \delta^{\mu_3\nu_1},\qquad   
{p_1}^{\mu_4} {p_3}^{\mu_2} {p_3}^{\nu_2} {p_4}^{\mu_1} \delta^{\mu_3\nu_1},\qquad   
{p_3}^{\mu_1} {p_3}^{\mu_2} {p_3}^{\nu_1} {p_3}^{\nu_2} \delta^{\mu_3\mu_4},\\
{p_3}^{\mu_1} {p_3}^{\mu_2} {p_3}^{\nu_1} {p_4}^{\nu_2} \delta^{\mu_3\mu_4},\qquad   
{p_3}^{\mu_1} {p_3}^{\nu_1} {p_4}^{\mu_2} {p_4}^{\nu_2} \delta^{\mu_3\mu_4},\qquad   
{p_3}^{\mu_1} {p_3}^{\mu_2} {p_4}^{\nu_1} {p_4}^{\nu_2} \delta^{\mu_3\mu_4} .
\end{split}
\end{equation}
Finally, the last sector $pppppp$ is described by the representatives
\begin{equation}
\begin{split}
{p_3}^{\mu_1} {p_3}^{\mu_2} {p_1}^{\mu_3} {p_1}^{\mu_4} {p_3}^{\nu_1} {p_3}^{\nu_2},  \qquad
{p_3}^{\mu_1} {p_3}^{\mu_2} {p_1}^{\mu_3} {p_1}^{\mu_4} {p_3}^{\nu_1} {p_4}^{\nu_2},  \qquad
{p_3}^{\mu_1} {p_3}^{\mu_2} {p_1}^{\mu_3} {p_1}^{\mu_4} {p_4}^{\nu_1} {p_3}^{\nu_2},  \\
{p_3}^{\mu_1} {p_4}^{\mu_2} {p_1}^{\mu_3} {p_1}^{\mu_4} {p_3}^{\nu_1} {p_4}^{\nu_2}, \qquad      
{p_3}^{\mu_1} {p_3}^{\mu_2} {p_1}^{\mu_3} {p_1}^{\mu_4} {p_4}^{\nu_1} {p_4}^{\nu_2}, \qquad      
{p_3}^{\mu_1} {p_3}^{\mu_2} {p_1}^{\mu_3} {p_2}^{\mu_4} {p_3}^{\nu_1} {p_3}^{\nu_2}, \\         
{p_3}^{\mu_1} {p_3}^{\mu_2} {p_1}^{\mu_3} {p_2}^{\mu_4} {p_3}^{\nu_1} {p_4}^{\nu_2}, \qquad      
{p_3}^{\mu_1} {p_3}^{\mu_2} {p_1}^{\mu_3} {p_2}^{\mu_4} {p_4}^{\nu_1} {p_3}^{\nu_2}, \qquad      
{p_3}^{\mu_1} {p_4}^{\mu_2} {p_1}^{\mu_3} {p_2}^{\mu_4} {p_3}^{\nu_1} {p_4}^{\nu_2}, \\    
{p_3}^{\mu_1} {p_3}^{\mu_2} {p_1}^{\mu_3} {p_2}^{\mu_4} {p_4}^{\nu_1} {p_4}^{\nu_2}, \qquad      
{p_4}^{\mu_1} {p_3}^{\mu_2} {p_1}^{\mu_3} {p_2}^{\mu_4} {p_4}^{\nu_1} {p_3}^{\nu_2}.
\end{split}
\end{equation}

As already mentioned in \secref{OrbitsSec}, we directly obtain the structure of the transverse traceless part through these representatives of the orbits. Indeed, one has to consider independent form factors times each representative. Then, applying the permutation operator, one finds the entire structure of $X^{\alpha_1\dots\alpha_4}$ with the symmetry condition on the form factors. 
\subsection{The case of $d=4$}
In this section we list the representatives of the orbits for each sector of tensor structures used to construct the transverse traceless part when $d=4$ and we use the $n$-$p$ basis. In particular for the $nnnnpp$ sector we have chosen the two representatives
\begin{equation}
{p_3}^{\mu_1}{p_3}^{\mu_2} n^{\nu_1} n^{\nu_2} n^{\mu_3} n^{\mu_4} \qquad 
{p_3}^{\mu_1}{p_4}^{\mu_2} n^{\nu_1} n^{\nu_2} n^{\mu_3} n^{\mu_4} ;
\end{equation}
and for the $nn pp pp$ sector we have
\begin{align}
{p_1}^{\mu_3} {p_1}^{\mu_4} {p_3}^{\mu_1} {p_3}^{\mu_2} n^{\nu_1} n^{\nu_2}, \qquad   
{p_1}^{\mu_3} {p_1}^{\mu_4} {p_3}^{\mu_1} {p_4}^{\mu_2} n^{\nu_1} n^{\nu_2}, \qquad   
{p_1}^{\mu_3} {p_2}^{\mu_4} {p_3}^{\mu_1} {p_3}^{\mu_2} n^{\nu_1} n^{\nu_2}, \notag\\     
{p_1}^{\mu_3} {p_2}^{\mu_4} {p_3}^{\mu_1} {p_4}^{\mu_2} n^{\nu_1} n^{\nu_2},\qquad   
{p_1}^{\mu_3} {p_2}^{\mu_4} {p_3}^{\mu_2} {p_4}^{\mu_1} n^{\nu_1} n^{\nu_2},\qquad   
{p_1}^{\mu_3} {p_3}^{\mu_1} {p_3}^{\mu_2} {p_3}^{\nu_1} n^{\mu_4} n^{\nu_2},\notag\\     
{p_1}^{\mu_3} {p_3}^{\mu_1} {p_3}^{\mu_2} {p_3}^{\nu_2} n^{\mu_4} n^{\nu_1},\qquad   
{p_1}^{\mu_4} {p_3}^{\mu_1} {p_3}^{\mu_2} {p_3}^{\nu_1} n^{\mu_3} n^{\nu_2},\qquad   
{p_1}^{\mu_4} {p_3}^{\mu_1} {p_3}^{\mu_2} {p_3}^{\nu_2} n^{\mu_3} n^{\nu_1},\notag\\     
{p_1}^{\mu_3} {p_3}^{\mu_1} {p_3}^{\nu_1} {p_4}^{\mu_2} n^{\mu_4} n^{\nu_2},\qquad   
{p_1}^{\mu_3} {p_3}^{\mu_1} {p_3}^{\mu_2} {p_4}^{\nu_1} n^{\mu_4} n^{\nu_2},\qquad   
{p_1}^{\mu_3} {p_3}^{\mu_1} {p_3}^{\mu_2} {p_4}^{\nu_2} n^{\mu_4} n^{\nu_1},\notag\\      
{p_1}^{\mu_3} {p_3}^{\mu_2} {p_3}^{\nu_2} {p_4}^{\mu_1} n^{\mu_4} n^{\nu_1},\qquad   
{p_1}^{\mu_4} {p_3}^{\mu_1} {p_3}^{\nu_1} {p_4}^{\mu_2} n^{\mu_3} n^{\nu_2},\qquad   
{p_1}^{\mu_4} {p_3}^{\mu_1} {p_3}^{\mu_2} {p_4}^{\nu_1} n^{\mu_3} n^{\nu_2},\notag\\      
{p_1}^{\mu_4} {p_3}^{\mu_1} {p_3}^{\mu_2} {p_4}^{\nu_2} n^{\mu_3} n^{\nu_1},\qquad   
{p_1}^{\mu_4} {p_3}^{\mu_2} {p_3}^{\nu_2} {p_4}^{\mu_1} n^{\mu_3} n^{\nu_1},\qquad   
{p_3}^{\mu_1} {p_3}^{\mu_2} {p_3}^{\nu_1} {p_3}^{\nu_2} n^{\mu_3} n^{\mu_4},\notag\\      
{p_3}^{\mu_1} {p_3}^{\mu_2} {p_3}^{\nu_1} {p_4}^{\nu_2} n^{\mu_3} n^{\mu_4},\qquad   
{p_3}^{\mu_1} {p_3}^{\nu_1} {p_4}^{\mu_2} {p_4}^{\nu_2} n^{\mu_3} n^{\mu_4},\qquad   
{p_3}^{\mu_1} {p_3}^{\mu_2} {p_4}^{\nu_1} {p_4}^{\nu_2} n^{\mu_3} n^{\mu_4} ;
\end{align}
and finally the last sector $pppppp$ is identified, as in the general case, by the representatives
\begin{align}
{p_3}^{\mu_1} {p_3}^{\mu_2} {p_1}^{\mu_3} {p_1}^{\mu_4} {p_3}^{\nu_1} {p_3}^{\nu_2},  \qquad
{p_3}^{\mu_1} {p_3}^{\mu_2} {p_1}^{\mu_3} {p_1}^{\mu_4} {p_3}^{\nu_1} {p_4}^{\nu_2},  \qquad
{p_3}^{\mu_1} {p_3}^{\mu_2} {p_1}^{\mu_3} {p_1}^{\mu_4} {p_4}^{\nu_1} {p_3}^{\nu_2},  \notag\\
{p_3}^{\mu_1} {p_4}^{\mu_2} {p_1}^{\mu_3} {p_1}^{\mu_4} {p_3}^{\nu_1} {p_4}^{\nu_2}, \qquad      
{p_3}^{\mu_1} {p_3}^{\mu_2} {p_1}^{\mu_3} {p_1}^{\mu_4} {p_4}^{\nu_1} {p_4}^{\nu_2}, \qquad      
{p_3}^{\mu_1} {p_3}^{\mu_2} {p_1}^{\mu_3} {p_2}^{\mu_4} {p_3}^{\nu_1} {p_3}^{\nu_2}, \notag\\         
{p_3}^{\mu_1} {p_3}^{\mu_2} {p_1}^{\mu_3} {p_2}^{\mu_4} {p_3}^{\nu_1} {p_4}^{\nu_2}, \qquad      
{p_3}^{\mu_1} {p_3}^{\mu_2} {p_1}^{\mu_3} {p_2}^{\mu_4} {p_4}^{\nu_1} {p_3}^{\nu_2}, \qquad      
{p_3}^{\mu_1} {p_4}^{\mu_2} {p_1}^{\mu_3} {p_2}^{\mu_4} {p_3}^{\nu_1} {p_4}^{\nu_2}, \notag\\         
{p_3}^{\mu_1} {p_3}^{\mu_2} {p_1}^{\mu_3} {p_2}^{\mu_4} {p_4}^{\nu_1} {p_4}^{\nu_2}, \qquad      
{p_4}^{\mu_1} {p_3}^{\mu_2} {p_1}^{\mu_3} {p_2}^{\mu_4} {p_4}^{\nu_1} {p_3}^{\nu_2};  
\end{align}

\subsection{The case of $d=3$}
In this section we list the representatives of the orbits for each sector of tensor structures used to construct the transverse traceless part when $d=3$.
\begin{align}
{p_1}^{\mu_3} {p_1}^{\mu_4} {p_3}^{\mu_1} {p_3}^{\mu_2} {p_3}^{\nu_1} {p_3}^{\nu_2}, \qquad 
{p_1}^{\mu_3} {p_1}^{\mu_4} {p_3}^{\mu_1} {p_3}^{\nu_1} {p_4}^{\mu_2} {p_4}^{\nu_2}, \qquad
{p_1}^{\mu_3} {p_2}^{\mu_4} {p_3}^{\mu_1} {p_3}^{\mu_2} {p_3}^{\nu_1} {p_3}^{\nu_2}, \notag\\
{p_1}^{\mu_3} {p_2}^{\mu_4} {p_3}^{\mu_1} {p_3}^{\nu_1} {p_4}^{\mu_2} {p_4}^{\nu_2}, \qquad
{p_1}^{\mu_3} {p_2}^{\mu_4} {p_3}^{\mu_2} {p_3}^{\nu_2} {p_4}^{\mu_1} {p_4}^{\nu_1} .
\end{align}

\section{More on the reconstruction}\label{moreReconstr}
In this section we derived the anomalous part of the correlator after the renormalization in more detail. From \eqref{tlocTJJ} we have computed the $\braket{t_{loc}TJJ}_{anom}$ part of the correlator after the renomalization as
\begin{align}
	&\braket{ t_{loc}^{\mu_1 \nu_1}T^{\mu_2 \nu_2} J^{\mu_3} J^{\mu_4}}_{anom}^{(d=4)}=\notag\\
	&=\beta_C\,\bigg\{\mathcal{I}^{(d=4)\,\mu_1\nu_1}_{\alpha_1}\bigg[2\ {p_2}_{\lambda_1}  \frac{\delta^{\alpha_1 (\mu_2}\pi^{\nu_2) \lambda_1} (p_1+p_2)}{3}\left[F^2\right]^{\mu_3\mu_4}(p_3,p_4) - 
	{p_2}^{\alpha_1}  \frac{\pi^{\mu_2 \nu_2} (p_1+p_2) }{3}\left[F^2\right]^{\mu_3\mu_4}(p_3,p_4) \notag\\ 
	&
	+ \frac{\pi^{\mu_2\nu_2}(p_2)}{3}\bigg( \delta^{\mu_3 \alpha_1} p_{3\lambda_1} \left[F^2\right]^{\lambda_1\mu_4}(p_1+p_3,p_4)- p_3^{\alpha_1 } \left[F^2\right]^{\mu_3\mu_4}(p_1+p_3,p_4)\bigg)\notag\\
	&+ \frac{\pi^{\mu_2\nu_2}(p_2)}{3}\bigg( \delta^{\mu_4 \alpha_1} p_{4\lambda_1} \left[F^2\right]^{\lambda_1\mu_3}(p_1+p_4,p_3)- p_4^{\alpha_1 } \left[F^2\right]^{\mu_3\mu_4}(p_3,p_1+p_4)\bigg)\bigg]\notag\\
	&-2\,\frac{\pi^{\mu_1\nu_1}(p_1)}{3}\frac{\pi^{\mu_2\nu_2}(p_1+p_2)}{3} \left[F^2\right]^{\mu_3\mu_4}(p_3,p_4)+2\frac{\pi^{\mu_1\nu_1}(p_1)}{3}\big[\sqrt{g}\,F^2\big]^{\mu_2\nu_2\mu_3\mu_4}(p_2,p_3,p_4)\bigg\}\notag\\
	&=\beta_C\,\bigg\{\mathcal{I}^{(d=4)\,\mu_1\nu_1}_{\alpha_1}\bigg[2\ {p_2}_{\lambda_1}  \bigg(\frac{\delta^{\alpha_1 (\mu_2}\pi^{\nu_2) \lambda_1} (p_1+p_2)}{3}- \frac{\pi^{\alpha_1 \lambda_1} (p_2+p_1)}{3}\frac{\pi^{\mu_2\nu_2}(p_2)}{3}\bigg)\left[F^2\right]^{\mu_3\mu_4}(p_3,p_4) \notag\\
	&\qquad- {p_2}^{\alpha_1} \bigg( \frac{\pi^{\mu_2 \nu_2} (p_1+p_2) }{3}-\frac{\pi^{\mu_2\nu_2}(p_2)}{3}\bigg)\left[F^2\right]^{\mu_3\mu_4}(p_3,p_4)\bigg]\notag\\
	&+\mathcal{I}^{(d=4)\,\mu_1\nu_1}_{\alpha_1}\frac{\pi^{\mu_2\nu_2}(p_2)}{3}\,p_{1\beta_1}\left(2\Big[\sqrt{g}\,F^2\Big]^{\alpha_1\beta_1\mu_3\mu_4}(p_1,p_3,p_4)-\frac{2}{3}\pi^{\alpha_1\beta_1}(p_1+p_2)\Big[F^2\Big]^{\mu_3\mu_4}(p_3,p_4)\right)\notag\\
	&-2\,\frac{\pi^{\mu_1\nu_1}(p_1)}{3}\frac{\pi^{\mu_2\nu_2}(p_1+p_2)}{3} \left[F^2\right]^{\mu_3\mu_4}(p_3,p_4)+2\frac{\pi^{\mu_1\nu_1}(p_1)}{3}\big[\sqrt{g}\,F^2\big]^{\mu_2\nu_2\mu_3\mu_4}(p_2,p_3,p_4)\bigg\},
\end{align}
where we have considered the relation
\begin{align}
	&\mathcal{I}^{(d=4)\,\mu_1\nu_1}_{\alpha_1}\bigg[2\ {p_2}_{\lambda_1}  \frac{\pi^{\alpha_1 \lambda_1} (p_2+p_1)}{3}\frac{\pi^{\mu_2\nu_2}(p_2)}{3}\left[F^2\right]^{\mu_3\mu_4}(p_3,p_4) - 
	{p_2}^{\alpha_1}\frac{\pi^{\mu_2\nu_2}(p_2)}{3}\left[F^2\right]^{\mu_3\mu_4}(p_3,p_4) \notag\\ 
	&
	+\frac{\pi^{\mu_2\nu_2}(p_2)}{3}\bigg(\delta^{\mu_3 \alpha_1} p_{3\lambda_1} \left[F^2\right]^{\lambda_1\mu_4}(p_1+p_3,p_4)- p_3^{\alpha_1 } \left[F^2\right]^{\mu_3\mu_4}(p_1+p_3,p_4)\bigg)\notag\\
	&+\frac{\pi^{\mu_2\nu_2}(p_2)}{3}\bigg(  \delta^{\mu_4 \alpha_1} p_{4\lambda_1} \left[F^2\right]^{\lambda_1\mu_3}(p_2+p_4,p_3)- p_4^{\alpha_1 } \left[F^2\right]^{\mu_3\mu_4}(p_3,p_1+p_4)\bigg)\bigg]\notag\\
	&=\mathcal{I}^{(d=4)\,\mu_1\nu_1}_{\alpha_1}\frac{\pi^{\mu_2\nu_2}(p_2)}{3}\,p_{1\beta_1}\left(2\Big[\sqrt{g}\,F^2\Big]^{\alpha_1\beta_1\mu_3\mu_4}(p_1,p_3,p_4)-\frac{2}{3}\pi^{\alpha_1\beta_1}(p_1+p_2)\Big[F^2\Big]^{\mu_3\mu_4}(p_3,p_4)\right).\label{Res3}
\end{align}
 We consider here the other part, $\braket{t_{loc}t_{loc}JJ}_{anom}$, that is explicitly given by
\begin{align}
	&\braket{ t_{loc}^{\mu_1 \nu_1} t_{loc}^{\mu_2 \nu_2}J^{\mu_3} J^{\mu_4}}_{anom}^{(d=4)}=\notag\\
	&=\beta_C\,\bigg\{\mathcal{I}^{(d=4)\,\mu_1\nu_1}_{\alpha_1}\mathcal{I}^{(d=4)\,\mu_2\nu_2}_{\alpha_2}\bigg[\frac{1}{3}\,{p_2}_{\lambda_1}  {p_2}_{\beta_2} \delta^{\alpha_1 \alpha_2}\pi^{\beta_2 \lambda_1} (p_1+p_2)\left[F^2\right]^{\mu_3\mu_4}(p_3,p_4) \bigg]\notag\\
	&-\frac{2}{3}\,\pi^{\mu_1\nu_1}(p_1)\,\mathcal{I}^{(d=4)\,\mu_2\nu_2}_{\alpha_2}{p_2}_{\beta_2}\bigg[\frac{\pi^{\alpha_2\beta_2}(p_1+p_2)}{3} \left[F^2\right]^{\mu_3\mu_4}(p_3,p_4)-\big[\sqrt{g}\,F^2\big]^{\alpha_2\beta_2\mu_3\mu_4}(p_2,p_3,p_4)\bigg]\bigg\}\notag\\
	&+\bigg\{\mathcal{I}^{(d=4)\,\mu_1\nu_1}_{\alpha_1}\frac{\pi^{\mu_2\nu_2}(p_2)}{3}\bigg[\frac{2}{3}{p_2}_{\lambda_1} \pi^{\alpha_1 \lambda_1} (p_1+p_2)\left[F^2\right]^{\mu_3\mu_4}(p_3,p_4) - 
	{p_2}^{\alpha_1} \left[F^2\right]^{\mu_3\mu_4}(p_3,p_4) \notag\\ 
	&
	+ \bigg( \delta^{\mu_3 \alpha_1} p_{3\lambda_1} \left[F^2\right]^{\lambda_1\mu_4}(p_1+p_3,p_4)- p_3^{\alpha_1 } \left[F^2\right]^{\mu_3\mu_4}(p_1+p_3,p_4)\bigg)\notag\\
	&+\bigg( \delta^{\mu_4 \alpha_1} p_{4\lambda_1} \left[F^2\right]^{\lambda_1\mu_3}(p_1+p_4,p_3)- p_4^{\alpha_1 } \left[F^2\right]^{\mu_3\mu_4}(p_3,p_1+p_4)\bigg)\bigg]-2\,\frac{\pi^{\mu_1\nu_1}(p_1)}{3}\frac{\pi^{\mu_2\nu_2}(p_2)}{3} \left[F^2\right]^{\mu_3\mu_4}(p_3,p_4)\bigg\}\notag\\
	&=\beta_C\,\bigg\{\mathcal{I}^{(d=4)\,\mu_1\nu_1}_{\alpha_1}\mathcal{I}^{(d=4)\,\mu_2\nu_2}_{\alpha_2}\bigg[\frac{1}{3}\,{p_2}_{\lambda_1}  {p_2}_{\beta_2} \delta^{\alpha_1 \alpha_2}\pi^{\beta_2 \lambda_1} (p_1+p_2)\left[F^2\right]^{\mu_3\mu_4}(p_3,p_4) \bigg]\notag\\
	&-\frac{2}{3}\,\pi^{\mu_1\nu_1}(p_1)\,\mathcal{I}^{(d=4)\,\mu_2\nu_2}_{\alpha_2}{p_2}_{\beta_2}\bigg[\frac{\pi^{\alpha_2\beta_2}(p_1+p_2)}{3} \left[F^2\right]^{\mu_3\mu_4}(p_3,p_4)-\big[\sqrt{g}\,F^2\big]^{\alpha_2\beta_2\mu_3\mu_4}(p_2,p_3,p_4)\bigg]\notag\\
	&-\frac{2}{3}\,\pi^{\mu_2\nu_2}(p_2)\,\mathcal{I}^{(d=4)\,\mu_1\nu_1}_{\alpha_1}{p_1}_{\beta_1}\bigg[\frac{\pi^{\alpha_1\beta_1}(p_1+p_2)}{3} \left[F^2\right]^{\mu_3\mu_4}(p_3,p_4)-\big[\sqrt{g}\,F^2\big]^{\alpha_1\beta_1\mu_3\mu_4}(p_1,p_3,p_4)\bigg]\notag\\
	&-2\,\frac{\pi^{\mu_1\nu_1}(p_1)}{3}\frac{\pi^{\mu_2\nu_2}(p_2)}{3} \left[F^2\right]^{\mu_3\mu_4}(p_3,p_4)\bigg\}\notag\\
\end{align}
where we have considered the relation \eqref{Res3}.
The trace anomalous part of the correlator is then 
\begin{align}
\braket{ T^{\mu_1 \nu_1} T^{\mu_2 \nu_2}J^{\mu_3} J^{\mu_4}}_{anom}^{(d=4)}=\braket{ t_{loc}^{\mu_1 \nu_1} T^{\mu_2 \nu_2}J^{\mu_3} J^{\mu_4}}_{anom}^{(d=4)}+\braket{ T^{\mu_1 \nu_1} t_{loc}^{\mu_2 \nu_2}J^{\mu_3} J^{\mu_4}}_{anom}^{(d=4)}-\braket{ t_{loc}^{\mu_1 \nu_1} t_{loc}^{\mu_2 \nu_2}J^{\mu_3} J^{\mu_4}}_{anom}^{(d=4)},
\end{align}
or explicitly
\begin{align}
&\braket{ T^{\mu_1 \nu_1} T^{\mu_2 \nu_2}J^{\mu_3} J^{\mu_4}}_{anom}^{(d=4)}=\notag\\
&=\beta_C\,\bigg\{\mathcal{I}^{(d=4)\,\mu_1\nu_1}_{\alpha_1}\bigg[2\ {p_2}_{\lambda_1}  \bigg(\frac{\delta^{\alpha_1 (\mu_2}\pi^{\nu_2) \lambda_1} (p_1+p_2)}{3}- \frac{\pi^{\alpha_1 \lambda_1} (p_2+p_1)}{3}\frac{\pi^{\mu_2\nu_2}(p_2)}{3}\bigg)\left[F^2\right]^{\mu_3\mu_4}(p_3,p_4) \notag\\
&\qquad- {p_2}^{\alpha_1} \bigg( \frac{\pi^{\mu_2 \nu_2} (p_1+p_2) }{3}-\frac{\pi^{\mu_2\nu_2}(p_2)}{3}\bigg)\left[F^2\right]^{\mu_3\mu_4}(p_3,p_4)\bigg]\notag\\
&+\mathcal{I}^{(d=4)\,\mu_2\nu_2}_{\alpha_2}\bigg[2\ {p_1}_{\lambda_2}  \bigg(\frac{\delta^{\alpha_2 (\mu_1}\pi^{\nu_1) \lambda_1} (p_1+p_2)}{3}- \frac{\pi^{\alpha_2 \lambda_2} (p_2+p_1)}{3}\frac{\pi^{\mu_1\nu_1}(p_1)}{3}\bigg)\left[F^2\right]^{\mu_3\mu_4}(p_3,p_4) \notag\\
&\qquad- {p_1}^{\alpha_2} \bigg( \frac{\pi^{\mu_1 \nu_1} (p_1+p_2) }{3}-\frac{\pi^{\mu_1\nu_1}(p_1)}{3}\bigg)\left[F^2\right]^{\mu_3\mu_4}(p_3,p_4)\bigg]\notag\\
&-\mathcal{I}^{(d=4)\,\mu_1\nu_1}_{\alpha_1}\mathcal{I}^{(d=4)\,\mu_2\nu_2}_{\alpha_2}\bigg[\frac{1}{3}\,{p_2}_{\lambda_1}  {p_2}_{\beta_2} \delta^{\alpha_1 \alpha_2}\pi^{\beta_2 \lambda_1} (p_1+p_2)\left[F^2\right]^{\mu_3\mu_4}(p_3,p_4) \bigg]\notag\\
&-\frac{2}{3}\pi^{\mu_2\nu_2}(p_2)\bigg(\frac{\pi^{\mu_1\nu_1}(p_1+p_2)}{3} \left[F^2\right]^{\mu_3\mu_4}(p_3,p_4)-\big[\sqrt{g}\,F^2\big]^{\mu_1\nu_1\mu_3\mu_4}(p_1,p_3,p_4)\bigg)\notag\\
&-\frac{2}{3}\,\frac{\pi^{\mu_1\nu_1}(p_1)}{3}\bigg(\frac{\pi^{\mu_2\nu_2}(p_1+p_2)}{3} \left[F^2\right]^{\mu_3\mu_4}(p_3,p_4)-\big[\sqrt{g}\,F^2\big]^{\mu_2\nu_2\mu_3\mu_4}(p_2,p_3,p_4)\bigg)\notag\\
&+2\,\frac{\pi^{\mu_1\nu_1}(p_1)}{3}\frac{\pi^{\mu_2\nu_2}(p_2)}{3} \left[F^2\right]^{\mu_3\mu_4}(p_3,p_4)\bigg\}.
\end{align}
Using the relation 
\begin{align}
\delta^{(\mu_2}_{\alpha_2}\delta^{\nu_2)}_{\beta_2}=\Pi^{\mu_2\nu_2}_{\alpha_2\beta_2}+\mathcal{I}^{\mu_2\nu_2}_{(\alpha_2}p_{2\beta_2)}+\frac{1}{3}\pi^{\mu_2\nu_2}\delta_{\alpha_2\beta_2}
\end{align}
on the first lines we obtain
\begin{align}
	&\braket{ T^{\mu_1 \nu_1} T^{\mu_2 \nu_2}J^{\mu_3} J^{\mu_4}}_{anom}^{(d=4)}=\notag\\
	&=\beta_C\,\bigg\{\mathcal{I}^{(d=4)\,\mu_1\nu_1}_{\alpha_1}\,\Pi^{(d=4)\,\mu_2\nu_2}_{\alpha_2\beta_2}(p_2)\bigg[\frac{2}{3}\ {p_2}_{\lambda_1} \delta^{\alpha_1 \alpha_2}\pi^{\beta_2 \lambda_1} (p_1+p_2)-\frac{1}{3} {p_2}^{\alpha_1} \pi^{\alpha_2 \beta_2} (p_1+p_2)\bigg]\left[F^2\right]^{\mu_3\mu_4}(p_3,p_4)\notag\\
	&\hspace{0.9cm}+\Pi^{(d=4)\,\mu_1\nu_1}_{\alpha_1\beta_1}(p_1)\,\mathcal{I}^{(d=4)\,\mu_2\nu_2}_{\alpha_2}\bigg[\frac{2}{3}\ {p_1}_{\lambda_2} \delta^{\alpha_1 \alpha_2}\pi^{\beta_1 \lambda_2} (p_1+p_2)-\frac{1}{3} {p_1}^{\alpha_2} \pi^{\alpha_1 \beta_1} (p_1+p_2)\bigg]\left[F^2\right]^{\mu_3\mu_4}(p_3,p_4)\notag\\
	&\hspace{0.9cm}+\mathcal{I}^{(d=4)\,\mu_1\nu_1}_{\alpha_1}\mathcal{I}^{(d=4)\,\mu_2\nu_2}_{\alpha_2}\bigg[\frac{1}{3}\,{p_2}_{\lambda_1}  {p_2}_{\beta_2} \delta^{\alpha_1 \alpha_2}\pi^{\beta_2 \lambda_1} (p_1+p_2)\left[F^2\right]^{\mu_3\mu_4}(p_3,p_4) \bigg]\notag\\
	&\hspace{0.9cm}-\frac{2}{3}\pi^{\mu_2\nu_2}(p_2)\bigg(\frac{\pi^{\mu_1\nu_1}(p_1+p_2)}{3} \left[F^2\right]^{\mu_3\mu_4}(p_3,p_4)-\big[\sqrt{g}\,F^2\big]^{\mu_1\nu_1\mu_3\mu_4}(p_1,p_3,p_4)\bigg)\notag\\
	&\hspace{0.9cm}-\frac{2}{3}\,\pi^{\mu_1\nu_1}(p_1)\bigg(\frac{\pi^{\mu_2\nu_2}(p_1+p_2)}{3} \left[F^2\right]^{\mu_3\mu_4}(p_3,p_4)-\big[\sqrt{g}\,F^2\big]^{\mu_2\nu_2\mu_3\mu_4}(p_2,p_3,p_4)\bigg)\notag\\
	&\hspace{0.9cm}+2\,\frac{\pi^{\mu_1\nu_1}(p_1)}{3}\frac{\pi^{\mu_2\nu_2}(p_2)}{3} \left[F^2\right]^{\mu_3\mu_4}(p_3,p_4)\bigg\}.
\end{align}
From this equation, the identification of $\braket{TTJJ}_{0-trace}$ and $\braket{TTJJ}_{pole}$ is manifest in accordance with \eqref{pole} and \eqref{0-residue}.

\section{Metric variations involving the Vielbein\label{VielbMetr}}

We present in this section all the useful relations related to the metric variation of the vielbein. In particular we have that the variation with respect to the metric tensor can be written as
\begin{equation}
	\frac{\delta}{\delta\,g_{\mu\nu}}\equiv\,\frac{1}{4}\left(V^{\mu}_a\,\frac{\delta}{\delta\,V_{\nu\,a}}+V^{\nu}_a\,\frac{\delta}{\delta\,V_{\mu\,a}}\right)
\end{equation}
\begin{equation}
	\frac{\delta\,V^\mu_a}{\delta\,g_{\mu_1\nu_1}}=-\frac{1}{2}\,g^{\mu(\mu_1}V^{\nu_1)}_a,\quad \frac{\delta\,V_\mu^a}{\delta\,g_{\mu_1\nu_1}}=\frac{1}{2}\,\delta_{\mu}^{(\mu_1}V^{\nu_1)}_a\delta^{ab},\quad 
	\frac{\delta\,V^\mu_a}{\delta\,g^{\mu_1\nu_1}}=\frac{1}{2}\,\delta^{\mu}_{(\mu_1}V_{\nu_1)}^b\delta_{ab}
\end{equation}	
and these relations are in accordance with the well known variation of the curved metric
\begin{align}
	\frac{\delta\,g^{\mu\nu}}{\delta\,g_{\mu_1\nu_1}}=\eta^{ab}\,\frac{\delta\,(V^\mu_aV^\nu_b)}{\delta\,g_{\mu_1\nu_1}}&=-\frac{1}{2}\,\eta^{ab}\,g^{\mu(\mu_1}V^{\nu_1)}_aV^\nu_b-\frac{1}{2}\,\eta^{ab}\,g^{\nu(\mu_1}V^{\nu_1)}_bV^\mu_a=-g^{\mu(\mu_1}g^{\nu_1)\nu}\\
	\frac{\delta\,g^{\mu\nu}}{\delta\,g^{\mu_1\nu_1}}=\eta^{ab}\,\frac{\delta\,(V^\mu_aV^\nu_b)}{\delta\,g^{\mu_1\nu_1}}&=\frac{1}{2}\,\eta^{ab}\,\delta^{\mu}_{(\mu_1}V_{\nu_1)}^c\eta_{ac}V^\nu_b+\frac{1}{2}\,\eta^{ab}\,\delta^{\nu}_{(\mu_1}V_{\nu_1)}^c\eta_{bc}V^\mu_a=\delta^\mu_{(\mu_1}\delta^\nu_{\nu_1)}.
\end{align}	
Other useful relations in order to get the vertices are written as
\begin{align}
	\frac{\delta\,\left(V\,V^\mu_a\right)}{\delta\,g_{\mu_1\nu_1}}&=\frac{V}{2}\left(g^{\mu_1\nu_1}V^\mu_a-g^{\mu(\mu_1}V^{\nu_1)}_a\right)\\
	\frac{\delta^2\,\left(V\,V^\mu_a\right)}{\delta\,g_{\mu_1\nu_1}\delta g_{\mu_2\nu_2}}&=\frac{V}{2}\left(-g^{\mu_1(\mu_2}g^{\nu_2)\nu_1}V^\mu_a+g^{\mu(\mu_2}g^{\nu_2)(\mu_1}V^{\nu_1)}_a\right)+\frac{V}{4}g^{\mu_1\nu_1}\left(g^{\mu_2\nu_2}V^\mu_a-g^{\mu(\mu_2}V^{\nu_2)}_a\right)\notag\\
	&\qquad-\frac{V}{4}g^{\mu(\mu_1}V^{\nu_1)}_ag^{\mu_2\nu_2}+\frac{V}{4}g^{\mu(\mu_1}g^{\nu_1)(\mu_2}V^{\nu_2)}_a
\end{align}
and we observe that
\begin{align}
	\frac{\delta^2\,\left(V\,V^\mu_a\right)}{\delta\,g_{\mu_1\nu_1}\delta g_{\mu_2\nu_2}}\neq
	\frac{\delta^2\,\left(V\,V^\mu_a\right)}{\delta\,g_{\mu_2\nu_2}\delta g_{\mu_1\nu_1}}
\end{align}
We can obtain the metric variation of the spin connection starting from its definition
\begin{equation} 
	\nabla_\mu V_{b \nu} = \partial_\mu V_{b \nu} - \omega_{\mu \ b}^{\ a} V_{\nu a} - \Gamma^\lambda_{\mu \nu} V_{b \nu} = 0 ,
\end{equation}
\begin{equation} 
	\omega_{\mu a b}= V^\nu_a (\partial_\mu V_{b \nu} - \Gamma^\lambda_{\mu \nu} V_{b \lambda} ) . 
\end{equation} 
Then its variation variation under a metric variation is given as
\begin{align} 
	\delta \omega_{\mu a b} &=
	\delta V^\rho_a \underbrace{V^\nu_c V^c_\rho}_{\delta^\nu_\rho} (\partial_\mu V_{b \nu} - \Gamma^\lambda_{\mu \nu} V_{b \lambda} ) + V^\nu_a (\partial_\mu \delta V_{b \nu} - \Gamma^\lambda_{\mu \nu} \delta V_{b \lambda} ) - V^\nu_a V_{b \lambda} \delta \Gamma^\lambda_ {\mu \nu} \notag\\
	&= - V^\rho_a \delta V_\rho^c \underbrace{V^\nu_c( \partial_\mu V_{b \nu} - \Gamma^\lambda_{\mu \nu} V_{b \lambda} )}_{\omega_{\mu cb}} +V^\nu_a (\partial_\mu \delta V_{b \nu} - \Gamma^\lambda_{\mu \nu} \delta V_{b \lambda} ) - V^\nu_a V_{b \lambda} \delta \Gamma^\lambda_ {\mu \nu} \notag\\
	&= V^\nu_a (\partial_\mu \delta V_{b \nu} - \omega_{\mu \ b}^{\ c} \delta V_{c \nu} - \Gamma^\lambda_{\mu \nu} \delta V_{b \lambda} ) - V^\nu_a V_{b \lambda} \delta \Gamma^{\lambda}_{\mu \nu} \notag\\
	&= V_a^\nu \nabla_\mu \delta V_{b \nu} - V^\nu_a V_{b \lambda} \delta \Gamma^\lambda_{\mu \nu} .  \end{align}
Then the variation with respect to the metric is obtained as
\begin{equation} \dfun{\omega_{\mu a b} (x)}{g_{\mu_1 \nu_1} (y) } 
	= \delta^{(\mu_1}_\mu V^{\nu_1)}_{[a} V^\lambda_{b]} \, \nabla_\lambda \delta(x-y). \end{equation}
It is worth mentioning that, since the spin connection always appears contracted with the antisymmetric tensor $\Sigma^{ab}$,  we can omit the antisymmetrization on the indices $[a,b]$. 
\section{Vertices}\label{AppendixB}
We have presented in \figref{vertici} a list of all the vertices which we needed for the momentum space computation of the $TTJJ$ correlator. They are obtained by taking functional derivatives of the action. We consider all the momenta incoming into the vertex. 
In the case of QED we consider the action
\begin{equation} 
	S = \int d^d x \, V \left( {i \over 2} \bar{\psi} \gamma^\lambda \overleftrightarrow{\partial_\lambda} \psi - e \bar{\psi} \gamma^\lambda A_\lambda \psi + {i \over 4} \omega_{\mu a b } V^\mu_c \bar{\psi} \gamma^{abc} \psi \right) 
\end{equation}
where 
\begin{equation}
	\bar{\psi} \gamma^\lambda \overleftrightarrow{\partial_\lambda} \psi =\bar \psi \gamma^\lambda  \partial_\lambda  \psi - \partial_\lambda \bar \psi \gamma^\lambda \psi \qquad 
	\gamma^{ab \, \dots \, c} = \gamma^{[a}\gamma^{b} \dots \gamma^{c]},
\end{equation}
and we use the Lorentz generator
\begin{equation}
	\Sigma^{ab} = {1 \over 4} [\gamma^a, \gamma^b] = {1 \over 2} \gamma^{ab}.
\end{equation}
The variations of the action with respect to the metric in the flat limit are given by
\begin{align}
	\left. \dfun{S}{g_{\mu_1 \nu_1} (x_1)} \right|_{g=\delta} &=
	{1 \over 2} A^{\mu_1 \nu_1 \rho \lambda} \left( {i \over 2} \bar{\psi} \gamma_\lambda \overleftrightarrow{\partial_\rho} \psi - e \bar{\psi} \gamma_\alpha A_\beta \psi \right) (x_1), 
	\notag\\
	\left. \dfun{^2 S}{g_{\mu_2 \nu_2} (x_2) \delta g_{\mu_1 \nu_1} (x_1)} \right|_{g=\delta} &=
	{1 \over 4} (B^{\mu_1 \nu_1 \mu_2 \nu_2 \rho \lambda} - C^{\mu_1 \nu_1 \mu_2 \nu_2 \rho \lambda} + D^{\mu_1 \nu_1 \mu_2 \nu_2 \rho \lambda}) \left( {i \over 2} \bar{\psi} \gamma_\lambda \frac{\overleftrightarrow{\partial}}{\partial x_1^\rho} \psi - e \bar{\psi} \gamma_\lambda A_\rho \psi \right) \delta_{x_1, x_2} \notag\\ 
	& - \frac{i}{8}\delta^{(\mu_1}_\alpha \delta^{\nu_1 ) (\mu_2} \delta^{\nu_2)}_\beta \, \bar{\psi} \, \gamma^{\alpha \beta \lambda} \, \psi \, \dpar{}{x_1^\lambda}\delta_{x_1, x_2},\label{S2}
\end{align}
where we have defined 
\begin{align} A^{\mu \nu \rho \sigma} &= \delta^{\mu \nu} \delta^{\rho \sigma} -  \delta^{\mu (\rho} \delta^{\sigma) \nu} \notag\\
	B^{\mu \nu \rho \sigma \alpha \beta} &= \delta^{\mu \nu} \delta^{\rho \sigma} \delta^{\alpha \beta} - 2 \delta^{\mu (\rho} \delta^{\sigma) \nu} \delta^{\alpha \beta} \notag\\
	\tilde B^{\mu \nu \rho \sigma} &=   \delta^{\mu \nu} \delta^{\rho \sigma} -  2\delta^{\mu (\rho} \delta^{\sigma) \nu} \notag\\
	C^{\mu \nu \rho \sigma \alpha \beta} &= \delta^{\mu \nu} \delta^{\alpha (\rho} \delta^{\sigma) \beta} + \delta^{\rho \sigma} \delta^{\alpha (\mu} \delta^{\nu) \beta}  \notag\\
	\tilde C^{\mu \nu \rho \sigma \alpha \beta} &=  \delta^{\mu \nu} \delta^{\alpha (\rho} \delta^{\sigma) \beta} \notag\\
	D^{\mu \nu \rho \sigma \alpha \beta} &= \delta^{\alpha (\mu} \delta^{\nu)(\rho} \delta^{\sigma) \beta} + 2 \delta^{\alpha (\rho} \delta^{\sigma)(\mu} \delta^{\nu) \beta} \notag\\
	\tilde D^{\mu \nu \rho \sigma \alpha \beta} &= \delta^{\alpha (\mu} \delta^{\nu)(\rho} \delta^{\sigma) \beta} + \delta^{\alpha (\rho} \delta^{\sigma)(\mu} \delta^{\nu) \beta} \notag\\
	E^{\mu \nu \rho \sigma \alpha \beta} & =  \delta^{\alpha (\mu} \delta^{\nu)(\rho} \delta^{\sigma) \beta}.
\end{align}
and we have used the results in \appref{VielbMetr}. 
From \eqref{S2} we extract the vertices in direct space involving the fields $\psi(y_1)$,  $\bar{\psi}(y_2)$, and $T^{\mu_2 \nu_2}(x_2)$, $T^{\mu_1 \nu_1}(x_1)$,  $J^{\alpha_1} (z) $ as
\begin{align}
	 \label{gne} V^{\alpha_1}_{\bar{\psi} \psi J} (y_1, y_2, z ) &= -e \gamma^{\alpha_1} \delta_{y_2,z} \delta_{y_1,z} ,\\
	V_{T \bar{\psi} \psi}^{\mu_1 \nu_1} (x_1,  y_1, y_2 )&= - {i \over 4} A^{\mu_1 \nu_1 \alpha \beta} \gamma_\alpha \left( \dpar{}{y_1^\beta} - \dpar{}{y_2^\beta} \right) [\delta_{y_1,x_1}\delta_{y_2,x_1}],
	\\
	V_{T  \bar{\psi} \psi J }^{\mu_1 \nu_1 \alpha_1}(x_1,  y_1, y_2, z ) &= -{e \over 2} A^{\mu_1 \nu_1 \alpha_1 \lambda} \gamma_\lambda \delta_{y_1,x_1}\delta_{y_2,x_1} ,\\
	V^{\mu_1 \nu_1 \mu_2 \nu_2}_{TT\bar{\psi} \psi} (x_1,  x_2, y_1, y_2 ) &= -{i \over 8} (B^{\mu_1 \nu_1 \mu_2 \nu_2 \alpha \beta} - C^{\mu_1 \nu_1 \mu_2 \nu_2 \alpha \beta} + D^{\mu_1 \nu_1 \mu_2 \nu_2 \alpha \beta}) \gamma_\alpha \left( \dpar{}{y_1^\beta} - \dpar{}{y_2^\beta} \right) [\delta_{y_1,x_1}\delta_{y_2,x_1} \delta_{x_2,x_1}]  \notag\\ 
	&- {i \over 8} \delta^{\alpha (\mu_1} \delta^{\nu_1) (\mu_2} \delta^{\nu_2) \beta}  \gamma_{\alpha \beta}^{\ \ \lambda} \dpar{}{x^\lambda_2} [\delta_{y_1,x_1}\delta_{y_2,x_1} \delta_{x_2,x_1}] 
	\\
	V^{\mu_1 \nu_1 \mu_2 \nu_2 \alpha_1}_{TT \bar{\psi}\psi J} (x_1, x_2,  y_1, y_2, z) &= {1 \over 4} (B^{\mu_1 \nu_1 \mu_2 \nu_2 \alpha_1 \lambda} - C^{\mu_1 \nu_1 \mu_2 \nu_2 \alpha_1 \lambda} + D^{\mu_1 \nu_1 \mu_2 \nu_2 \alpha_1 \lambda} ) \gamma_\lambda \delta_{y_1,x_1}\delta_{y_2,x_1} \delta_{x_2,x_1}\delta_{z,x_1}.
\end{align}

In momentum space these vertices read as
\begin{align} 
	V^{\mu_3}_{J \bar{\psi} \psi} (k_1, k_2,  p_3) &= - e \gamma^{\mu_3} ,\\
	V_{T \bar{\psi} \psi}^{\mu_1 \nu_1} (p_1 ,  k_1, k_2)&=  {1 \over 4} A^{\mu_1 \nu_1 \alpha \beta} \,  \gamma_\alpha \, (k_1 +k_2)_\beta ,
	\\
	V_{T J \bar{\psi} \psi}^{\mu_1 \nu_1 \mu_3} (p_1,  p_3 ,  k_1, k_2)&= - {e \over 2} A^{\mu_1 \nu_1 \mu_3 \lambda} \gamma_\lambda ,\\
	V^{\mu_1 \nu_1 \mu_2 \nu_2}_{TT\bar{\psi} \psi} (p_1,  p_2,  k_1, k_2) &=  {1 \over 8} (B^{\mu_1 \nu_1 \mu_2 \nu_2 \rho \lambda} - C^{\mu_1 \nu_1 \mu_2 \nu_2 \rho \lambda} + D^{\mu_1 \nu_1 \mu_2 \nu_2 \rho \lambda}) \, \gamma_\lambda \, (k_1 + k_2)_\rho \notag\\ 
	& - {1 \over 8}\, \delta^{\alpha (\mu_1} \delta^{\nu_1) (\mu_2} \delta^{\nu_2) \beta} \, \gamma_{\alpha \beta \lambda} \, p_2^\lambda
	\\
	V^{\mu_1 \nu_1 \mu_2 \nu_2 \mu_3}_{TTJ\bar{\psi}\psi} (p_1,  p_2,  p_3 ,  k_1, k_2) &= - {e \over 4} (B^{\mu_1 \nu_1 \mu_2 \nu_2 \mu_3 \lambda} - C^{\mu_1 \nu_1 \mu_2 \nu_2 \mu_3 \lambda} + D^{\mu_1 \nu_1 \mu_2 \nu_2 \mu_3 \lambda}) \gamma_\lambda ,
\end{align}
where $k_1$ is outgoing.

In the case of scalar QED with the action
\begin{equation}
S=\int d^d x \, \sqrt{-g} \, \left( \partial^\mu {\phi^\dagger} \partial_\mu \phi + ieA^\mu (\partial_\mu {\phi^\dagger} \, \phi - {\phi^\dagger} \, \partial_\mu \phi) + e^2 A^\mu A_\mu {\phi^\dagger} \phi + \chi R \, {\phi^\dagger} \phi \right),
\end{equation}
the variations of the action with respect to the metric are given as
\begin{align} 
	\left. \dfun{S}{g_{\mu_1 \nu_1}(x_1) }  \right|_{g=\delta} =& 
	{1 \over 2} \tilde B^{\mu_1 \nu_1 \alpha \beta} (\partial_\alpha {\phi^\dagger} \partial_\beta \phi 
	- ie A_\alpha \,  {\phi^\dagger} \partial_\beta \phi
	+ e^2 A_{\alpha} A_{\beta} | \phi |^2) 
	- \chi A^{\mu_1 \nu_1 \alpha \beta} \partial_\alpha \partial_\beta |\phi|^2 
	\notag\\
	\left. \dfun{^2 S}{g_{\mu_2 \nu_2}(x_2)\, \delta g_{\mu_1 \nu_1} (x_1) }  \right|_{g=\delta} =
	& \left( 
	{1 \over 4} B^{\mu_1 \nu_1 \mu_2 \nu_2 \alpha \beta} + \tilde D^{\mu_1 \nu_1 \mu_2 \nu_2 \alpha \beta} -{1 \over 2}C^{\mu_1 \nu_1 \mu_2 \nu_2 \alpha \beta} 
	\right) (D_\alpha \phi)^\dagger( D_\beta \phi ) \delta_{x_1 x_2} \notag\\
	& + \chi \Big[ {1 \over 2} \left( 
	\delta^{\mu_1 \nu_1} A^{\mu_2 \nu_2 \alpha \beta} + C^{\alpha \beta \mu_1 \nu_1 \mu_2 \nu_2} -2 E^{\mu_1 \nu_1 \mu_2 \nu_2 \beta \alpha}
	\right) \partial_\alpha \partial_\beta \delta_{x_1 x_2} |\phi|^2  \notag\\
	& - {1 \over 2} \left(
	\delta^{\mu_1 \nu_1} B^{\mu_2 \nu_2 \alpha} - \tilde C^{\alpha \beta \mu_1 \nu_1 \mu_2 \nu_2}- E^{\mu_1 \nu_1 \mu_2 \nu_2 \alpha \beta} 
	\right) \partial_\alpha \partial_\beta |\phi|^2  \delta_{x_1 x_2}  \notag\\
	& + \left( 
	{1 \over 2} A^{\mu_1 \nu_1 \alpha \beta} \delta^{\mu_2 \nu_2} + \tilde D^{\mu_1 \nu_1 \mu_2 \nu_2 \alpha \beta}  - C^{\alpha \beta \mu_1 \nu_1 \mu_2 \nu_2} 
	\right) \partial_\alpha |\phi|^2 \, \partial_\beta \delta_{x_1 x_2} \Big] .
\end{align}
From the previous equation we extract the vertices in direct space involving the fields $\phi(y_1)$ and ${\phi^\dagger}(y_2)$, and $T^{\mu_2 \nu_2}(x_2)$, $T^{\mu_1 \nu_1}(x_1)$, $J^{\mu_3} (x_3) $, $J^{\mu_4} (x_4) $ as
\begin{align} V_{J{\phi^\dagger}\phi}^{\mu_3} &= ie \left( \frac{\partial}{\partial {y_1}_{\mu_3}} - \frac{\partial}{\partial {y_2}_{\mu_3}} \right) \delta_{x_3 y_1} \delta_{x_3 y_2} \\ 
	V_{J J {\phi^\dagger}\phi}^{\mu_3 \mu_4} &= 2  e^2 \delta^{\mu_3 \mu_4} . \\
	V_{T {\phi^\dagger} \phi}^{\mu_1 \nu_1} &=  \left[ -{1\over 2} \tilde B^{\mu_1 \nu_1 \alpha \beta} \frac{\partial}{\partial y_2^\alpha} \frac{\partial}{\partial y_1^\beta} + i \chi A^{\mu_1 \nu_1 \alpha \beta} \left( \frac{\partial}{\partial y_1^\alpha} + \frac{\partial}{\partial y_2^\alpha} \right)  \left( \frac{\partial}{\partial y_1^\beta} + \frac{\partial}{\partial y_2^\beta} \right) \right] \delta_{x_1 y_1} \delta_{x_1 y_2} \notag\\
	V^{\mu_1 \nu_1 \mu_3}_{T J {\phi^\dagger} \phi} & = -{i e \over 2} A^{\mu_1 \nu_1 \mu_3 \lambda}  \left( \frac{\partial}{\partial {y_1}^{\lambda}} - \frac{\partial}{\partial {y_2}^{\lambda}} \right) \delta_{x_3 y_1} \delta_{x_3 y_2}  \notag\\
	V^{\mu_1 \nu_1 \mu_3 \mu_4}_{T J J {\phi^\dagger} \phi} &= e^2 \tilde B^{\mu_1 \nu_1 \mu_3 \mu_4}\\
	V_{TT {\phi^\dagger} \phi}^{\mu_1 \nu_1 \mu_2 \nu_2} 
	&=  \left( 
	{1 \over 4} B^{\mu_1 \nu_1 \mu_2 \nu_2 \alpha \beta} + \tilde D^{\mu_1 \nu_1 \mu_2 \nu_2 \alpha \beta} - 	{1 \over 2} C^{\mu_1 \nu_1 \mu_2 \nu_2 \alpha 	\beta} 	
	\right) \frac{\partial}{\partial {y_1}^\alpha} \frac{\partial}{\partial {y_2}^\beta} (\delta_{x_1 y_2} \delta_{x_1 x_2}) \notag\\
	& + \chi \Big[ {1 \over 2} \left( 
	\delta^{\mu_1 \nu_1} A^{\mu_2 \nu_2 \alpha \beta} + C^{\alpha \beta \mu_1 \nu_1 \mu_2 \nu_2} -2 E^{\mu_1 \nu_1 \mu_2 \nu_2 \beta \alpha}
	\right) (\partial_\alpha \partial_\beta \delta_{x_1 x_2} )\delta_{x_1 y_1}\delta_{x_1 y_2}  \notag\\
	& - {1 \over 2} \left(
	\delta^{\mu_1 \nu_1} B^{\mu_2 \nu_2 \alpha} - \tilde C^{\alpha \beta \mu_1 \nu_1 \mu_2 \nu_2}- E^{\mu_1 \nu_1 \mu_2 \nu_2 \alpha \beta} 
	\right) ( \partial_\alpha \delta_{x_1 y_1} \delta_{x_1 y_2}) \, \partial_\beta \delta_{x_1 x_2} \Big] 	\notag\\
	& + \left( 
	{1 \over 2} A^{\mu_1 \nu_1 \alpha \beta} \delta^{\mu_2 \nu_2} + \tilde D^{\mu_1 \nu_1 \mu_2 \nu_2 \alpha \beta}  - C^{\alpha \beta \mu_1 \nu_1 \mu_2 \nu_2} 
	\right) ( \partial_\alpha \delta_{x_1 y_1} \delta_{x_1 y_2}) \, \partial_\beta \delta_{x_1 x_2} \Big] 
\end{align}

\begin{align} 
	V^{\mu_1 \nu_1 \mu_2 \nu_2 \mu_3}_{TTJ{\phi^\dagger} \phi} 
	& =  ie \left( 
	{1 \over 4} B^{\mu_1 \nu_1 \mu_2 \nu_2 \mu_3 \lambda} + \tilde D^{\mu_1 \nu_1 \mu_2 \nu_2 \mu_3 \lambda} -{1 \over 2}C^{\mu_1 \nu_1 \mu_2 \nu_2 \mu_3 \lambda} 
	\right) \left( \frac{\partial}{\partial {y_1}^{\lambda}} - \frac{\partial}{\partial {y_2}^{\lambda}} \right) \delta_{x_1 y_1} \delta_{x_1 y_2} \notag\\
	V^{\mu_1 \nu_1 \mu_2 \nu_2 \mu_3 \mu_4 }_{TTJJ{\phi^\dagger} \phi} 
	& =  -2e^2 \left( {1 \over 4} B^{\mu_1 \nu_1 \mu_2 \nu_2 \mu_3 \mu_4} + \tilde D^{\mu_1 \nu_1 \mu_2 \nu_2 \mu_3 \mu_4} -{1 \over 2}C^{\mu_1 \nu_1 \mu_2 \nu_2 \mu_3 \mu_4} \right).
\end{align}
In momentum space these vertices are

\begin{align} 
	V_{J{\phi^\dagger}\phi}^{\mu_3} (k_1,k_2)  &= -e (k_1 + k_2)^{\mu_3}, \\
	V_{J J {\phi^\dagger}\phi}^{\mu_3 \mu_4} &= -2e^2 \delta^{\mu_3 \mu_4},\\
V_{T {\phi^\dagger} \phi}^{\mu_1 \nu_1}  (k_1,k_2) &=  
	-{1 \over 2} \tilde B^{\mu_1 \nu_1 \alpha \beta} {k_1}_\alpha {k_2}_\beta -\chi A^{\mu_1 \nu_1 \alpha \beta} (k_1-k_2)_\alpha (k_1-k_2)_\beta ,\\
	V^{\mu_1 \nu_1 \mu_3}_{T J {\phi^\dagger} \phi} (k_1,k_2) & = - { e \over 2} \tilde B^{\mu_1 \nu_1 \mu_3 \lambda}  (k_1+k_2)_\lambda ,\\
	V^{\mu_1 \nu_1 \mu_3 \mu_4}_{T J J {\phi^\dagger} \phi} &= -e^2 \tilde B^{\mu_1 \nu_1 \mu_3 \mu_4}\\
	V_{TT {\phi^\dagger} \phi}^{\mu_1 \nu_1 \mu_2 \nu_2} (p_2, k_1,k_2)
	&= - \left( 
	{1 \over 4} B^{\mu_1 \nu_1 \mu_2 \nu_2 \alpha \beta} + \tilde D^{\mu_1 \nu_1 \mu_2 \nu_2 \alpha \beta} - 	{1 \over 2} C^{\mu_1 \nu_1 \mu_2 \nu_2 \alpha 	\beta} 	
	\right) {k_1}_\alpha {k_2}_\beta \notag\\
	& -\chi \Big[ \left( 
	{1 \over 2} A^{\mu_1 \nu_1 \alpha \beta} \delta^{\mu_2 \nu_2} + \tilde D^{\mu_1 \nu_1 \mu_2 \nu_2 \alpha \beta}  - C^{\alpha \beta \mu_1 \nu_1 \mu_2 \nu_2} 
	\right) (k_1-k_2)_\alpha (k_1-k_2)_\beta  \notag\\
	& - {1 \over 2} \left(
	\delta^{\mu_1 \nu_1} B^{\mu_2 \nu_2 \alpha} - \tilde C^{\alpha \beta \mu_1 \nu_1 \mu_2 \nu_2}- E^{\mu_1 \nu_1 \mu_2 \nu_2 \alpha \beta} 
	\right)  \, {p_2}_\alpha \, (k_1-k_2)_\beta \notag\\
	& + {1 \over 2} \left( 
	\delta^{\mu_1 \nu_1} A^{\mu_2 \nu_2 \alpha \beta} + C^{\alpha \beta \mu_1 \nu_1 \mu_2 \nu_2} -2 E^{\mu_1 \nu_1 \mu_2 \nu_2 \beta \alpha}
	\right)  {p_2}_\alpha {p_2}_\beta \Big]  \\
	V^{\mu_1 \nu_1 \mu_2 \nu_2 \mu_3}_{TTJ{\phi^\dagger} \phi} 
	& =  -e \left( {1 \over 4} B^{\mu_1 \nu_1 \mu_2 \nu_2 \alpha \beta} + \tilde D^{\mu_1 \nu_1 \mu_2 \nu_2 \mu_3 \lambda} -{1 \over 2}C^{\mu_1 \nu_1 \mu_2 \nu_2 \mu_3 \lambda} \right) (k_1+k_2)_\lambda \\
	V^{\mu_1 \nu_1 \mu_2 \nu_2 \mu_3 \mu_4 }_{TTJJ{\phi^\dagger} \phi} 
	& =  -2e^2 \left( {1 \over 4} B^{\mu_1 \nu_1 \mu_2 \nu_2 \mu_3 \mu_4} +\tilde D^{\mu_1 \nu_1 \mu_2 \nu_2 \mu_3 \mu_4} -{1 \over 2}C^{\mu_1 \nu_1 \mu_2 \nu_2 \mu_3 \mu_4} \right).
\end{align}
\section{Functional Variations\label{FunctionalVariations}}
We list some functional variations in momentum space needed for the identification of the anomaly part of the correlator $\braket{TTJJ}$. 
\begin{align}
	\big[R^{\mu \nu \rho \sigma} \big]^{\mu_1 \nu_1}  (p_1) &= {1 \over 2} \, \Big[ \delta^{\mu(\mu_1} \delta^{\nu_1)\rho} \, p_1^\nu \, p_1^\sigma   
	+ \delta^{\nu(\mu_1}\delta^{\nu_1)\sigma} \, p_1^\mu \, p_1^\rho  
	- \delta^{\mu(\mu_1}\delta^{\nu_1)\sigma}\, p_1^\nu \,  p_1^\rho  
	- \delta^{\nu(\mu_1} \delta^{\nu_1)\rho} \, p_1^\mu \, p_1^\sigma \Big] \\[1ex]
	\	\big[R^{\mu \nu}\big ]^{\mu_1 \nu_1}  (p_1) &= {1 \over 2} \, \Big[  \delta^{\mu(\mu_1} \delta^{ \nu_1 )\nu} \, p_1^2
	+ \delta^{\mu_1 \nu_1} p_1^\mu p_1^\nu 
	- 2  p_1^{ (\mu_1} \delta^{\nu_1)( \mu } p_1^{\nu) }  \Big] \\[1ex]
	[R]^{\mu_1\nu_1}(p_1)&=p_1^2\left(\delta^{\mu_1\nu_1}-\frac{p_1^{\mu_1}p_1^{\nu_1}}{p_1^2}\right)=p_1^2\,\pi^{\mu_1\nu_1}(p_1),\\[1ex]
	[\square_1]^{\mu_1\nu_1}(p_1,p_2)&=-\frac{1}{2}\delta^{\mu_1\nu_1}(p_1\cdot p_2)+\frac{1}{2}\big[p_2^{\mu_1}(p_1+p_2)^{\nu_1}+(p_1+p_2)^{\mu_1}p_2^{\nu_1}\big]\\[1ex]
	\big[\sqrt{-g}\,\square_1\big]^{\mu_1\nu_1}(p_1,p_2)&=-\frac{1}{2}\delta^{\mu_1\nu_1}p_2^2+\big[\square_1\big]^{\mu_1\nu_1}(p_1,p_2)\\[1ex]
	[R]^{\mu_1\nu_1\mu_2\nu_2}(p_1,p_2)&= 2(p_1+p_2)^{(\mu_1}\delta^{\nu_1)(\mu_2}(p_1+p_2)^{\nu_2)}-p_2^{(\mu_1}\delta^{\nu_1)(\mu_2}p_1^{\nu_2)}\notag\\
	&-\delta^{\mu_1\nu_1}p_1^{(\mu_2}(p_1+p_2)^{\nu_2)}-\delta^{\mu_2\nu_2}p_2^{(\mu_1}(p_1+p_2)^{\nu_1)}-\delta^{\mu_1(\mu_2}\delta^{\nu_2)\nu_1}(p_1+p_2)^2\notag\\
	&+\frac{1}{2}\left(\delta^{\mu_1(\mu_2}\delta^{\nu_2)\nu_1}+\delta^{\mu_1\nu_1}\delta^{\mu_2\nu_2}\right)(p_1\cdot p_2)
	\\[1ex]
	\big[F^2\big]^{\mu_3 \mu_4} (p_3, p_4) &= 
		4 \left(p_3^{\mu_4} p_4^{\mu_3}- \delta^{\mu_3 \mu_4} p_3 \cdot p_4 \right) \label{Fmu3mu4}\\[1ex]
	\big[\sqrt{-g}\,F^2\big]^{\mu_1 \mu_2 \mu_3 \mu_4} (p_1 , p_3, p_4) &=\frac{1}{2}\delta^{\mu_1\nu_1}\,\big[F^2\big]^{\mu_3\mu_4}(p_3,p_4)\notag\\
	&\quad+4\delta^{\mu(\mu_1}\delta^{\nu_1)\nu}\bigg[\delta^{\mu_3\mu_4}\,p_{3\mu}\,p_{4\nu}+\delta_\mu^{\mu_3}\delta_\nu^{\mu_4}(p_3\cdot p_4)-\delta_\mu^{\mu_3}p_3^{\mu_4}p_{4\nu}-\delta_\nu^{\mu_4}p_4^{\mu_3}p_{3\mu}\bigg]\label{gFmu3mu4}
\end{align}
These results can be re-written using the projectors $\Pi$, $\mathcal{I}$ and $\pi$ in the form 
\begin{align}
	[\square_1]^{\mu_1\nu_1}(p_1,p_2)&=\Pi^{\mu_1\nu_1}_{\alpha_1\beta_1}(p_1)\,p_2^{\alpha_1}\,p_2^{\beta_1}+\frac{1}{2}\mathcal{I}^{\mu_1\nu_1}_{\hspace{0.5cm}\alpha_1}\,p_2^{\alpha_1}\Big(p_1^2+2(p_1\cdot p_2)\Big)+\frac{1}{3}\pi^{\mu_1\nu_1}(p_1)\Big(p_2^2-(p_1\cdot p_2)\Big)
\end{align}
\begin{align}
	[R]^{\mu_1\nu_1\mu_2\nu_2}(p_1,p_2)&=\Pi^{\mu_1\nu_1}_{\alpha_1\beta_1}(p_1)\Pi^{\mu_2\nu_2}_{\alpha_2\beta_2}(p_2)\left[-(p_1+p_2)^2\delta^{\alpha_1\alpha_2}\pi^{\beta_1\beta_2}(p_1+p_2)+\frac{1}{2}\delta^{\alpha_1\alpha_2}\delta^{\beta_1\beta_2}(p_1\cdot p_2)\right]\notag\\
	&\hspace{-2cm}+\left[\mathcal{I}^{\mu_1\nu_1}_{\hspace{0.5cm}\alpha_1}(p_1)\Pi^{\mu_2\nu_2}_{\alpha_2\beta_2}(p_2)\,\frac{p_1^{\beta_2}}{2}\,\Big(-(p_1+p_2)^2\,\pi^{\alpha_1\alpha_2}(p_1+p_2)+p_1^2\,\pi^{\alpha_1\alpha_2}(p_1)-p_2^2\,\delta^{\alpha_1\alpha_2}\Big)+(1\leftrightarrow2)\right]\notag\\
	&\hspace{-2cm}+\mathcal{I}^{\mu_1\nu_1}_{\hspace{0.5cm}\alpha_1}(p_1)\mathcal{I}^{\mu_2\nu_2}_{\hspace{0.5cm}\alpha_2}(p_2)\left[-\frac{1}{2}\delta^{\alpha_1\alpha_2}\big((p_1\cdot p_2)^2-p_1^2p_2^2\big)\right]-\frac{2}{3}\bigg[\Pi^{\mu_1\nu_1}_{\alpha_1\beta_1}(p_1)\pi^{\mu_2\nu_2}(p_2)\,p_2^{\alpha_1}\,p_2^{\beta_1}+(1\leftrightarrow 2)\bigg]\notag\\
	&\hspace{-2cm}+\frac{1}{3}\bigg\{\bigg[\mathcal{I}^{\mu_1\nu_1}_{\hspace{0.5cm}\alpha_1}\pi^{\mu_2\nu_2}\bigg(-\frac{1}{2}p_2^{\alpha_1}\big(4(p_1\cdot p_2)+p_1^2\big)-p_1^{\alpha_1}\big((p_1\cdot p_2)+p_2^2\big)\bigg)\bigg]+[1\leftrightarrow2]\bigg\}\notag\\
	&\hspace{-2cm}-\frac{1}{3}\,\pi^{\mu_1\nu_1}(p_1)\pi^{\mu_2\nu_2}(p_2)\bigg[(p_1\cdot p_2)+2p_1^2+2p_2^2\bigg].
\end{align}

\bibliographystyle{jhep}

\begin{thebibliography}{10}
	
	\bibitem{Duff:1993wm}
	M.~J. Duff, {\it {Twenty years of the Weyl anomaly}},  {\em Class. Quant.
		Grav.} {\bf 11} (1994) 1387--1404,
	[\href{http://xxx.lanl.gov/abs/hep-th/9308075}{{\tt hep-th/9308075}}].
	
	\bibitem{Coriano:2017mux}
	C.~Corian\`o, M.~M. Maglio, and E.~Mottola, {\it {TTT in CFT: Trace Identities
			and the Conformal Anomaly Effective Action}},  {\em Nucl. Phys.} {\bf B942}
	(2019) 303--328, [\href{http://xxx.lanl.gov/abs/1703.0886}{{\tt
			arXiv:1703.0886}}].
	
	\bibitem{Coriano:2021nvn}
	C.~Corian\`o, M.~M. Maglio, and D.~Theofilopoulos, {\it {The Conformal Anomaly
			Action to Fourth Order (4T) in $d=4$ in Momentum Space}},
	\href{http://xxx.lanl.gov/abs/2103.1395}{{\tt arXiv:2103.1395}}.
	
	\bibitem{Osborn:1993cr}
	H.~Osborn and A.~C. Petkou, {\it {Implications of Conformal Invariance in Field
			Theories for General Dimensions}},  {\em Ann. Phys.} {\bf 231} (1994)
	311--362, [\href{http://xxx.lanl.gov/abs/hep-th/9307010}{{\tt
			hep-th/9307010}}].
	
	\bibitem{Erdmenger:1996yc}
	J.~Erdmenger and H.~Osborn, {\it {Conserved currents and the energy momentum
			tensor in conformally invariant theories for general dimensions}},  {\em
		Nucl.Phys.} {\bf B483} (1997) 431--474,
	[\href{http://xxx.lanl.gov/abs/hep-th/0103237}{{\tt hep-th/0103237}}].
	
	\bibitem{Coriano:2013jba}
	C.~Corian\`o, L.~Delle~Rose, E.~Mottola, and M.~Serino, {\it {Solving the
			Conformal Constraints for Scalar Operators in Momentum Space and the
			Evaluation of Feynman's Master Integrals}},  {\em JHEP} {\bf 1307} (2013)
	011, [\href{http://xxx.lanl.gov/abs/1304.6944}{{\tt arXiv:1304.6944}}].
	
	\bibitem{Bzowski:2013sza}
	A.~Bzowski, P.~McFadden, and K.~Skenderis, {\it {Implications of conformal
			invariance in momentum space}},  {\em JHEP} {\bf 03} (2014) 111,
	[\href{http://xxx.lanl.gov/abs/1304.7760}{{\tt arXiv:1304.7760}}].
	
	\bibitem{Bzowski:2020kfw}
	A.~Bzowski, P.~McFadden, and K.~Skenderis, {\it {Conformal correlators as
			simplex integrals in momentum space}},  {\em JHEP} {\bf 01} (2021) 192,
	[\href{http://xxx.lanl.gov/abs/2008.0754}{{\tt arXiv:2008.0754}}].
	
	\bibitem{Bzowski:2017poo}
	A.~Bzowski, P.~McFadden, and K.~Skenderis, {\it {Renormalised 3-point functions
			of stress tensors and conserved currents in CFT}},
	\href{http://xxx.lanl.gov/abs/1711.0910}{{\tt arXiv:1711.0910}}.
	
	\bibitem{Coriano:2022ftl}
	C.~Corian\`o, M.~M. Maglio, and D.~Theofilopoulos, {\it {Topological
			corrections and conformal backreaction in the Einstein
			Gauss\textendash{}Bonnet/Weyl theories of gravity at $D=4$}},  {\em Eur.
		Phys. J. C} {\bf 82} (2022), no.~12 1121,
	[\href{http://xxx.lanl.gov/abs/2203.0421}{{\tt arXiv:2203.0421}}].
	
	\bibitem{Giannotti:2008cv}
	M.~Giannotti and E.~Mottola, {\it {The Trace Anomaly and Massless Scalar
			Degrees of Freedom in Gravity}},  {\em Phys. Rev.} {\bf D79} (2009) 045014,
	[\href{http://xxx.lanl.gov/abs/0812.0351}{{\tt arXiv:0812.0351}}].
	
	\bibitem{Armillis:2009pq}
	R.~Armillis, C.~Corian\`{o}, and L.~Delle~Rose, {\it {Conformal Anomalies and
			the Gravitational Effective Action: The $TJJ$ Correlator for a Dirac
			Fermion}},  {\em Phys. Rev.} {\bf D81} (2010) 085001,
	[\href{http://xxx.lanl.gov/abs/0910.3381}{{\tt arXiv:0910.3381}}].
	
	\bibitem{Armillis:2010qk}
	R.~Armillis, C.~Corian\`o, and L.~Delle~Rose, {\it {Trace Anomaly, Massless
			Scalars and the Gravitational Coupling of QCD}},  {\em Phys. Rev.} {\bf D82}
	(2010) 064023, [\href{http://xxx.lanl.gov/abs/1005.4173}{{\tt
			arXiv:1005.4173}}].
	
	\bibitem{Donoghue:2015xla}
	J.~F. Donoghue and B.~K. El-Menoufi, {\it {QED trace anomaly, non-local
			Lagrangians and quantum Equivalence Principle violations}},  {\em JHEP} {\bf
		05} (2015) 118, [\href{http://xxx.lanl.gov/abs/1503.0609}{{\tt
			arXiv:1503.0609}}].
	
	\bibitem{Donoghue:2015nba}
	J.~F. Donoghue and B.~K. El-Menoufi, {\it {Covariant non-local action for
			massless QED and the curvature expansion}},  {\em JHEP} {\bf 10} (2015) 044,
	[\href{http://xxx.lanl.gov/abs/1507.0632}{{\tt arXiv:1507.0632}}].
	
	\bibitem{Coriano:2020ees}
	C.~Corian\`o and M.~M. Maglio, {\it {Conformal field theory in momentum space
			and anomaly actions in gravity: The analysis of three- and four-point
			function}},  {\em Phys. Rept.} {\bf 952} (2022) 2198,
	[\href{http://xxx.lanl.gov/abs/2005.0687}{{\tt arXiv:2005.0687}}].
	
	\bibitem{Coriano:2018bsy}
	C.~Corian\`o and M.~M. Maglio, {\it {The general 3-graviton vertex ($TTT$) of
			conformal field theories in momentum space in $d =4$}},  {\em Nucl. Phys.}
	{\bf B937} (2018) 56--134, [\href{http://xxx.lanl.gov/abs/1808.1022}{{\tt
			arXiv:1808.1022}}].
	
	\bibitem{Coriano:2018bbe}
	C.~Corian\`o and M.~M. Maglio, {\it {Exact Correlators from Conformal Ward
			Identities in Momentum Space and the Perturbative $TJJ$ Vertex}},  {\em Nucl.
		Phys.} {\bf B938} (2019) 440--522,
	[\href{http://xxx.lanl.gov/abs/1802.0767}{{\tt arXiv:1802.0767}}].
	
	\bibitem{Chernodub:2017jcp}
	M.~N. Chernodub, A.~Cortijo, and M.~A.~H. Vozmediano, {\it {A Nernst current
			from the conformal anomaly in Dirac and Weyl semimetals}},
	\href{http://xxx.lanl.gov/abs/1712.0538}{{\tt arXiv:1712.0538}}.
	
	\bibitem{Chernodub:2021nff}
	M.~N. Chernodub, Y.~Ferreiros, A.~G. Grushin, K.~Landsteiner, and M.~A.~H.
	Vozmediano, {\it {Thermal transport, geometry, and anomalies}},  {\em Phys.
		Rept.} {\bf 977} (2022) 1--58, [\href{http://xxx.lanl.gov/abs/2110.0547}{{\tt
			arXiv:2110.0547}}].
	
	\bibitem{Chernodub:2019tsx}
	M.~N. Chernodub, C.~Corian\`o, and M.~M. Maglio, {\it {Anomalous Gravitational
			TTT Vertex, Temperature Inhomogeneity, and Pressure Anisotropy}},  {\em Phys.
		Lett.} {\bf B802} (2020) 135236,
	[\href{http://xxx.lanl.gov/abs/1910.1372}{{\tt arXiv:1910.1372}}].
	
	\bibitem{Tutschku:2020rjq}
	C.~Tutschku, F.~S. Nogueira, C.~Northe, J.~van~den Brink, and E.~M. Hankiewicz,
	{\it {Temperature and chemical potential dependence of the parity anomaly in
			quantum anomalous Hall insulators}},  {\em Phys. Rev. B} {\bf 102} (2020),
	no.~20 205407, [\href{http://xxx.lanl.gov/abs/2007.1185}{{\tt
			arXiv:2007.1185}}].
	
	\bibitem{Fruchart:2013tza}
	M.~Fruchart and D.~Carpentier, {\it {An introduction to topological
			insulators}},  {\em Comptes Rendus Physique} {\bf 14} (2013) 779--815,
	[\href{http://xxx.lanl.gov/abs/1310.0255}{{\tt arXiv:1310.0255}}].
	
	\bibitem{Arouca:2022psl}
	R.~Arouca, A.~Cappelli, and T.~H. Hansson, {\it {Quantum Field Theory Anomalies
			in Condensed Matter Physics}},  \href{http://xxx.lanl.gov/abs/2204.0215}{{\tt
			arXiv:2204.0215}}.
	
	\bibitem{Landsteiner:2013sja}
	K.~Landsteiner, {\it {Anomalous transport of Weyl fermions in Weyl
			semimetals}},  {\em Phys. Rev.} {\bf B89} (2014), no.~7 075124,
	[\href{http://xxx.lanl.gov/abs/1306.4932}{{\tt arXiv:1306.4932}}].
	
	\bibitem{Mottola:2019nui}
	E.~Mottola and A.~V. Sadofyev, {\it {Chiral Waves on the Fermi-Dirac Sea:
			Quantum Superfluidity and the Axial Anomaly}},
	\href{http://xxx.lanl.gov/abs/1909.0197}{{\tt arXiv:1909.0197}}.
	
	\bibitem{Luttinger:1964zz}
	J.~M. Luttinger, {\it {Theory of Thermal Transport Coefficients}},  {\em Phys.
		Rev.} {\bf 135} (1964) A1505--A1514.
	
	\bibitem{Capozziello:2021krv}
	S.~Capozziello and F.~Bajardi, {\it {Nonlocal gravity cosmology: An overview}},
	{\em Int. J. Mod. Phys. D} {\bf 31} (2022), no.~06 2230009,
	[\href{http://xxx.lanl.gov/abs/2201.0451}{{\tt arXiv:2201.0451}}].
	
	\bibitem{Belgacem:2019lwx}
	E.~Belgacem, Y.~Dirian, A.~Finke, S.~Foffa, and M.~Maggiore, {\it {Nonlocal
			gravity and gravitational-wave observations}},  {\em JCAP} {\bf 11} (2019)
	022, [\href{http://xxx.lanl.gov/abs/1907.0204}{{\tt arXiv:1907.0204}}].
	
	\bibitem{Bzowski:2019kwd}
	A.~Bzowski, P.~McFadden, and K.~Skenderis, {\it {Conformal 4-point functions in
			momentum space}},  \href{http://xxx.lanl.gov/abs/1910.1016}{{\tt
			arXiv:1910.1016}}.
	
	\bibitem{Coriano:2019nkw}
	C.~Corian\`o, M.~M. Maglio, and D.~Theofilopoulos, {\it {Four-Point Functions
			in Momentum Space: Conformal Ward Identities in the Scalar/Tensor case}},
	\href{http://xxx.lanl.gov/abs/1912.0190}{{\tt arXiv:1912.0190}}.
	
	\bibitem{Caloro:2022zuy}
	F.~Caloro and P.~McFadden, {\it {Shift operators from the simplex
			representation in momentum-space CFT}},
	\href{http://xxx.lanl.gov/abs/2212.0388}{{\tt arXiv:2212.0388}}.
	
	\bibitem{Bautista:2019qxj}
	T.~Bautista and H.~Godazgar, {\it {Lorentzian CFT 3-point functions in momentum
			space}},  \href{http://xxx.lanl.gov/abs/1908.0473}{{\tt arXiv:1908.0473}}.
	
	\bibitem{Gillioz:2019lgs}
	M.~Gillioz, {\it {Conformal 3-point functions and the Lorentzian OPE in
			momentum space}},  \href{http://xxx.lanl.gov/abs/1909.0087}{{\tt
			arXiv:1909.0087}}.
	
	\bibitem{Gillioz:2018mto}
	M.~Gillioz, {\it {Momentum-space conformal blocks on the light cone}},
	\href{http://xxx.lanl.gov/abs/1807.0700}{{\tt arXiv:1807.0700}}.
	
	\bibitem{Arkani-Hamed:2018kmz}
	N.~Arkani-Hamed, D.~Baumann, H.~Lee, and G.~L. Pimentel, {\it {The Cosmological
			Bootstrap: Inflationary Correlators from Symmetries and Singularities}},
	\href{http://xxx.lanl.gov/abs/1811.0002}{{\tt arXiv:1811.0002}}.
	
	\bibitem{Arkani-Hamed:2017fdk}
	N.~Arkani-Hamed, P.~Benincasa, and A.~Postnikov, {\it {Cosmological Polytopes
			and the Wavefunction of the Universe}},
	\href{http://xxx.lanl.gov/abs/1709.0281}{{\tt arXiv:1709.0281}}.
	
	\bibitem{Baumann:2020dch}
	D.~Baumann, C.~Duaso~Pueyo, A.~Joyce, H.~Lee, and G.~L. Pimentel, {\it {The
			Cosmological Bootstrap: Spinning Correlators from Symmetries and
			Factorization}},  \href{http://xxx.lanl.gov/abs/2005.0423}{{\tt
			arXiv:2005.0423}}.
	
	\bibitem{Benincasa:2022gtd}
	P.~Benincasa, {\it {Amplitudes meet Cosmology: A (Scalar) Primer}},
	\href{http://xxx.lanl.gov/abs/2203.1533}{{\tt arXiv:2203.1533}}.
	
	\bibitem{Shapiro:2008sf}
	I.~L. Shapiro, {\it {Effective Action of Vacuum: Semiclassical Approach}},
	{\em Class. Quant. Grav.} {\bf 25} (2008) 103001,
	[\href{http://xxx.lanl.gov/abs/0801.0216}{{\tt arXiv:0801.0216}}].
	
	\bibitem{Asorey:2022ebz}
	M.~Asorey, W.~C.~e. Silva, I.~L. Shapiro, and P.~R. B.~d. Vale, {\it {Trace
			anomaly and induced action for a metric-scalar background}},
	\href{http://xxx.lanl.gov/abs/2202.0015}{{\tt arXiv:2202.0015}}.
	
	\bibitem{Edgar:2001vv}
	S.~B. Edgar and A.~Hoglund, {\it {Dimensionally dependent tensor identities by
			double antisymmetrization}},  {\em J. Math. Phys.} {\bf 43} (2002) 659--677,
	[\href{http://xxx.lanl.gov/abs/gr-qc/0105066}{{\tt gr-qc/0105066}}].
	
	\bibitem{lovelock_1970}
	D.~Lovelock, {\it Dimensionally dependent identities},  {\em Mathematical
		Proceedings of the Cambridge Philosophical Society} {\bf 68} (1970), no.2,
	345-350.
	
	\bibitem{Coriano:2018zdo}
	C.~Corian\`o and M.~M. Maglio, {\it {Renormalization, Conformal Ward Identities
			and the Origin of a Conformal Anomaly Pole}},  {\em Phys. Lett.} {\bf B781}
	(2018) 283--289, [\href{http://xxx.lanl.gov/abs/1802.0150}{{\tt
			arXiv:1802.0150}}].
	
	\bibitem{Barvinsky:1995it}
	A.~O. Barvinsky, A.~G. Mirzabekian, and V.~V. Zhytnikov, {\it {Conformal
			decomposition of the effective action and covariant curvature expansion}},
	in {\em {6th Moscow Quantum Gravity}}, 6, 1995.
	\newblock \href{http://xxx.lanl.gov/abs/gr-qc/9510037}{{\tt gr-qc/9510037}}.
	
	\bibitem{Fradkin:1978yw}
	E.~S. Fradkin and G.~A. Vilkovisky, {\it {Conformal Off Mass Shell Extension
			and Elimination of Conformal Anomalies in Quantum Gravity}},  {\em Phys.
		Lett. B} {\bf 73} (1978) 209--213.
	
	\bibitem{Riegert:1984kt}
	R.~J. Riegert, {\it {A Nonlocal Action for the Trace Anomaly}},  {\em Phys.
		Lett.} {\bf 134B} (1984) 56--60.
	
	\bibitem{Stergiou:2022qqj}
	A.~Stergiou, G.~P. Vacca, and O.~Zanusso, {\it {Weyl covariance and the energy
			momentum tensors of higher-derivative free conformal field theories}},  {\em
		JHEP} {\bf 06} (2022) 104, [\href{http://xxx.lanl.gov/abs/2202.0470}{{\tt
			arXiv:2202.0470}}].
	
	\bibitem{Brust:2016gjy}
	C.~Brust and K.~Hinterbichler, {\it {Free \ensuremath{\square}$^{k}$ scalar
			conformal field theory}},  {\em JHEP} {\bf 02} (2017) 066,
	[\href{http://xxx.lanl.gov/abs/1607.0743}{{\tt arXiv:1607.0743}}].
	
	\bibitem{Nesterov:2010jh}
	D.~Nesterov and S.~N. Solodukhin, {\it {Short-distance regularity of Green's
			function and UV divergences in entanglement entropy}},  {\em JHEP} {\bf 09}
	(2010) 041, [\href{http://xxx.lanl.gov/abs/1008.0777}{{\tt
			arXiv:1008.0777}}].
	
\end{thebibliography}
\providecommand{\href}[2]{#2}\begingroup\raggedright\endgroup
\end{document}